\documentclass[twocolumn,showpacs,preprintnumbers,
               superscriptaddress,floatfix]{revtex4}
\usepackage{epsf}

\def\simm#1{\mathop{\vtop{\ialign{##\crcr
        $\hfil\displaystyle{#1}\hfil$\crcr\noalign{\kern0.5pt\nointerlineskip}
        $\sim$\crcr\noalign{\kern0.5pt}}}}\limits}
\def\lsim{\raise0.3ex\hbox{$<$\kern-0.75em\raise-1.1ex\hbox{$\sim$}}}
\def\gsim{\raise0.3ex\hbox{$>$\kern-0.75em\raise-1.1ex\hbox{$\sim$}}}

\begin{document}

\preprint{UTHEP-470}
\preprint{UTCCP-P-137}
\preprint{YAMAGATA-HEP-03-28}

\title{Phase structure of lattice QCD 
for general number of flavors}

\newcommand{\Tsukuba}%
{Institute of Physics, University of Tsukuba, 
 Tsukuba, Ibaraki 305-8571, Japan}

\newcommand{\CCP}%
{Center for Computational Physics, University of Tsukuba, 
 Tsukuba, Ibaraki 305-8577, Japan}

\newcommand{\Yamagata}%
{Faculty of Education, Yamagata University, 
 Yamagata 990-8560, Japan}

\author{Y.~Iwasaki}
\affiliation{\Tsukuba}
\affiliation{\CCP}

\author{K.~Kanaya}
\affiliation{\Tsukuba}
\affiliation{\CCP} 

\author{S.~Kaya\footnote{Present address: Information and Mathematical 
Laboratory, Inc., Tokyo 171-0014, Japan}}
\affiliation{\Tsukuba} 

\author{S. Sakai}
\affiliation{\Yamagata} 

\author{T.~Yoshi\'{e}}
\affiliation{\Tsukuba}
\affiliation{\CCP}

\date{\today}


\begin{abstract}
We investigate the phase structure of lattice QCD for the general
number of flavors in the parameter space of gauge coupling constant
and quark mass,
employing the one-plaquette gauge action and the standard Wilson quark
action.
Performing a series of simulations for the number of flavors $N_F=6$--360
with degenerate-mass quarks,
we find that when $N_F \ge 7$
there is a line of a bulk first order phase transition between the
confined phase
and a deconfined phase at a finite current quark mass
in the strong coupling region and the intermediate coupling region.
The massless quark line exists only in the deconfined phase.

Based on these numerical results
in the strong coupling limit and in the intermediate coupling region,
we propose the following phase structure,
depending on the number of flavors whose masses
are less than $\Lambda_d$ which is
the physical scale characterizing the phase transition in the
weak coupling region:
When $N_F \ge 17$, there is only
a trivial IR fixed point and therefore the theory in the continuum limit
is free. On the other hand, when $16 \ge N_F \ge 7$,
there is a non-trivial IR fixed point and therefore the theory is
non-trivial with anomalous dimensions, however, without quark confinement.
Theories which satisfy both quark confinement and spontaneous
chiral symmetry breaking in the continuum limit
exist only for $N_F \le 6$.
\end{abstract}

\maketitle

\section{Introduction}
\label{sec:introduction}
The fundamental properties of QCD are quark confinement,
asymptotic freedom and spontaneous breakdown of chiral symmetry.
Among them, the asymptotic freedom is lost
when the number of flavors exceeds  $16 \frac{1}{2}$.
Thus the question which naturally arises is what
is the constraint on the number of flavors for
quark confinement and/or the spontaneous breakdown of chiral symmetry.

As asymptotic freedom is the nature at short distances, one can apply
the perturbation theory to investigate the critical number for it.
However, because quark confinement and spontaneous chiral symmetry breaking
are due to non-perturbative effects, one has to apply a non-perturbative
method throughout the investigation of the critical numbers for them.
We employ lattice QCD for the investigation in this work,
since lattice QCD is the only known theory of QCD which is constructed
non-perturbatively.

Lattice QCD is a theory with fundamental parameters, the gauge coupling
constant $g$ and quark masses $m_q$, defined on a lattice with lattice
spacing $a$. The inverse of the lattice spacing $a^{-1}$ plays a role of
an UV cutoff.
In order to investigate properties of the theory in the
continuum limit, one has to first clarify the phase structure of lattice 
QCD at zero temperature for general number of flavors, and
then identify an UV fixed point and/or
an IR fixed point of an RG (Renormalization Group) transformation.
When the existence of such fixed points is established,
one is able to conclude what kind of theory exists in the continuum limit. 

In our previous work \cite{previo},
it was shown that, even in the strong coupling limit, 
when the number of flavors $N_F$ is greater than or equal to seven,
quarks are deconfined and chiral symmetry is restored at zero temperature,
if the quark mass is lighter than a critical value which is of order $a^{-1}$.

In this work we extend the region of the gauge coupling constant 
to weaker ones, and investigate the phase diagram for general
number of flavors $N_F$ at zero temperature.
We mainly consider the case where $N_F$ quark masses are degenerate. 
However, we also discuss the non-degenerate case.

We employ the simplest form of the action in this work, that is,
the Wilson quark action and the standard one-plaquette gauge
action.
At finite lattice spacings, 
the Wilson quark action is not chirally symmetric even at vanishing bare
quark mass due to the Wilson term which is added to a naive discretized
Dirac action in order to lift the doublers. Because of the fact that
the action does not hold chiral symmetry, the phase diagram becomes 
complicated in general.

As the phase diagram turns out to be complicated on the lattice,
we first give a brief summary in terms of the $\beta$ function in 
Sec.~\ref{sec:betafunc}. 
Further, the phase diagram for general number of flavors
when the theory would be chirally symmetric is conjectured 
in Sec.~\ref{sec:generic}. This conjecture is based on our proposal
for the phase diagram for the lattice action we employ, and is given 
because it may help the reader to understand the whole structure of
the phase diagram.
After showing the brief summary in this way,
the action we employ is given with
some basic notations in Sec.~\ref{sec:fundamentals}.
Here some important facts related to chiral property of the quark are
also discussed.
Then the main part of the paper, our proposal for the phase structure
for the general number of flavors in the case of the Wilson quark action,
is summarized in Sec.~\ref{sec:phasediagram}, before giving the 
detailed numerical results which lead to our proposal.
We also comment on previous results obtained with the staggered quark action.
After giving numerical parameters for our simulations in 
Sec.~\ref{sec:parameters}, numerical results in the strong coupling limit 
and at finite coupling constants
are given in Secs.~\ref{sec:strong} and \ref{sec:finite}, respectively.
All the investigations up to this point are limited to the degenerate quark
mass case. 
In Sec.~\ref{sec:non-degenerate}, we extend the study to the 
non-degenerate case to discuss implication for physics.
We give a summary in Sec.~\ref{sec:conclusions}. 
In an Appendix, we study the case of $N_c=2$ in the strong coupling limit. 
Preliminary results of our study have been presented 
in Refs.~\cite{lat94,report97}.

\section{Phase structure in terms of beta function}
\label{sec:betafunc}

The perturbative beta function of QCD with $N_F$ flavors of massless quarks
is universal up to two-loop order:
\begin{equation}
\tilde{\beta}(g) = - b_0 g^3 - b_1 g^5 + \cdots,
\end{equation}
where
\begin{equation}
b_0 = \frac{1}{16\pi^2} \left( 11-\frac{2}{3}N_F \right)
\end{equation}
and
\begin{equation}
b_1 = \frac{1}{(16\pi^2)^2} \left( 102-\frac{38}{3}N_F \right).
\end{equation}
The coefficient $b_0$ changes its sign at $N_F = 16\frac{1}{2}$.
Hence, for $N_F \ge 17$,
asymptotic freedom is lost and the point $g=0$ is an IR fixed point.
The theory governed by this IR fixed point is a free theory 
and the quark is not confined.
The second coefficient $b_1$ changes its sign
at $N_F \approx 8.05$.
This implies,
if one would
take the two-loop form of the beta-function, 
that an IR fixed point appears at finite coupling constant
for $9 \le N_F \le 16$.
Of course we cannot trust the two-loop form at finite coupling constants.
However, since the IR fixed point from the two-loop beta function locates 
in a weak coupling region for $N_F \sim 16$, it is plausible 
that the beta function has a non-trivial
IR fixed point for $N' \le N_F \le 16$ with some $N' \le 16$.
When such an IR fixed point exists, the coupling constant cannot become
arbitrarily large in the IR region.
This implies that quarks are not confined.
The long distance behavior of the theory is governed by the non-trivial IR 
fixed point: It is a theory with an anomalous dimension.

\begin{figure}[tb]
\centerline{
a)\epsfxsize=4cm\epsfbox{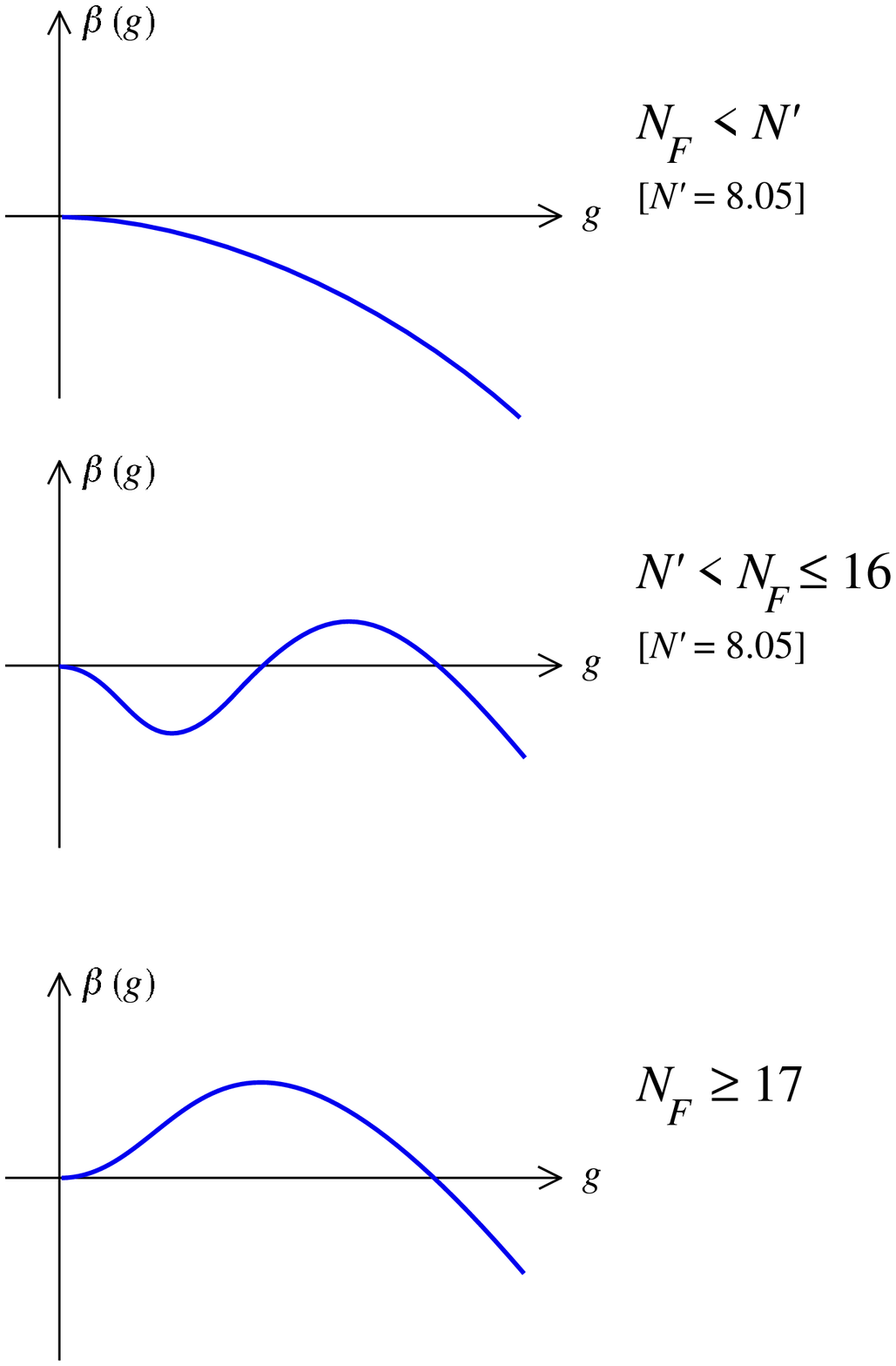}
b)\epsfxsize=4cm\epsfbox{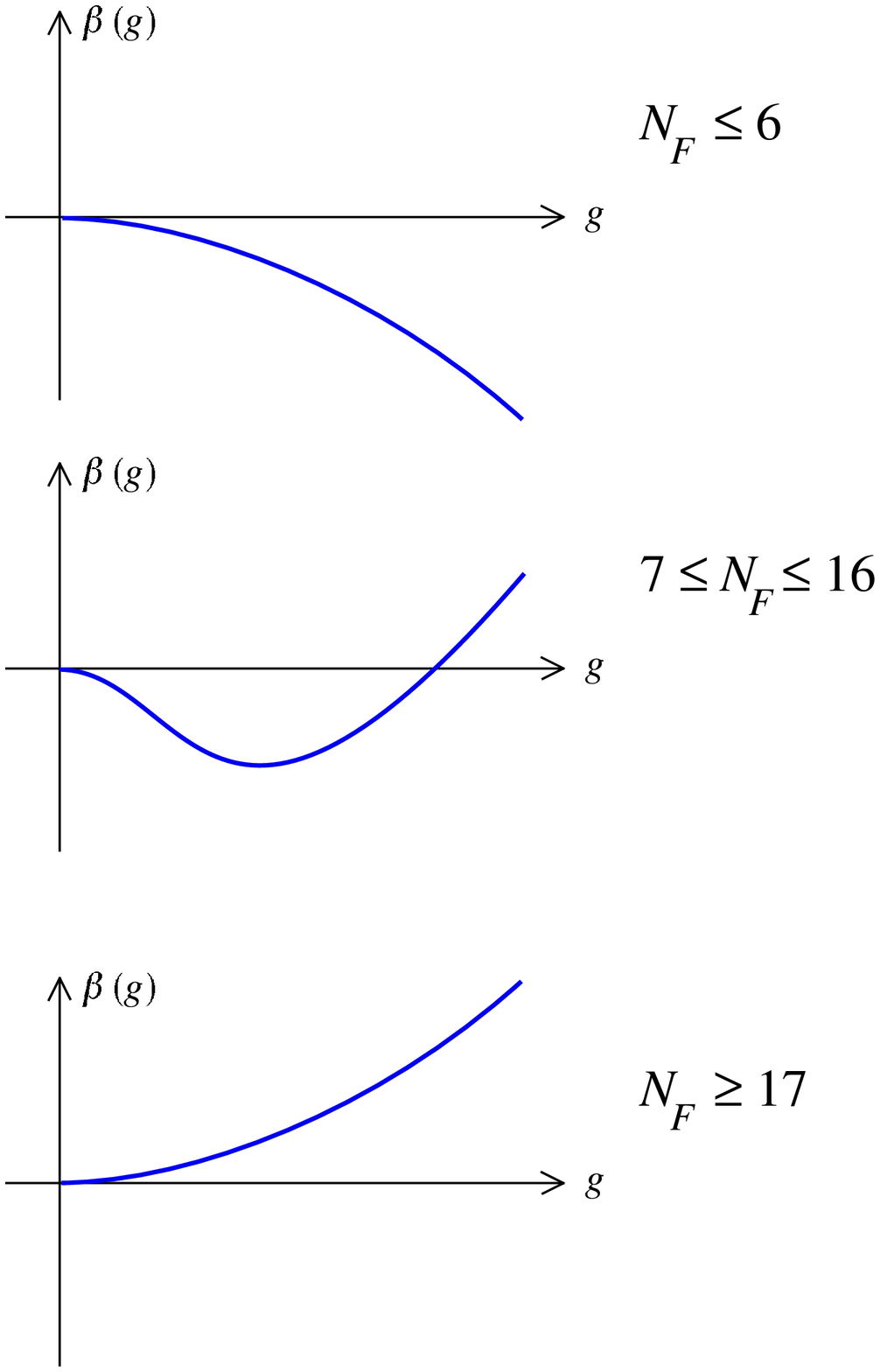}
}
\vspace{-0.1cm}
\caption{Renormalization group beta function.
(a) Conjecture by Banks and Zaks \protect\cite{Banks1}
assuming confinement in the strong coupling limit for all $N_F$.
(b) Our conjecture deduced from the results of lattice simulations.
}
\label{BetaFunc}
\end{figure}

A pioneering study on the $N_F$ dependence of the QCD vacuum was
made by Banks and Zaks in 1982 \cite{Banks1}.
Based on the result of the quark confinement
in a pure gauge theory~\cite{Kogut79},
they assumed that the quark is confined
and that the  beta function is negative in the strong
coupling limit for all $N_F$.
Using the perturbative results mentioned above,
they conjectured Fig.~\ref{BetaFunc}(a) as the simplest $N_F$ dependence
of the beta function,
and studied the phase structure of QCD based on this $N_F$
dependence of the beta function.
Because of an additional non-trivial UV fixed point for $N_F > N'$,
their conjecture for the phase structure is complicated.

In the argument of Banks and Zaks,
the assumption of confinement and negative beta function in the strong
coupling limit plays an essential role.
However, 
there exist no proofs of confinement in QCD for general $N_F$ even
in the strong coupling limit.
A non-perturbative investigation on the lattice is required.
Therefore, in our previous work \cite{previo},
we performed numerical simulations of QCD
in the strong coupling limit for various $N_F$,
using the Wilson fermion formalism for lattice quarks.
We found that, when $N_F \ge 7$,
quarks are deconfined and chiral
symmetry is restored at zero temperature
even in the strong coupling limit, when the quark mass is lighter
than a critical value.

In some previous literatures, quark confinement in the strong
coupling limit has been assumed explicitly or implicitly. However,
the arguments for the confinement in the strong coupling limit are 
based on either the large $N_c$ limit \cite{SCENc},
a meanfield approximation ($1/d$ expansion) \cite{SCEmf,SCEmfMq},
or a heavy quark mass expansion \cite{SCEmfMq,wilson77}.
Because we are interested in the theory
with dynamical quarks with $N_c$ kept 3, and
effects of dynamical quarks become significant only when they are light,
the results of these approximations cannot be applied.
Actually, our numerical result at $g=\infty$ explicitly shows 
violation of quark confinement at small quark masses when $N_F\geq7$.

Here we extend the study to weaker couplings.
Based on the numerical results obtained on the lattice
(see the following sections for details) 
combined with the results of the perturbation theory,
we conjecture Fig.~\ref{BetaFunc}(b) for the $N_F$ dependence of the
beta function:
When $N_F \le 6$,
the beta function is negative for all values of $g$.
Quarks are confined and the chiral symmetry is spontaneously broken
at zero temperature.
(Corresponding critical $N_F$ is 2 for the case $N_c=2$ \cite{lat94}.
See the discussion in the Appendix.)
On the other hand, when $N_F$ is equal or larger than 17,
we conjecture that the beta function is positive for all $g$,
in contrast to
the conjecture by Banks and Zaks shown in Fig.~\ref{BetaFunc}(a).
The theory is trivial in this case.
When $N_F$ is between 7 and 16, the beta function changes
sign from negative to positive with increasing $g$.
Therefore, the theory has a non-trivial IR fixed point.

Here we note that there are several related works which do not use lattice 
formulation \cite{Oehme,Nishijima,Appelquist,sigma}.
These works gave interesting results for the quark confinement condition. 
However, the critical
number for the confinement is at most suggestive, because we need
fully non-perturbative method to investigate the problem.
That is, the theory should be constructed from the beginning non-perturbatively.
In this regards, lattice QCD is the only known theory constructed
non-perturbatively.

\section{Conjecture for the Phase diagram in Chirally Symmetric Case}
\label{sec:generic}

\begin{figure}[tb]
\centerline{
a) \epsfxsize=7.5cm\epsfbox{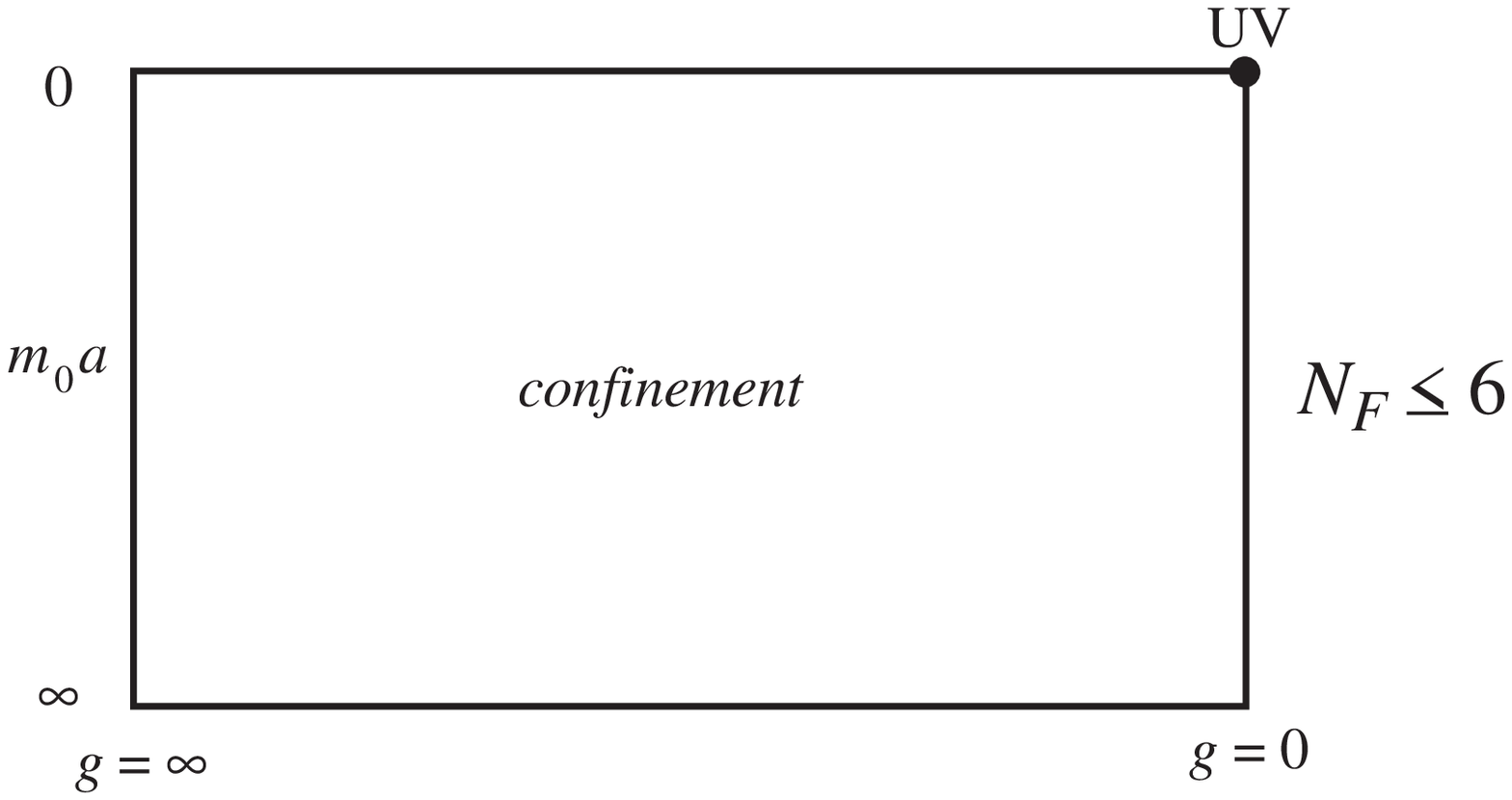}
\hspace*{2.5mm}
}
\vspace{2mm} 
\centerline{
b) \epsfxsize=7.8cm\epsfbox{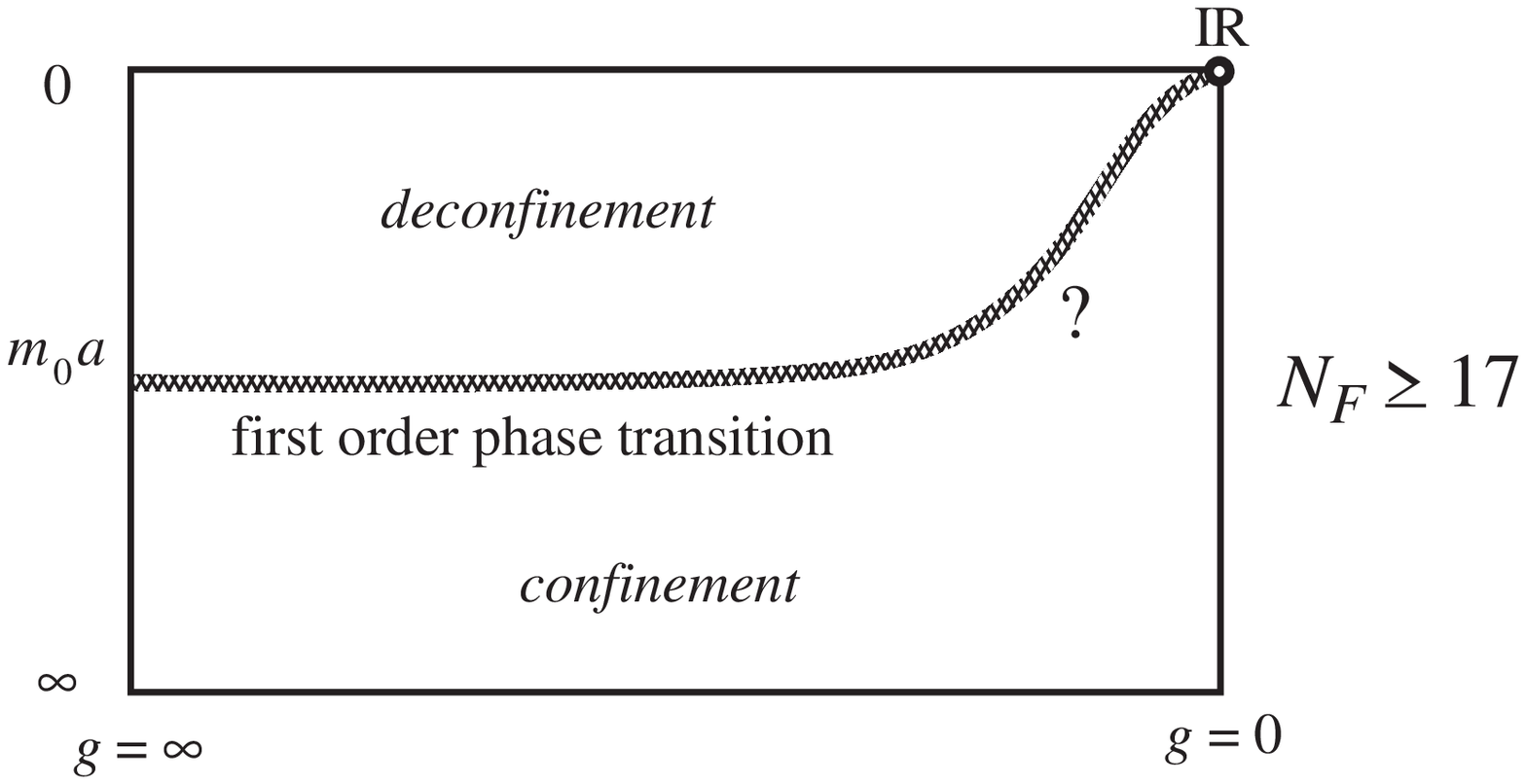}
}
\vspace{2mm} 
\centerline{
c) \epsfxsize=7.8cm\epsfbox{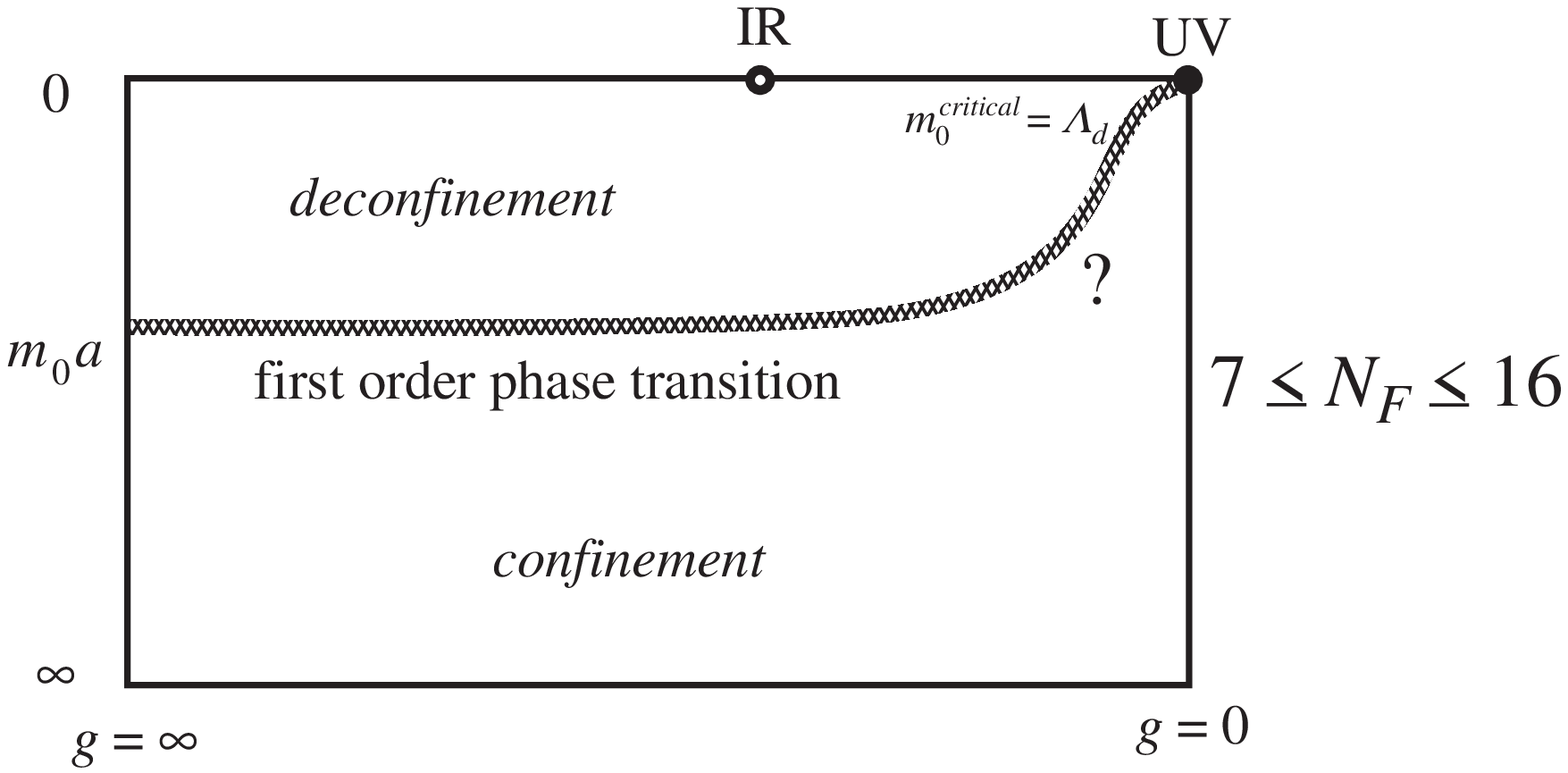}
}
\vspace{-0.1cm}
\caption{Phase diagram for a chirally symmetric case.
(a) $N_F \le 6$, (b) $N_F \ge 17$, (c) $7\le N_F \le 16$.
}
\label{chisym}
\end{figure}

In this section, we present our conjecture for the phase diagram of QCD, 
for the case when the theory is
chiral symmetric. The conjecture is based on the phase diagram we propose
for the case of the Wilson fermion action, where the chiral symmetry is
violated. The reason why we make this kind of conjecture is
that it may help the reader to understand the phase diagram of
the Wilson fermion action which is more complicated.
The phase diagrams we conjecture in the $g-m_0$ plane are presented in 
Fig.~\ref{chisym}.

\subsection{$N_F \le 6$}
The phase diagram is simple. At zero temperature all the region is in
the confined phase. See Fig.~\ref{chisym}(a).

\subsection{$N_F \ge 17$}

At zero temperature,
there is a first-order phase transition
which separates a deconfined phase from the confined phase.
The transition occurs at $m_0^{critical}$ 
which is of the same order of magnitude 
as the inverse of a typical correlation length $\xi$ of gluon dynamics,
such as the plaquette-plaquette correlation length.

In the confined phase, the nature of the system is essentially same as that 
for $N_F \leq 6$:
When $m_0 \gg \xi^{-1}$, dynamical effects by quark loops can be neglected.
In this case, the system is equivalent to the quenched QCD.
That is, for $m_0 \gg \xi^{-1}$ the phase is the confined phase.
When dynamical effects of quark loops change the phase from that
in the quenched case, there will be a phase transition in general,
and this kind of the transition occurs at $m_0 \sim \xi^{-1}$.

When the gauge coupling decreases towards zero, the dimension-less
correlation length $\xi/a$ 
increases and diverges at zero coupling constant in the confined
phase. Therefore, 
along the phase transition line, the dimension-less 
critical quark mass should behave as $m_0^{critical} a=\Lambda_d a$.
Here $a$ vanishes towards $g=0$, and $\Lambda_d \sim \xi^{-1}$ is
the physical scale which characterizes the phase transition 
in the weak coupling region.

The ``?'' mark in the weak coupling region means that
we have performed numerical simulation
for only the strong and intermediate coupling regions
because of technical reason. In this sense, our proposal for 
the strong and intermediate coupling regions is based
on our numerical results, while our conjecture in the weak region is 
based on the assumption 
that the transition occurs at $m_0^{critical} \sim \xi^{-1}$.

In the upper region above the phase transition line quarks are not
confined.
There is only an IR fixed point at $g=0$. That is, the theory in the
continuum limit is free.

We would like to make a comment that there is
an alternative possibility that the transition occurs 
at $m_0^{critical} \sim a^{-1}$,
where $a^{-1}$ is the inverse of the lattice spacing.
In the strong and intermediate coupling regions $\xi^{-1}$ is of order
$a^{-1}$. Therefore, there is no essential difference between the two 
possibilities whether  $m_0^{critical} \sim a^{-1}$ or
$m_0^{critical} \sim \xi^{-1}$.

In the usual argument of the decoupling theorem for QCD, the effect
of the particle
whose mass is much heavier than the QCD scale parameter $\Lambda_{QCD}$ 
can be absorbed by renormalization of physical quantities.
Based on a similar consideration, one may argue that
the particles whose masses are much heavier than $\Lambda_{QCD}$ are 
irrelevant for quark confinement.
This is the reason why we assume that
the transition occurs at $m_0^{critical} \sim \xi^{-1}$.
However, whether a theory exists in a sense of constructive field 
theory is a different problem.
Therefore, we think that it is not possible  
to disregard the alternative possibility of $m_0^{critical} \sim a^{-1}$ 
only from a theoretical argument. 

Nonetheless we think that the assumption that 
the transition occurs at $m_0^{critical} \sim \xi^{-1}$ is more plausible 
than the assumption $m_0^{critical} \sim a^{-1}$.
In the following we assume that the phase transition occurs
at $m_0^{critical} \sim \xi^{-1}$. This is our conjecture.

However, we also refer to the alternative 
possibility of $m_0^{critical} \sim a^{-1}$
when it is appropriate to refer to.
In this alternative case,
the transition point at $g=0$ is off the massless point 
and therefore all finite massive quarks belong to the 
deconfined phase.

It will be possible to see in future which of $m_0^{critical} \sim a^{-1}$ 
or $m_0^{critical} \sim \xi^{-1}$ is realized
by numerical simulations.

\subsection{$7 \le N_F \le 16$}
The phase diagram at zero temperature shown in Fig.~\ref{chisym}(c) is similar
to that in the case of $N_F \ge 17$. 
The phase transition line locates around $m_0 \sim \xi^{-1}$,
due to the same reason as in the case of $N_F \ge 17$.

However, the flow of RG transformation
is different. The $g=0$ point in this case is an UV fixed point, 
contrary to the case $N_F \ge 17$. 
On the other hand, 
quarks are deconfined in the upper region above the phase 
transition line as in the case of $N_F \ge 17$.
Therefore there should be an IR fixed point at finite coupling constant.
Otherwise gauge  coupling would become arbitrary large at large distance,
and quarks would be confined in the strong coupling limit contrary to the
result of the numerical simulation.

\section{Fundamentals of Lattice QCD and the Action}
\label{sec:fundamentals}

Lattice QCD is defined on a hypercubic lattice in 4-dimensional euclidean space
with lattice spacing $a$.
A site is denoted by a vector $n=(n_1,n_2,n_3,n_4)$,
where $n_i$'s are integers.
A link with end points at the sites $n$ and $n+\hat\mu$
is specified by a pair $(n,\mu)$,
where $\hat\mu$ denotes a unit vector in the $\mu$ direction.

\subsection{Action}
\label{sec:action}

The gauge variable $U_{n,\mu}$ which is an element of SU(3) gauge group
is defined on the link $(n,\mu)$.
The action for gluons which we adopt is given by
\begin{equation}
S_{gauge}=\frac{1}{g^2} \sum_{n,\mu\neq\nu}{\rm Tr}
(U_{n,\mu}U_{n+\hat\mu,\nu}
U_{n+\hat\mu+\hat\nu,\mu}^{\dag} U_{n+\hat\nu,\nu}^{\dag}),
\label{eqn:gaction}
\end{equation}
where g is the gauge coupling constant.
This action is called the standard one-plaquette gauge action.
We usually use, instead of the bare gauge coupling constant $g$,
$\beta$ defined by
\begin{equation}
\beta=\frac{6}{g^2}.
\end{equation}

The quark variable is a Grassmann number defined on each site.
It is well-known that a naive discretization of the Dirac action
\begin{eqnarray} 
S_{fermion}&=&
\frac{a^3}{2}\sum_{n,\mu}(\bar{q}_{n}\gamma_{\mu}q_{n+\hat{\mu}}
-\bar{q}_{n+\hat{\mu}}\gamma_{\mu}q_{n})
\nonumber\\ 
&&+m_0a^4\sum_n \bar{q}_n q_n,
\label{eqn:naive}
\end{eqnarray}
with $m_0$ being the bare fermion mass,
leads to $16$ poles instead of one pole. This problem is called the species
doubling. To avoid this problem K. Wilson proposed to add
a dimension 5 operator called the Wilson term
\begin{equation}
-a^3\sum_{n,\mu}(\bar{q}_{n}q_{n+\hat{\mu}}
-2\bar{q}_n q_n +\bar{q}_{n+\hat{\mu}}q_{n}),
\label{eq:WilsonTerm}
\end{equation}
to the naive discretized action. 
Rescaling the fermion field by
$q=\sqrt{2K}\psi$ and making the action gauge invariant,
we obtain the Wilson fermion action
\begin{eqnarray}
S_{Wilson}&=&a^3\sum_{n}[\bar{\psi}_n \psi_n
-K\{\bar{\psi}_{n}(1-\gamma_{\mu})U_{n,\mu}\psi_{n+\hat{\mu}} 
\nonumber\\
&&+\bar{\psi}_{n+\hat{\mu}}(1+\gamma_{\mu})U_{n,\mu}^{\dag}\psi_{n}\}].
\label{eq:Wilson}
\end{eqnarray}
where
\begin{equation}
K=\frac{1}{2(m_0a+4)},
\end{equation}
which is called the hopping parameter.

The full action $S$ is given by the sum of the gauge part
$S_{gauge}$ and the fermion part $S_{Wilson}$,
\begin{equation}
S=S_{gauge}+ \sum_{f=1}^{N_F} S_{Wilson},
\end{equation}
where $f$ is for flavors.

The expectation value of an operator ${\cal O}(U,\psi,\bar \psi)$ is
given by
\begin{equation}
\langle {\cal O} \rangle 
=\frac{1}{Z}\int \prod_{n,\mu}{\rm d}U_{n,\mu}
\prod_{n,f}{\rm d}\psi_n^{(f)}{\rm d}\bar{\psi}_n^{(f)} {\cal O}(U,\psi,\bar \psi)
{\rm exp}(S),
\end{equation}
where $Z$ is the partition function
\begin{equation}
Z=\int \prod_{n,\mu}{\rm d}U_{n,\mu}
\prod_{n,f}{\rm d}\psi_n^{(f)}{\rm d}\bar{\psi}_n^{(f)}
{\rm exp}(S),
\end{equation}
with ${\rm d} U_{n,\mu}$ being the Haar measure of SU(3).

\subsection{Fundamental parameters}
In the case of degenerate $N_F$ flavors,
lattice QCD contains two parameters:
the gauge coupling constant $\beta =6/g^2$
and the bare quark mass or the hopping parameter $K=1/(m_0a+4)$.
In the non-degenerate case, we have, in general, $N_F$ independent
bare masses (hopping parameters).

In this work we consider mainly the case where $N_F$ quark masses are
degenerate, because it is simpler. However, the conjecture can be extended 
to non-degenerate cases. We will discuss this point 
in Sec.~\ref{sec:non-degenerate}.

\subsection{Continuum limit}
As mentioned in the Introduction, 
in order to see
what kind of theory in the continuum limit exists,
one has to investigate
the phase structure and identify UV fixed points and IR fixed points.

When the $g=0$ point is an UV fixed point, the theory governed by 
this fixed point is an asymptotically free theory.
On the other hand,
when the $g=0$ point is an IR fixed point, the theory
governed by this fixed point is
a free theory.
When there is a IR fixed point at finite coupling constant $g$, the theory
is a non-trivial theory with the long distance behavior determined
by the IR fixed point.

\subsection{Finite temperatures}
At finite temperatures the linear extension in the time direction $N_t$
is much smaller than those in the spatial directions ($N_x, N_y, N_z$).
The temperature $T$ is given by
$1/N_t a.$
For gluons the periodic boundary condition is imposed, while
for quarks an anti-periodic boundary condition is imposed in the
time direction.

Although we are interested in the phase structure of lattice QCD 
at zero temperature, 
we investigate phase transitions on lattices at finite temperatures.
Increasing the value of $N_t$, we carefully examine the $N_t$ dependence
of the phase transition. If the transition is $N_t$ independent for
sufficiently large $N_t$, then the transition is a bulk transition
(phase transition at zero temperature). 

\subsection{Quark mass and chiral symmetry} 
\label{sec:chiral}
In the Wilson quark formalism,
the flavor symmetry as well as C, P and T symmetries are exactly satisfied
on a lattice with a finite lattice spacing.
However, chiral symmetry is explicitly broken by the Wilson term
even for the vanishing bare quark mass $m_0=0$ at finite lattice spacings.
The lack of chiral symmetry causes much conceptual and technical
difficulties in numerical simulations and physics interpretation
of data. (See for more details Ref.~\cite{Stand26} and references cited there.)

The chiral property of the Wilson fermion action
was first systematically investigated
through Ward-Takahashi identities
by Bochicchio {\it et al.}~\cite{Bo}.
We also independently proposed \cite{ItohNP} to define the current quark mass
by

\begin{equation}
2 m_q \langle \,0\,|\,P\,|\,\pi\, \rangle
= - m_\pi \langle \,0\,|\,A_4\,|\,\pi\, \rangle,
\label{eq:quarkmass}
\end{equation}
where $P$ is the pseudoscalar density and $A_4$ the fourth component of the
local axial vector current. We use this definition of the current quark
mass as the quark mass
in this work. In general we need multiplicative normalization factors for
the pseudoscalar density and the axial current. 
In this study, we absorb these renormalization factors 
into the definition of the
quark mass, because this definition is sufficient for later use.
We note that the quark mass thus defined 
has an additive renormalization constant to the bare quark mass,
because the Wilson quark action does not hold chiral symmetry.

With this definition of quark mass,
when the quark is confined and the chiral symmetry
is spontaneously broken,
the pion mass vanishes in the chiral limit where the quark mass vanishes
at zero temperature. 

However, in the deconfined phase at finite temperatures 
the pion mass does not vanish in the chiral limit.
It is almost equal to twice the lowest
Matsubara frequency $\pi/N_t$. This implies that the pion state is
approximately a free two-quark state. 
The pion mass is nearly equal to the scalar meson mass, and the rho meson mass
to the axial vector meson mass; they are all nearly equal to the twice the lowest
Matsubara frequency $\pi/N_t$.
Thus the chiral symmetry is also manifest within
corrections due to finite lattice spacing.

In a previous study \cite{Wqmass} we found that, 
given the gauge coupling constant and bare quark masses,
the value of the quark mass does not depend on
whether the system is in the high or the low temperature phase
when the gauge coupling is not so strong: $\beta \ge 5.5$ for
$N_F=2$. 
However, this property is not guaranteed for smaller $\beta$.
In general, the quark mass in the deconfined phase differs from
that in the confined phase in the strong coupling region.

\subsection{Quark mass at $g=0$}

As mentioned earlier, the Wilson term (\ref{eq:WilsonTerm}) 
lifts the doublers and retains only
one pole around $m_0=0$ in the free case. On the other hand, there are
other poles at different values of the bare masses. They are remnants
of the doublers.

\begin{figure}[tb]
\centerline{
a) \epsfxsize=7cm\epsfbox{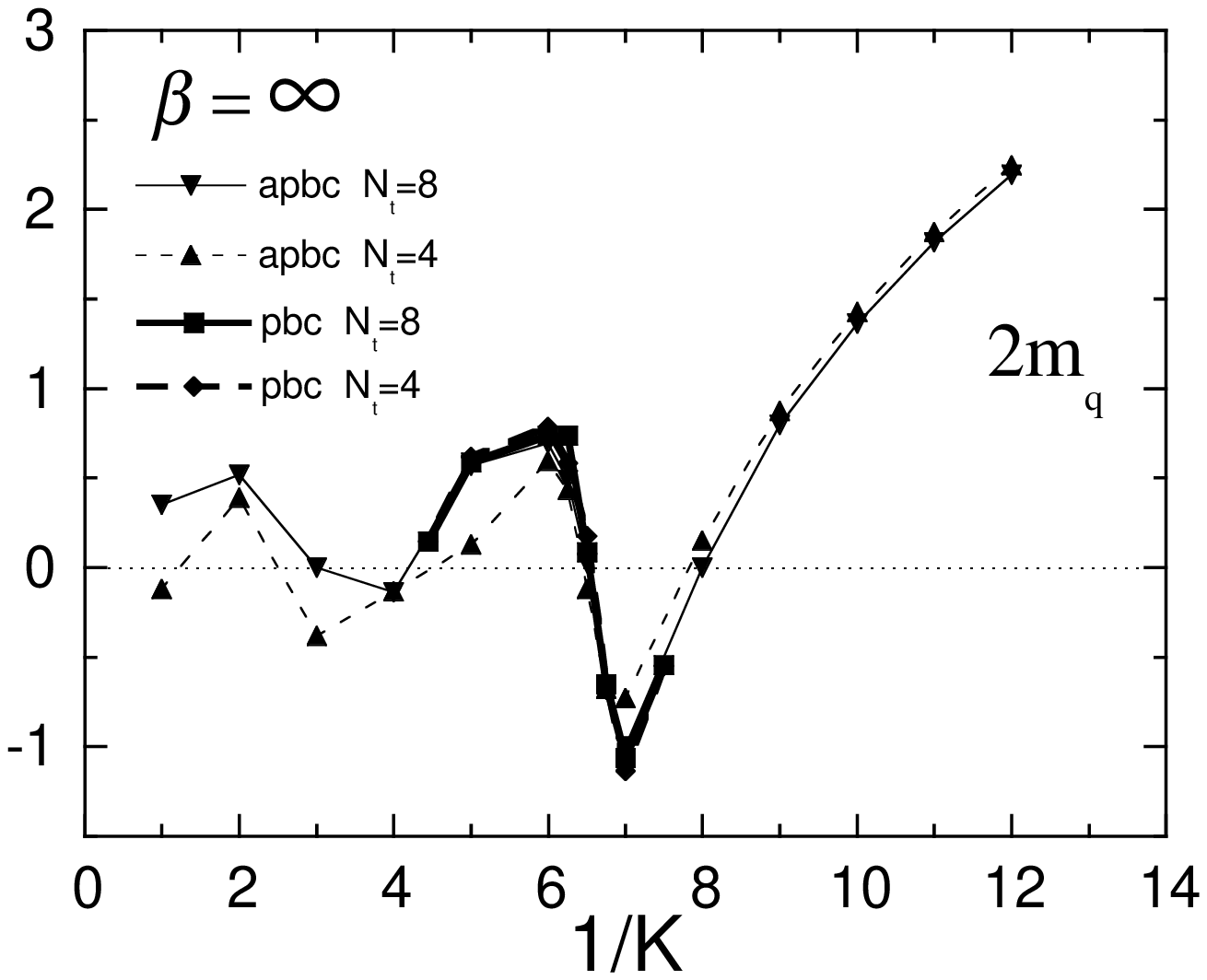}
}
\vspace{2mm} 
\centerline{ 
b) \epsfxsize=7cm\epsfbox{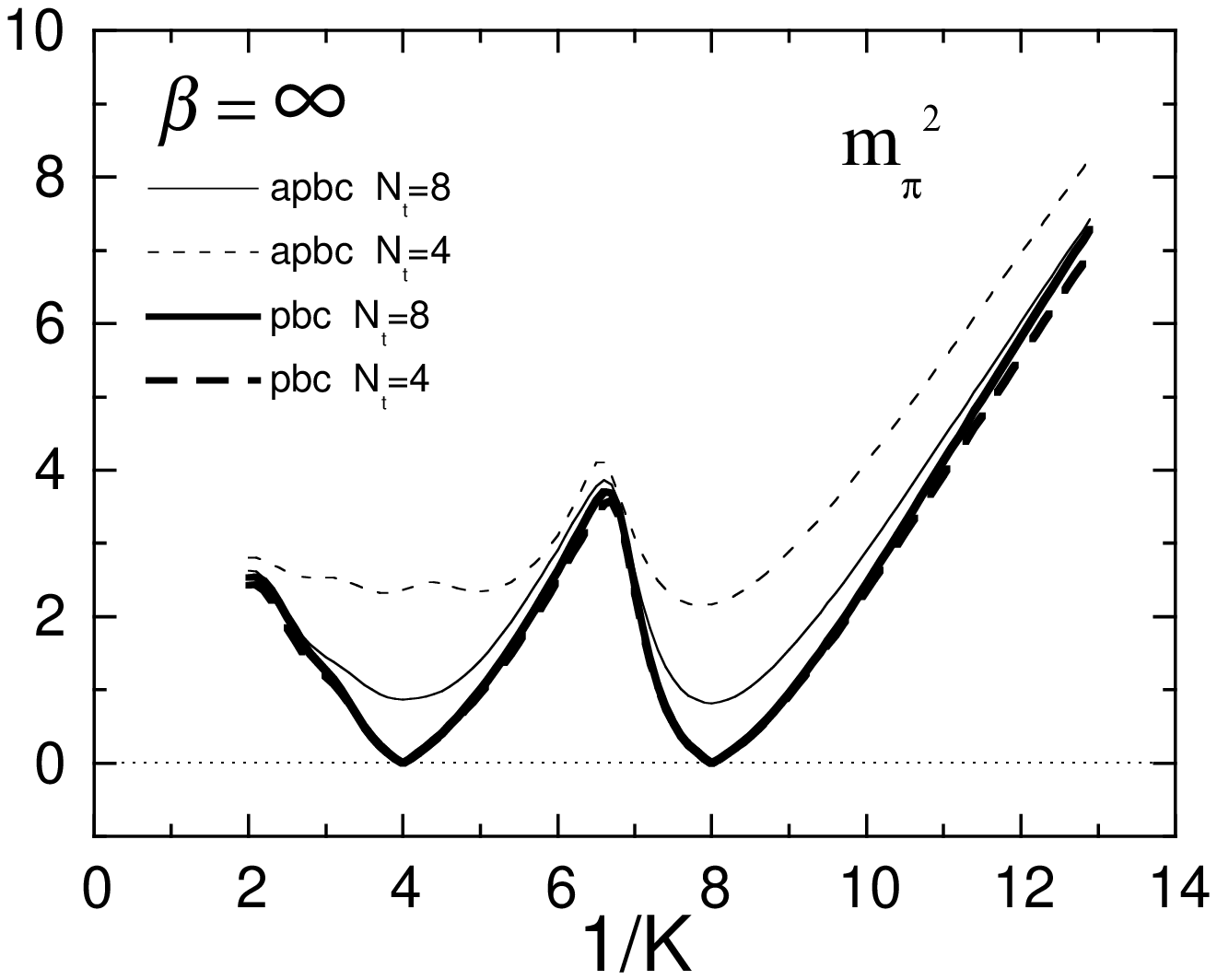}
}
\vspace{-0.1cm}
\caption{
(a) $m_q$ at $\beta=\infty$.
(b) $m_\pi$ at $\beta=\infty$.
Results with an anti-periodic boundary condition (apbc) in the $t$-direction 
and those with the periodic boundary condition (pbc) are compared 
on $N_t=8$ and 4 lattices.
}
\label{Binf}
\end{figure}

In Fig.~\ref{Binf}(a)
the quark mass defined by (\ref{eq:quarkmass}) is plotted
versus the bare quark mass for the case of free Wilson quark. 
At $m_0 \ge 0$ ($1/K \ge 8$)
the quark mass behaves as expected:
it monotonously increases with $m_0$.
On the other hand, at $m_0 \le 0$ ($1/K \le 8$)
the behavior of the quark mass is complicated:
it does not monotonously decrease with decreasing $m_0$, but increases after
some decrease and becomes zero at a finite negative value of $m_0$.
Usually, the region of negative bare quark mass is irrelevant for
numerical calculations of physical quantities. However, 
this region is also important for understanding the phase diagram.

We also plot in Fig.~\ref{Binf}(b)
the mass of the pion which is composed of two free
quarks with periodic boundary condition for the time direction and
with anti-periodic boundary condition. 
These results will be referred to later.

\begin{figure}[tb]
\centerline{
a) \epsfxsize=8cm\epsfbox{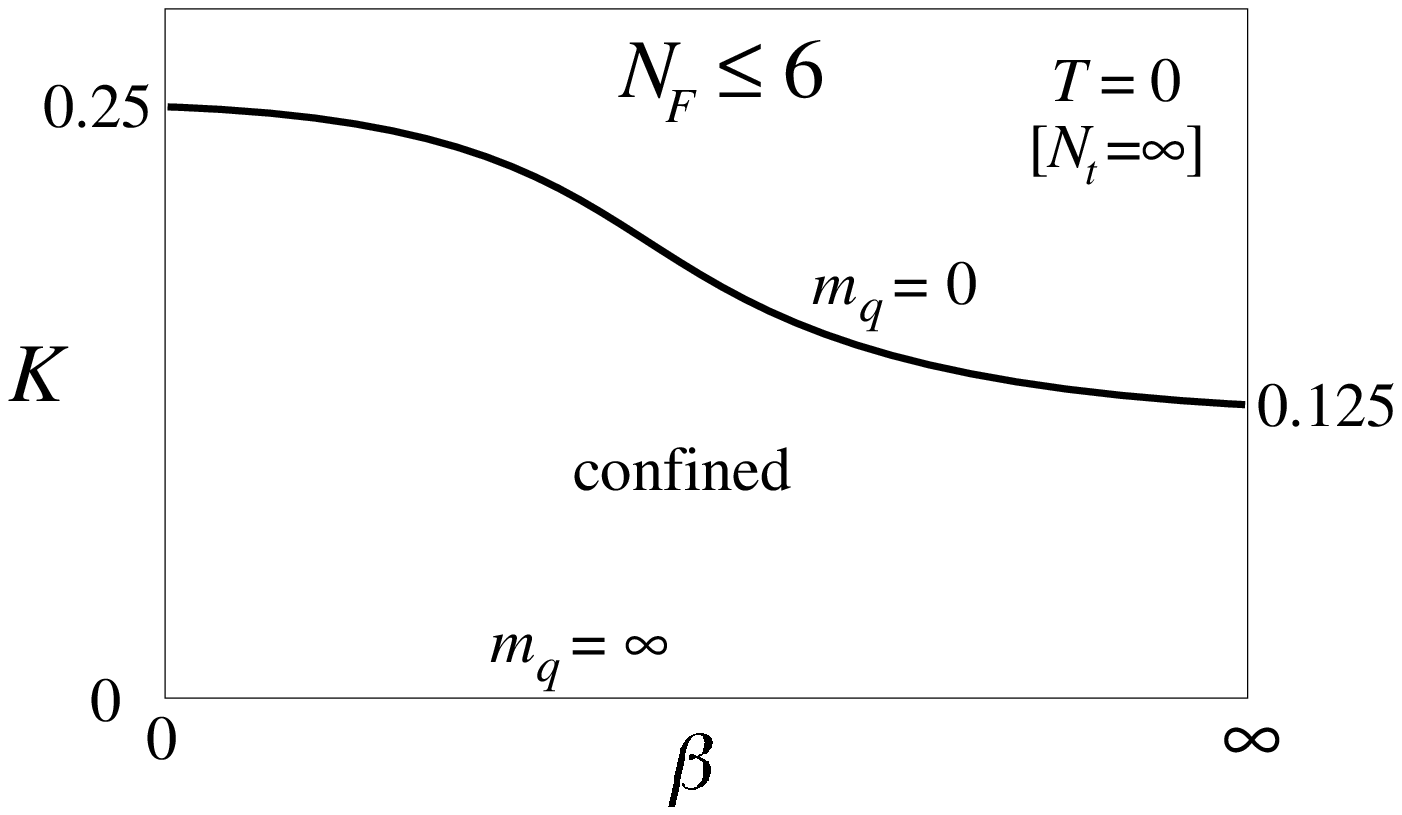}
}
\vspace{2mm}
\centerline{ 
b) \epsfxsize=8cm\epsfbox{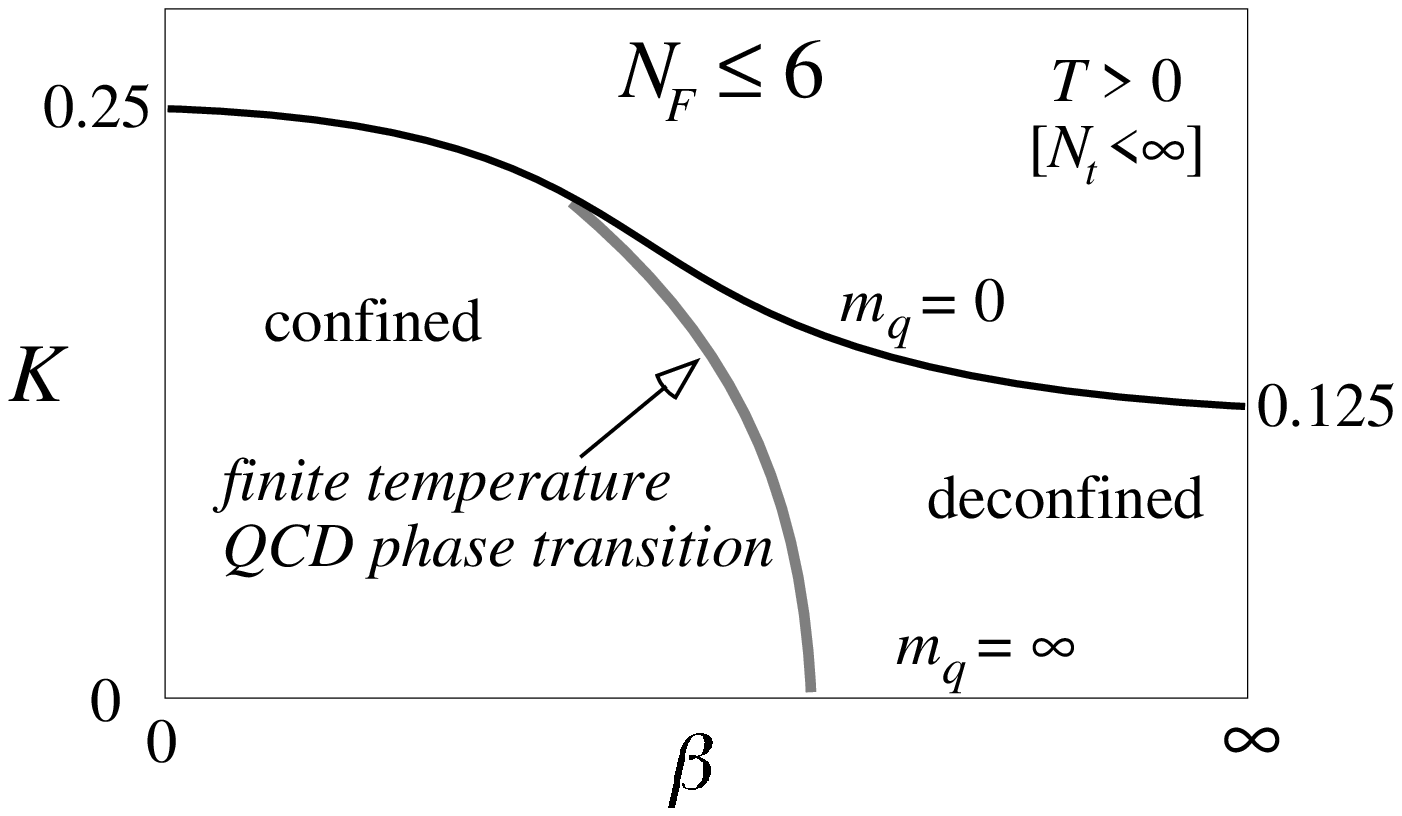}
}
\vspace{-0.1cm}
\caption{The phase structure for $N_F \leq 6$;
(a) at zero temperature, and
(b) at finite temperatures.
The chiral limit (massless quark limit) is shown by thick curves 
labeled by ``$m_q=0$'', and the finite temperature QCD transition at 
a fixed finite $N_t$ is shown by a shaded curve.
}
\label{NfSmall}
\end{figure}

\section{Phase diagram}
\label{sec:phasediagram}

Here we propose, based on our numerical results which will 
be shown later and the perturbation theory,
the phase diagram in the $(\beta,K)$ plane 
for the Wilson quark action coupled with the 
one-plaquette gauge action. 

\subsection{$N_F \le 6$}

For $N_F \le 6$,
the chiral line $K_c(\beta)$ where the current quark $m_q$ vanishes, 
exists in the confined phase(See Fig.~\ref{NfSmall}(a)).
The value of $K_c(\beta)$ at $\beta = \infty$ is 1/8, 
which corresponds to the vanishing bare quark mass $m_0=0$. 
As $\beta$ is decreased, the value of $K_c$ increases up to 1/4 at $\beta =0$.
If the action would be chiral symmetric, the chiral line should be a constant,
1/8, as in Sec.~\ref{sec:generic}.
The line $K=0$ corresponds to the case of infinitely heavy quarks.
Quarks are confined for any value of the current quark mass for all values
of $\beta$ at zero temperature ($N_t=\infty$).

On a lattice with a fixed finite $N_t$, we have the
finite temperature deconfining transition at finite $\beta$, 
because the temperature $T=1/N_t\, a$ becomes larger as $\beta$ 
increases in asymptotically free theories. 
At $K=0$ ($m_q=\infty$), the first order finite temperature phase transition 
of pure SU(3) gauge theory 
locates at $\beta_c=5.69254(24)$ and 5.89405(51) for $N_t=4$ and 6 
\cite{QCDPAX} and at $\beta_c \simeq 6.0625$ for $N_t=8$ \cite{Boyd96}.
This finite temperature transition turns into a crossover transition at 
intermediate values of $K$, and becomes stronger again towards the chiral 
limit $K_c$.
As $K$ is increased, 
the finite temperature transition line crosses the $K_c$ line 
at finite $\beta$ \cite{Stand26}.
We note that, for understanding the whole phase structure 
which includes the region above the $K_c$ line (negative values of the 
bare quark mass), the existence of the Aoki phase is important~\cite{Aoki}.
A schematic diagram of the phase structure for this case is shown
in Fig.~\ref{NfSmall}(b).
For simplicity, we omit the phase structure above the $K_c$ line.
It is known that the system is not singular on the $K_c$ line 
in the high-temperature phase (to the right of the finite temperature 
transition line) \cite{Stand26}.
The location of the finite temperature transition line moves toward
larger $\beta$ as $N_t$ is increased.
In the limit $N_t=\infty$,
the finite temperature transition line will shift to $\beta=\infty$
so that only the confined phase is realized at $T=0$. 

\begin{figure}[tb]
\centerline{
\epsfxsize=8cm\epsfbox{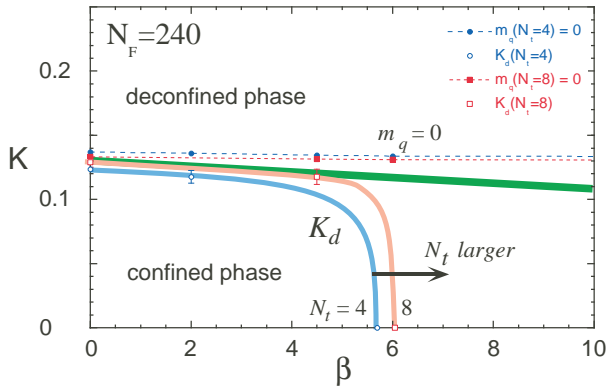}
}
\vspace{-0.1cm}
\caption{Phase diagram for $N_F=240$.
Dark shaded (green) lines represent our conjecture for the bulk transition
line in the limit $N_t=\infty$.
Points connected by dashed lines are the measured points where 
quarks are massless.
Light shaded lines are for the finite temperature transition at $N_t=4$ 
and 8.
}
\label{Nf240}
\end{figure}

\subsection{When $N_F$ is very large}

We present the result for the case of $N_F=240$ in Fig.~\ref{Nf240}. 
The reason why
we investigate the case where the number of flavor is so large 
as 240 is
the following: 
We have first investigated the case of $N_F=18$ as a generic case 
for $N_N \ge 17$. 
However it has turned out that 
the phase diagram looks complicated when $N_F=18$.
So, to understand the phase structure for $N_F \ge 17$,
we have increased the number of flavors
like 18, 60, 120, 180, 240, and 300, and systematically viewed the results of
the quark mass and the pion mass for all these numbers of flavors.
Then we have found that when the number of
flavors is very large as 240, the phase diagram is simple 
as the chirally symmetric case discussed in Sec.~\ref{sec:generic}.
Therefore we first show the result for the case of $N_F=240$.

At finite $N_t$ where numerical simulations have been performed,
the finite temperature transition occurs as shown in Fig.~\ref{Nf240}.
As $N_t$ increases, the transition line moves towards larger value of $\beta$.
The envelop of those finite temperature transition lines is 
the zero temperature phase transition line, shown by the dark shaded 
curve in the figure. 
Note that, at finite $N_t$, the system is not singular on the part of the 
dark shaded line in the right hand side of the finite temperature 
transition line (i.e., in the high temperature phase).

The salient fact is that, at zero temperature, 
the massless line exists only in the deconfined phase and passes through 
from $\beta=0$ down to $\beta=\infty$, i.e., it is quite similar to 
the chirally symmetric case of $N_F \ge 17$
shown in Fig.~\ref{chisym}(b). 
Thus the IR fixed point at $g=0$ governs the
long distance behavior of the theory and therefore
the theory is a free theory.

\begin{figure}[tb]
\centerline{
a)\epsfxsize=8cm\epsfbox{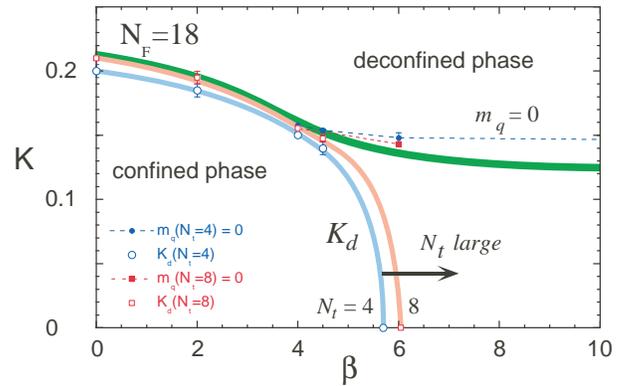}}
\vspace{2mm}
\centerline{
b)\epsfxsize=8cm\epsfbox{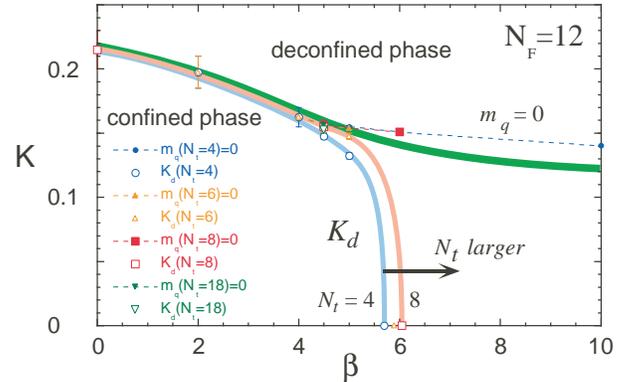}
}
\vspace{-0.1cm}
\caption{
The same as Fig.~\protect\ref{Nf240}, but for (a) $N_F=18$ and (b) $N_F=12$.
}
\label{Nf12_18}
\end{figure}

\subsection{$N_F \ge 17$, but not so large}

A typical phase diagram for $N_F \ge 17$ looks like that in 
Fig.~\ref{Nf12_18}(a),
where the case of $N_F=18$ is displayed.
The massless line in the deconfined phase which starts from $\beta=\infty$
hits the first-order phase transition around $\beta=4$,
and goes underground crossing the first-order phase transition line.

The place where the crossing occurs moves towards $\beta =0$ as the number
of flavors increases, 
and finally reaches the $\beta=0$ axis when $N_F \ge 240$.
The massless line exists only in the deconfined phase and it starts from
$\beta=\infty$ ($g=0$). Therefore the IR fixed point at $g=0$ governs the
long distance behavior of the theory. That is, the theory is free.

\subsection{$7 \le N_F \le 16$}

The phase diagram for $N_F=12$ is shown in Fig.~\ref{Nf12_18}(b),
which looks similar to that of $N_F=18$.
The salient fact is that the massless line exists only in the deconfined
phase. That is, there is no massless line in the confined phase.
Therefore, in the continuum limit, the quark is not confined.
  
The difference from the case of $N_F \ge 17$ is that 
the $g=0$ point is an UV fixed point in this case.
Thus there should be an IR fixed point at finite coupling constant.
If there would be no IR fixed point, we would encounter a contradiction that
on one side quarks are not confined, but on the other side the gauge coupling
constant becomes arbitrary large as the distance between quarks becomes
large.

When we increase the number of flavors from $N_F=6$ to 17, 
the form of the $\beta$ function changes from that in the upper frame
($N_F \le 6$)
in Fig.~\ref{BetaFunc}(b) to that in the lower frame
($N_F \le 17$)
through that in the middle frame ($7 \le N_F \le 16$). 
We safely assume that the form changes smoothly for varying $N_F$.
That is, for $N_F=7$ the IR fixed point should appear at very large coupling,
and gradually the position of the IR fixed point moves towards the $g=0$ point.
For $N_F = 16$, the IR fixed point is expected to be close to $g=0$.

We have been unable to identify numerically
the position of the IR fixed point. Technically it is not easy to do so.
It might be even one of the metastable states beyond the first-order phase 
transition line, i.e., under the first sheet of the phase diagram.

\begin{table}
\caption{Simulation parameters for $N_F=6$ 
on lattices $8^2\times10\times N_t$ ($N_t=4$, 6, 8), 
$12^3\times N_t$ ($N_t=12$) and $18^2\times24\times N_t$ ($N_t=18$).}
\begin{ruledtabular}
\begin{tabular}{ccll}
$N_t$ & $\beta$ & $\kappa$ & $\Delta\tau$ \\
\hline
4 & 
  0.0  & 0.2, 0.21, 0.22, 0.235, 0.24, 0.25 &  0.01--0.0025 \\
& 0.1  & 0.2495 & 0.01 \\
& 0.2  & 0.249, 0.24936 & 0.01 \\
& 0.3  & 0.2485, 0.249 & 0.01 \\
& 0.4  & 0.248  & 0.01 \\
& 0.5  & 0.23, 0.235, 0.24, 0.245, 0.2475 & 0.01 \\
& 1.0  & 0.2, 0.21, 0.22, 0.225, 0.23, 0.235, &\\
&      & 0.237, 0.238, 0.24, 0.245 & 0.01--0.005\\
& 2.0  & 0.24 & 0.01 \\
& 4.0  & 0.22 & 0.01 \\
& 4.5  & 0.15, 0.16, 0.165, 0.166, 0.167, 0.168, &\\
&      & 0.169, 0.17, 0.18, 0.19, 0.2143 & 0.01\\
\hline
6 &
  0.0  & 0.2, 0.21, 0.22 & 0.01 \\
& 0.5  & 0.2475 & 0.01 \\
& 1.0  & 0.245  & 0.01 \\
& 2.0  & 0.24  & 0.01 \\
& 4.0  & 0.22  & 0.01 \\
& 4.5  & 0.16, 0.165, 0.167, 0.168, 0.17, 0.171, &\\
&      & 0.1715, 0.1717, 0.1719, 0.172, 0.1725, &\\
&      & 0.173, 0.174, 0.175, 0.18 & 0.01 \\
\hline
8 &
  0.5  & 0.2475 & 0.01 \\
& 1.0  & 0.245  & 0.01 \\
& 2.0  & 0.24   & 0.01 \\
& 4.0  & 0.22   & 0.01 \\
& 4.5  & 0.16, 0.167, 0.168, 0.17, 0.172, 0.1723, &\\
&      & 0.1724, 0.1725, 0.173, 0.175, 0.18 & 0.01 \\
& 5.5  & 0.1615 & 0.01 \\
\hline
12 &
  4.5  & 0.175  & 0.01 \\
\hline
18 &
  0.3  & 0.2485 & 0.01 \\
& 0.4  & 0.248 & 0.01 \\
& 0.5  & 0.2475 & 0.01 \\
& 1.0  & 0.245  & 0.01 \\
& 4.5  & 0.172, 0.1725, 0.173, 0.175, 0.2143 & 0.01 \\
\end{tabular}
\label{tab:param6}
\end{ruledtabular}
\end{table}

\begin{table*}
\caption{The same as Table~\protect\ref{tab:param6}, but for $N_F=7$.}
\begin{ruledtabular}
\begin{tabular}{ccll}
$N_t$ & $\beta$ & $\kappa$ & $\Delta\tau$ \\
\hline
4 & 
  0.0  & 0.2, 0.21, 0.22, 0.23, 0.235, 0.24, 0.245, 0.25, 0.2857, 0.3333 &  0.01--0.005 \\
& 1.0  & 0.245 & 0.01 \\
& 2.0  & 0.125, 0.1429, 0.1667, 0.2, 0.2083, 0.2174, 0.2273, 0.24, 0.25 & 0.01 \\
& 3.0  & 0.235 & 0.01 \\
& 4.0  & 0.125, 0.1429, 0.1667, 0.178, 0.185, 0.195, 0.2, 0.2226, 0.25 & 0.01 \\
& 4.5  & 0.14, 0.15, 0.16, 0.161, 0.162, 0.163, 0.164, 0.165, 0.17, 0.18, 0.19, 0.2143, 0.25 & 0.01\\
& 5.0  & 0.12, 0.13, 0.14, 0.15, 0.16, 0.17, 0.18, 0.19, 0.1982, 0.21 & 0.01 \\
& 5.5  & 0.1, 0.11, 0.12, 0.13, 0.14, 0.15, 0.16, 0.161 & 0.01 \\
& 6.0  & 0.08, 0.09, 0.1, 0.11, 0.1111, 0.12, 0.125, 0.13, 0.14, 0.1429, 0.15, 0.1519, 0.152, &\\
&      & 0.155, 0.16, 0.1667, 0.1724, 0.1786, 0.2, 0.25, 0.3333 & 0.01 \\
\hline
6 &
  0.0  & 0.2, 0.21, 0.22, 0.23, 0.24, 0.245, 0.25 & 0.01 \\
& 2.0  & 0.193, 0.214, 0.24  & 0.01 \\
& 4.0  & 0.163, 0.178, 0.195, 0.217 & 0.01 \\
& 4.5  & 0.15, 0.16, 0.164, 0.165, 0.168, 0.1681, 0.1682, 0.169, 0.17, 0.18 & 0.01 \\
& 5.0  & 0.12, 0.13, 0.14, 0.15, 0.16, 0.17 & 0.01 \\
& 5.5  & 0.12, 0.13, 0.135, 0.138, 0.14, 0.15 & 0.01 \\
\hline
8 &
  0.0  & 0.25, 0.2857, 0.3333 & 0.01 \\
& 2.0  & 0.1667, 0.2, 0.2083, 0.2174, 0.2273, 0.24, 0.25 & 0.01 \\
& 4.5  & 0.15, 0.16, 0.164, 0.165, 0.168, 0.1683, 0.1684, 0.169, 0.18, 0.25 & 0.01 \\
& 5.0  & 0.12, 0.13, 0.14, 0.15, 0.16, 0.17 & 0.01 \\
& 5.5  & 0.1, 0.11, 0.12, 0.13, 0.135, 0.138, 0.14, 0.145, 0.15, 0.16 & 0.01 \\
& 6.0  & 0.1, 0.11, 0.12, 0.13, 0.14, 0.145, 0.1476, 0.1519, 0.1667, 0.1724, 0.1786, 0.2, 0.25 & 0.01 \\
\hline
18 &
  0.0  & 0.245, 0.25 & 0.01--0.005 \\
& 4.5  & 0.165, 0.1675, 0.1684, 0.17, 0.18, 0.19, 0.2143 & 0.01 \\
& 5.5  & 0.135, 0.15, 0.162 & 0.01 \\
& 6.0  & 0.1, 0.11, 0.12, 0.13, 0.14, 0.145, 0.1476, 0.1519 & 0.01 \\
\end{tabular}
\label{tab:param7}
\end{ruledtabular}
\end{table*}

\begin{table*}
\caption{The same as Table~\protect\ref{tab:param6}, but for $N_F=8$, 
10, 14, 17, 120, 180, and 360.}
\begin{ruledtabular}
\begin{tabular}{cccll}
$N_F$ & $N_t$ & $\beta$ & $\kappa$ & $\Delta\tau$ \\
\hline
8 &  4 & 0.0  & 0.22, 0.23, 0.24, 0.25 &  0.01 \\
  & 18 & 0.0  & 0.25 &  0.01 \\
\hline
10 &  4 & 0.0  & 0.25 &  0.01 \\
\hline
14 &  4 & 0.0  & 0.25 &  0.01 \\
\hline
17 &  4 & 6.0  & 0.145, 0.1473, 0.15 &  0.01 \\
   &  8 & 6.0  & 0.1473 &  0.01 \\
   & 18 & 6.0  & 0.1473, 0.15 &  0.01 \\
\hline
120 &  4 & 0.0  & 0.125, 0.1316, 0.1389, 0.1429, 0.1471 & 0.005--0.0025 \\
\hline
180 &  4 & 0.0  & 0.122, 0.125, 0.1274, 0.1303, 0.1333, 0.1429 & 0.005--0.0025 \\
\hline
360 &  4 & 0.0  & 0.125 & 0.00125 \\
    &  8 & 0.0  & 0.125 & 0.00125 \\
\end{tabular}
\label{tab:param8-360}
\end{ruledtabular}
\end{table*}

\begin{table*}
\caption{The same as Table~\protect\ref{tab:param6}, but for $N_F=12$.}
\begin{ruledtabular}
\begin{tabular}{ccll}
$N_t$ & $\beta$ & $\kappa$ & $\Delta\tau$ \\
\hline
4 & 
  0.0  & 0.18, 0.2, 0.205, 0.21, 0.215, 0.22, 0.23, 0.24, 0.25 &  0.01 \\
& 2.0  & 0.15, 0.17, 0.185, 0.19, 0.195, 0.2, 0.21, 0.25 & 0.01 \\
& 4.0  & 0.14, 0.15, 0.155, 0.16, 0.165, 0.17, 0.18, 0.2, 0.25 & 0.01 \\
& 4.5  & 0.13, 0.14, 0.145, 0.15, 0.16, 0.1667, 0.17, 0.1786, 0.1852, 0.2, 0.25 & 0.01\\
& 5.0  & 0.11, 0.12, 0.13, 0.135, 0.14, 0.15, 0.16, 0.1667, 0.1724, 0.1786, 0.2, 0.25 & 0.01 \\
& 6.0  & 0.1, 0.1111, 0.125, 0.1429, 0.1667, 0.1724, 0.1786, 0.2, 0.25, 0.3333 & 0.01 \\
& 10.0 & 0.13, 0.14, 0.15, 0.16 & 0.01 \\
\hline
6 &
  0.0  & 0.18, 0.2, 0.21, 0.215, 0.22, 0.23, 0.24, 0.25 & 0.01 \\
& 2.0  & 0.15, 0.17, 0.185, 0.19, 0.195, 0.2, 0.21, 0.2156, 0.2253, 0.236  & 0.01 \\
& 4.0  & 0.14, 0.145, 0.15, 0.155, 0.16, 0.165, 0.17, 0.18, 0.1949, 0.2028, 0.2114 & 0.01 \\
& 4.5  & 0.13, 0.14, 0.145, 0.15, 0.155, 0.16, 0.17  & 0.01 \\
& 5.0  & 0.11, 0.12, 0.13, 0.135, 0.14, 0.145, 0.15, 0.16 & 0.01 \\
\hline
8 &
  0.0  & 0.2, 0.205, 0.2083, 0.21, 0.215, 0.22, 0.2222, 0.225, 0.23, 0.24, 0.25 & 0.01--0.0025 \\
& 2.0  & 0.15, 0.17, 0.185, 0.19, 0.195, 0.2, 0.21, 0.225, 0.25 & 0.01 \\
& 4.5  & 0.13, 0.14, 0.15, 0.155, 0.16, 0.1667, 0.1786, 0.2, 0.25 & 0.01 \\
& 6.0  & 0.1429, 0.1667, 0.2, 0.25 & 0.01 \\
\hline
18 &
  4.5  & 0.14, 0.15, 0.155, 0.16  & 0.01 \\
\end{tabular}
\label{tab:param12}
\end{ruledtabular}
\end{table*}

\begin{table}
\caption{The same as Table~\protect\ref{tab:param6}, but for $N_F=16$.}
\begin{ruledtabular}
\begin{tabular}{ccll}
$N_t$ & $\beta$ & $\kappa$ & $\Delta\tau$ \\
\hline
4 & 
  0.0  & 0.18, 0.19, 0.2, 0.21, 0.22, 0.25 &  0.01 \\
& 4.5  & 0.125, 0.135, 14, 0.1429, 0.145, 0.15,  &\\
&      & 0.1563, 0.1667,0.2, 0.25 & 0.01\\
& 6.0  & 0.125, 0.1429, 0.1493, 0.1667, 0.2, 0.25 & 0.01 \\
& 10.0 & 0.1111, 0.125, 0.1429, 0.1667, 0.2, 0.25 & 0.005--0.0025 \\
& 100. & 0.1111, 0.125, 0.1429, 0.1667, 0.2, 0.25 & 0.005--0.0025 \\
\hline
8 &
  0.0  & 0.18, 0.19, 0.2, 0.21, 0.22, 0.25, 0.27 & 0.01 \\
& 4.5  & 0.125, 0.1429, 0.145, 0.15, 0.155, &\\
&      & 0.1563, 0.1667, 0.2, 0.25 & 0.01 \\
& 6.0  & 0.125, 0.1429, 0.1493, 0.1667, 0.2, 0.25 & 0.01 \\
& 10.0 & 0.1111, 0.125, 0.1429, 0.1667, 0.2, 0.25 & 0.005 \\
& 100. & 0.1111, 0.125, 0.1429, 0.1667, 0.2, 0.25 & 0.005 \\
\hline
18 &
  0.0  & 0.25 & 0.01 \\
& 4.5  & 0.15 & 0.01 \\
\end{tabular}
\label{tab:param16}
\end{ruledtabular}
\end{table}

\begin{table*}
\caption{The same as Table~\protect\ref{tab:param6}, but for $N_F=18$.}
\begin{ruledtabular}
\begin{tabular}{ccll}
$N_t$ & $\beta$ & $\kappa$ & $\Delta\tau$ \\
\hline
4 & 
  0.0  & 0.17, 0.175, 0.18, 0.19, 0.195, 0.2, 0.205, 0.21, 0.215, 0.22, 0.235, 0.245, 0.25 &  0.02--0.005 \\
& 2.0  & 0.16, 0.17, 0.18, 0.19, 0.2, 0.21, 0.25 & 0.01 \\
& 4.0  & 0.13, 0.14, 0.15, 0.152, 0.154, 0.16, 0.1613, 0.1667, 0.17, 0.18, 0.2, 0.25 & 0.01 \\
& 4.5  & 0.115, 0.125, 0.135, 0.14, 145, 0.15, 0.1613, 0.1667, 0.1724, 0.2, 0.2143, 0.25 & 0.01\\
& 6.0  & 0.1, 0.1111, 0.125, 0.1429, 0.1613, 0.1667, 0.1724, 0.2, 0.25 & 0.01 \\
& 10.0 & 0.13, 0.14, 0.15 & 0.01 \\
\hline
8 &
  0.0  & 0.18, 0.19, 0.2, 0.21, 0.22, 0.23, 0.25, 0.27 & 0.01 \\
& 2.0  & 0.16, 0.17, 0.18, 0.19, 0.2, 0.21 & 0.01 \\
& 4.0  & 0.13, 0.14, 0.15, 0.152, 0.154, 0.156, 0.158, 0.16, 0.17, 0.18
 & 0.01 \\
& 4.5  & 0.135, 0.14, 0.145, 0.15, 0.155, 0.1667, 0.2, 0.25 & 0.01 \\
& 6.0  & 0.125, 0.1389, 0.1429, 0.1667, 0.2, 0.25 & 0.01 \\
\hline
18 &
  0.0  & 0.25 & 0.01 \\
& 4.5  & 0.15, 0.2143 & 0.01 \\
\end{tabular}
\label{tab:param18}
\end{ruledtabular}
\end{table*}

\begin{table*}
\caption{The same as Table~\protect\ref{tab:param6}, but for $N_F=60$.}
\begin{ruledtabular}
\begin{tabular}{ccll}
$N_t$ & $\beta$ & $\kappa$ & $\Delta\tau$ \\
\hline
4 & 
  0.0  & 0.08333, 0.09091, 0.1, 0.1111, 0.125, 0.1429, 0.1538, 0.1613, 0.1667, 0.1724, 0.2, 0.25, 0.3333, 0.5, 1.0 &  0.01--0.00125 \\
& 6.0  & 0.08333, 0.09091, 0.1, 0.1111, 0.125, 0.1333, 0.1429, 0.1493, 0.1538, 0.16, 0.1667, 0.2, 0.25 & 0.005 \\
\hline
8 &
  6.0  & 0.125, 0.1429, 0.1667, 0.2, 0.25 & 0.005 \\
\end{tabular}
\label{tab:param60}
\end{ruledtabular}
\end{table*}

\begin{table*}
\caption{The same as Table~\protect\ref{tab:param6}, but for $N_F=240$.
For $N_t=16$ the spatial lattice is $16^2\times24$.}
\begin{ruledtabular}
\begin{tabular}{ccll}
$N_t$ & $\beta$ & $\kappa$ & $\Delta\tau$ \\
\hline
4 & 
  0.0  & 0.08333, 0.09091, 0.1, 0.1111, 0.122, 0.1234, 0.125, 0.1333, 0.137, 0.1429, 0.1493, 0.1538, &\\
&      & 0.1667, 0.2, 0.25 &  0.0025 \\
& 2.0  & 0.09091, 0.1, 0.1111, 0.125, 0.1359, 0.1429, 0.1538, 0.1667, 0.2, 0.25 & 0.0025 \\
& 4.5  & 0.08333, 0.09091, 0.1, 0.1111, 0.125, 0.1343, 0.1429, 0.1538, 0.1667, 0.2, 0.25 & 0.0025 \\
& 6.0  & 0.08333, 0.09091, 0.1, 0.1111, 0.125, 0.1337, 0.1429, 0.1538, 0.1667, 0.2, 0.25 &  0.0025--0.00125 \\
& 100. & 0.08333, 0.09091, 0.1, 0.1111, 0.125, 0.1293, 0.1429, 0.1538, 0.1667, 0.2, 0.25 &  0.0025--0.00125 \\
& 1000. & 0.125 &  0.0025 \\
\hline
8 &
  0.0  & 0.08333, 0.09091, 0.1, 0.1111, 0.125, 0.129, 0.1333, 0.137, 0.1429, 0.1538, 0.1667, 0.2, 0.25 &  0.0025 \\
& 4.5  & 0.08333, 0.09091, 0.1, 0.1111, 0.125, 0.1316, 0.1429, 0.1538, 0.1667, 0.2, 0.25 &  0.0025 \\
& 6.0  & 0.08333, 0.09091, 0.1, 0.1111, 0.125, 0.1309, 0.1429, 0.1538, 0.1667, 0.2, 0.25 &  0.0025 \\
& 100.  & 0.08333, 0.09091, 0.1, 0.1111, 0.125, 0.1264, 0.1429, 0.1538, 0.1667, 0.2, 0.25 &  0.0025 \\
\hline
16 &
  0.0  & 0.125, 0.129, 0.1333, 0.1429, 0.1538, 0.2 & 0.002--0.00125 \\
\end{tabular}
\label{tab:param240}
\end{ruledtabular}
\end{table*}

\begin{table*}
\caption{The same as Table~\protect\ref{tab:param6}, but for $N_F=300$.
For $N_t=16$ the spatial lattice is $16^2\times24$.}
\begin{ruledtabular}
\begin{tabular}{ccll}
$N_t$ & $\beta$ & $\kappa$ & $\Delta\tau$ \\
\hline
4 & 
  0.0  & 0.08333, 0.09091, 0.1, 0.1111, 0.1143, 0.1176, 0.1212, 0.125, 0.1429, 0.1493, 0.1538, 0.1667, 0.2, 0.25 &  0.00125 \\
& 4.5  & 0.08333, 0.09091, 0.1, 0.1111, 0.125, 0.1429, 0.1538, 0.1667, 0.2, 0.25 &  0.00125 \\
& 6.0  & 0.08333, 0.09091, 0.1, 0.1111, 0.125, 0.1429, 0.1493, 0.1538, 0.1667, 0.2, 0.25 &  0.00125 \\
& 100.  & 0.08333, 0.09091, 0.1, 0.1111, 0.125, 0.1429, 0.1493, 0.1538, 0.1667, 0.2, 0.25 &  0.00125 \\
\hline
8 &
  0.0  & 0.08333, 0.09091, 0.1, 0.1111, 0.1143, 0.1176, 0.1212, 0.125, 0.1429, 0.1667, 0.2, 0.25 &  0.0025--0.00125 \\
& 4.5  & 0.08333, 0.09091, 0.1, 0.1111, 0.125, 0.1429, 0.1538, 0.1667, 0.2, 0.25 &  0.0025 \\
& 6.0  & 0.08333, 0.09091, 0.1, 0.1111, 0.125, 0.1429, 0.1538, 0.1667, 0.2, 0.25 &  0.00125 \\
\hline
16 &
  0.0  & 0.1212, 0.1231, 0.125 & 0.002 \\
\end{tabular}
\label{tab:param300}
\end{ruledtabular}
\end{table*}

\subsection{Previous studies using staggered quarks}

We note that
there are several works investigating the critical number of flavors
for quark confinement using the staggered quarks.
The staggered fermion (Kogut-Susskind fermion) \cite{susskind77}
is a formulation of lattice fermions different from the Wilson fermion
we adopt. 
Unlike the Wilson fermion action (\ref{eq:Wilson}), 
the staggered fermion action explicitly violates the flavor symmetry 
due to flavor-mixing interactions at finite lattice spacings.
However, because a part of flavor-chiral symmetry is preserved on the lattice,
a numerical analysis of chiral properties is easier than the Wilson fermion.
On the other hand,
the staggered fermion action can describe quarks only when $N_F$ is 
a multiple of 4. 
A trick to study the cases $N_F \neq 4n$ 
is to modify by hand the power of the fermionic determinant in the 
numerical path-integration. 
This necessarily makes the action non-local 
which sometimes poses conceptually and technically difficult problems. 

Evidences of strong first order transition are reported
with staggered quarks for $N_F = 4$--18 \cite{KS1,KS2,KS3,KS4,KS5,KS6}.
Chiral symmetry is restored at the transition.
Furthermore, the transition was shown to be a bulk phase transition 
for several cases of $N_F=8$ \cite{KS2,KS5}, 12 \cite{KS2}
and 16 \cite{KS6}.

The most systematic study was done by the Columbia group in Ref.~\cite{KS5}.
They studied the case $N_F=8$ at a fixed bare quark mass $m_0 a = 0.015$ 
on $16^3\times N_t$ lattices where $N_t=4$--32.
They found that the transition point shifts towards larger $\beta$ 
as $N_t$ is increased from 4 to 8, but stays around $\beta_c = 4.73$ 
for $N_t=8$ and 16.
They concluded that this is a bulk transition, and proposed an interpretation
that this transition is an outgrowth of the crossover transition between 
the strong and weak coupling regions observed in pure SU(3) gauge theory.

Although their interpretation is in clear contrast with our proposal 
we note that their numerical results themselves are consistent with our 
phase diagram shown in Fig.~\ref{Nf12_18}:
As an illustration, let us suppose that we perform simulations 
at, say, $K = 0.14$ in Fig.~\ref{Nf12_18}(b).
For small $N_t$ we have the finite temperature transition at finite 
$\beta$.
This transition point moves towards larger $\beta$ as we increase $N_t$, 
but, eventually, stops on the dark shaded line when $N_t$ becomes larger than 
a critical value.
In order to clarify the whole phase structure in the case of staggered quarks
and to discriminate different possibilities, it is indispensable to perform
a more systematic study exploring a wider region of the parameter space
of quark mass and gauge coupling constant.

\section{Parameters for numerical simulations}
\label{sec:parameters}

We perform simulations on
lattices $8^2 \times 10 \times N_t$
($N_t =4$, 6 or 8), $16^2 \times 24 \times N_t$ ($N_t=16$)
and $18^2 \times 24 \times N_t$ ($N_t=18$).
We vary $N_F$ from 2 to 360, selecting some typical values of $N_F$.
For each $N_F$, we study the phase structure in the coupling
parameter space $(\beta,K)$. 
Simulations for $N_F\leq6$ are discussed in \cite{Stand26,Stand26a}.
We summarize simulation parameters for $N_F\geq6$ in 
Tables~\ref{tab:param6}-\ref{tab:param300}, where 
we list values of $N_t$, $\beta$, $\kappa$ and $\Delta\tau$.
(For readers who are interested in more details we will
provide them on request.)
We adopt an anti-periodic boundary condition for quarks
in the $t$ direction and periodic boundary conditions otherwise.
In the cases where a thermalized state is achieved,
typical statistics for hadronic measurements are about 10--100
configurations sampled every 1--5 trajectories, where the length of
one trajectory is one molecular-dynamics time. 
When the hadron spectrum is calculated, the lattice is duplicated
in the direction of lattice size 10 for $N_t \le 8$,
which we call the $z$ direction.
Statistical errors are estimated by the jack-knife method.

We use the hybrid R algorithm \cite{Ralgo} for the generation of
gauge configurations.
The R algorithm has discretization errors of $O(N_F \Delta\tau^2)$
for the step size $\Delta\tau$ of a molecular dynamic evolution.
As $N_F$ increases we have to decrease $\Delta\tau$, such as
$\Delta\tau$ =0.0025 for $N_F=240$, to reduce the errors.
We have checked that the errors in the physical observables we study
are sufficiently small
with our choices of $\Delta\tau$ for typical cases.

It should be noted that, in QCD with dynamical quarks,
there are no order parameters for quark confinement.
We discuss confinement by measuring the screening pion mass,
the screening quark mass, the values of the plaquette and the Polyakov loop.
See Sec.~\ref{sec:strong} for details.
In the following, we call the pion screening mass simply
the pion mass, and similarly for the quark mass.

The numerical calculations 
were performed on various computers including the dedicated parallel computer 
QCDPAX and Fujitsu VPP500 at the University of Tsukuba, and 
HITAC S820/80 at KEK.

\begin{figure}[tb]
\centerline{
a) \epsfxsize=7.5cm\epsfbox{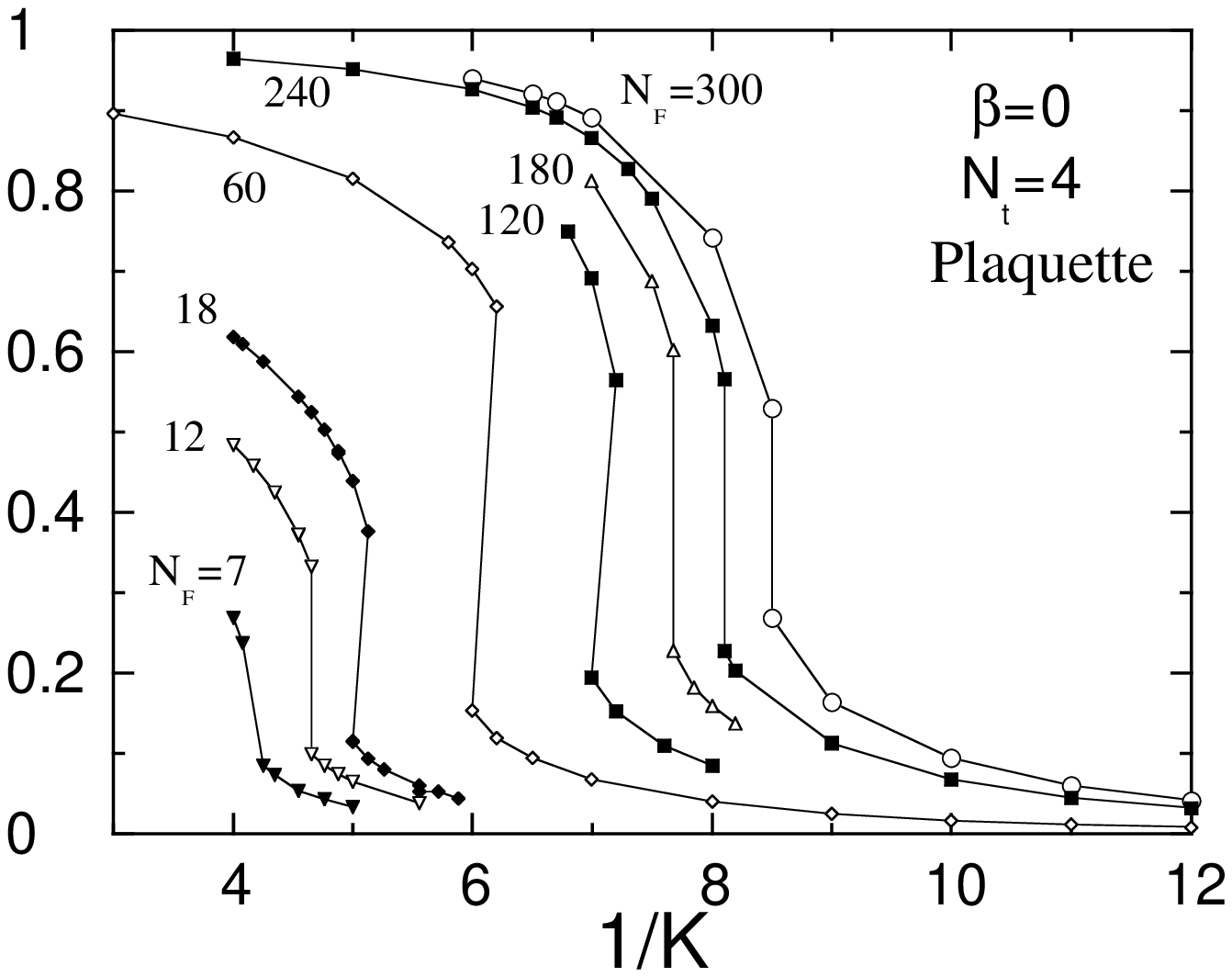}
}
\vspace{2mm}
\centerline{ 
b) \epsfxsize=7.5cm\epsfbox{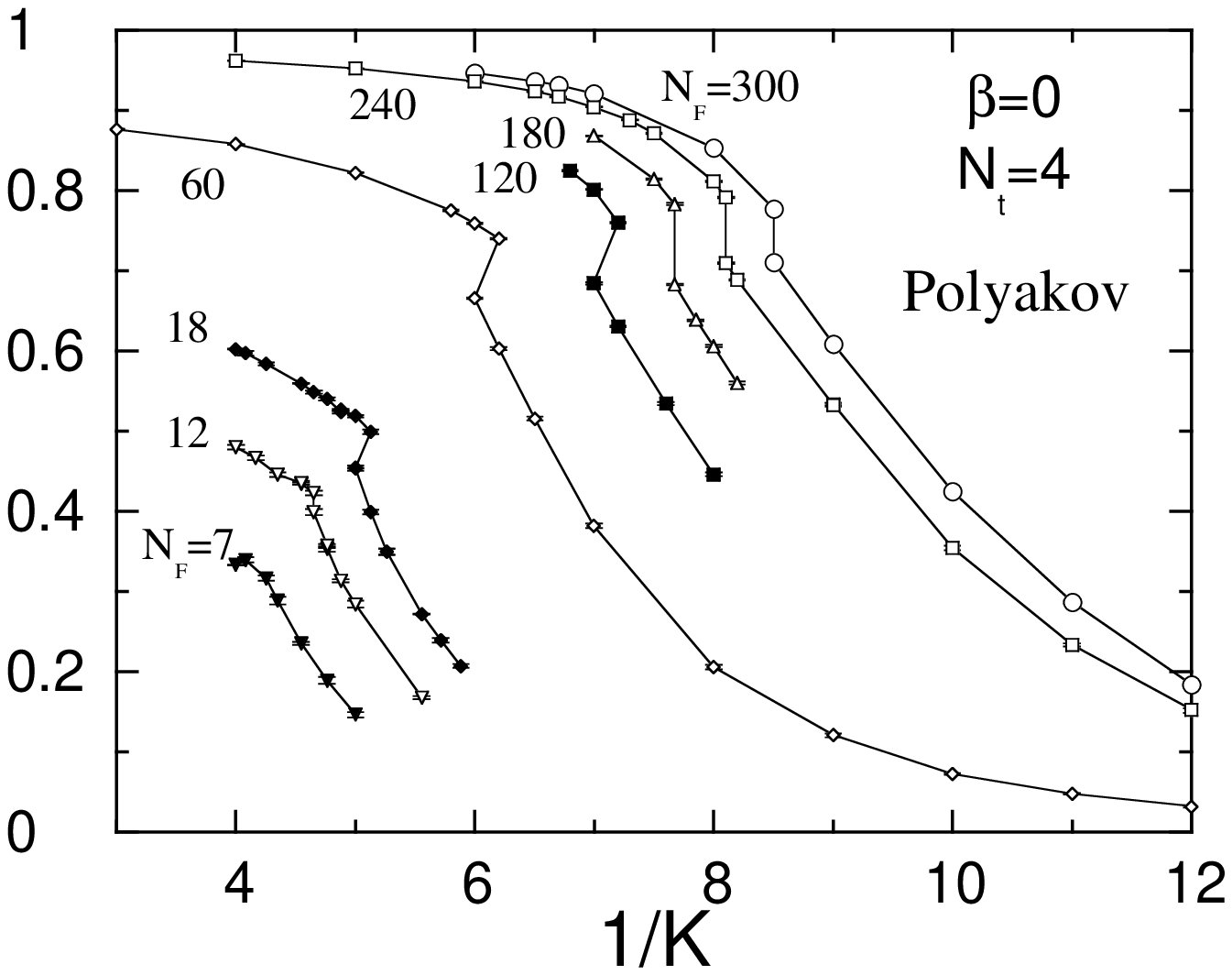}
}
\vspace{-0.1cm}
\caption{
Plaquette (a) and Polyakov loop (b) at $\beta=0$ as a function of $1/K$
for various $N_F$ on $N_t=4$ lattices.
Lines are to guide the eyes.
In this figure, approximately vertical lines leaning to the right 
mean that we have two state signals there.
}
\label{B0.W}
\end{figure}

\begin{figure}[tb]
\centerline{
a) \epsfxsize=7.5cm\epsfbox{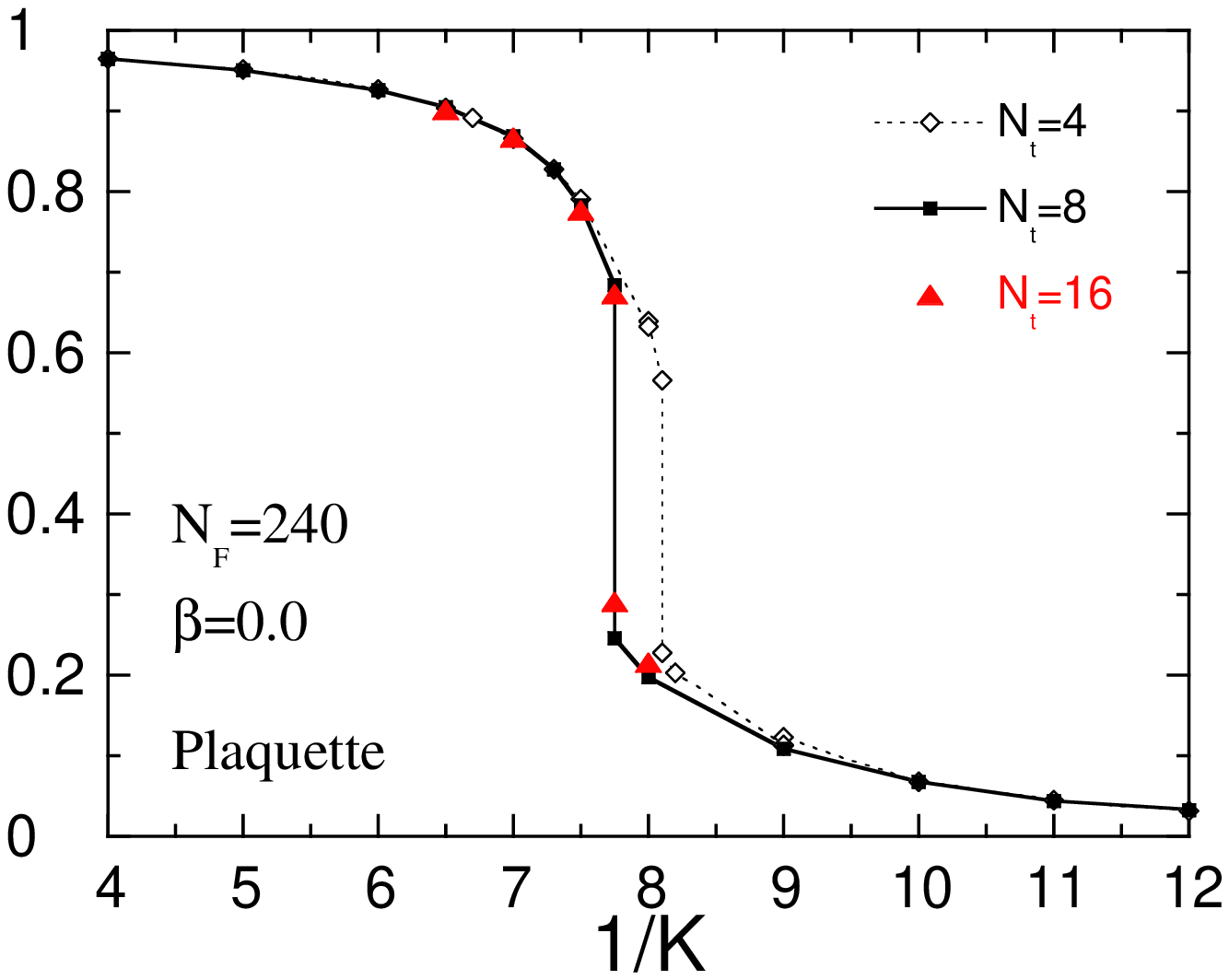}
}
\vspace{2mm}
\centerline{ 
b) \epsfxsize=7.5cm\epsfbox{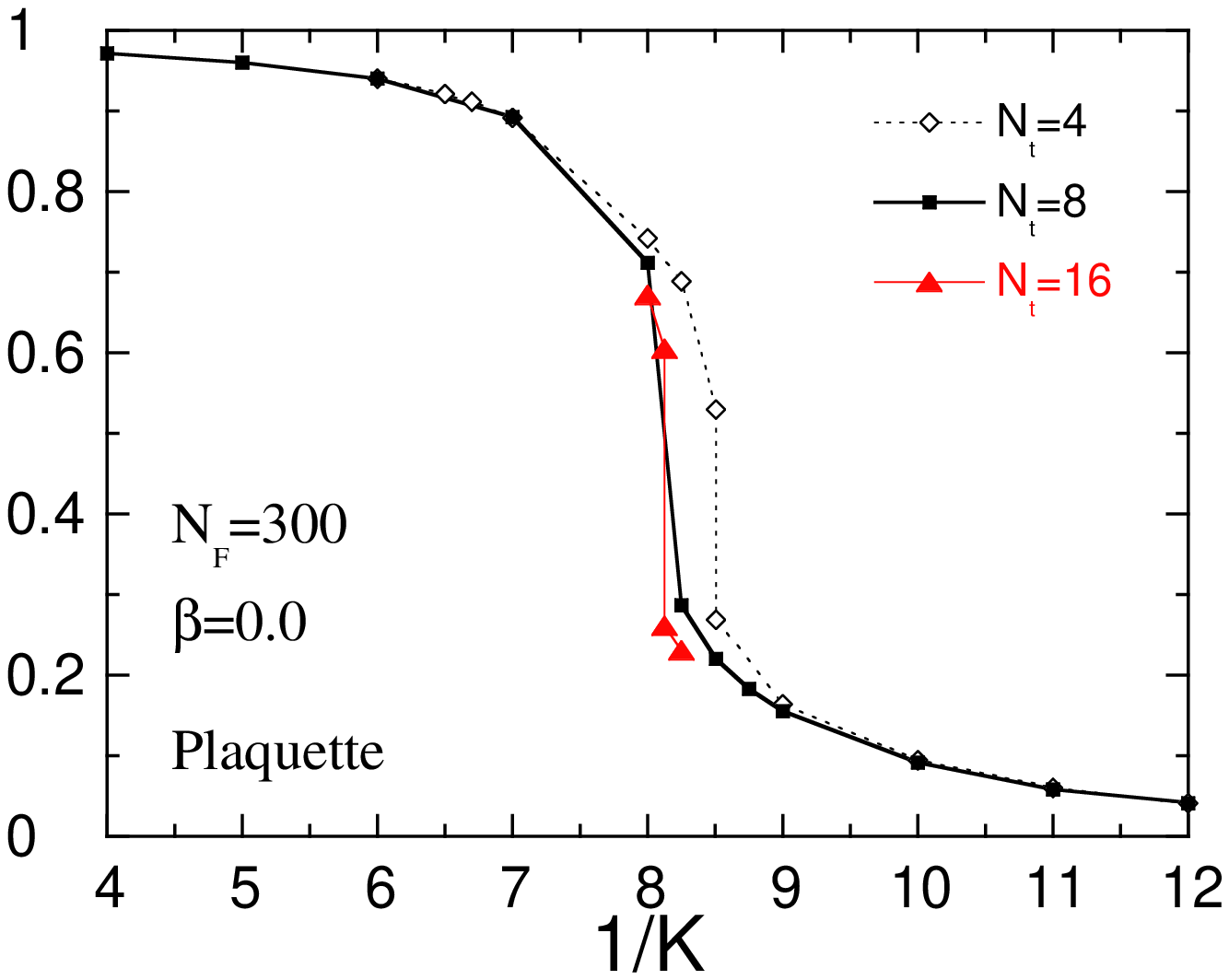}
}
\vspace{-0.1cm}
\caption{
The plaquette at $\beta=0$ as a function of $1/K$
for (a) $N_F=240$ and (b) $N_F=300$, at $N_t=4$, 8 and 16.
Lines are to guide the eyes.
}
\label{B0.W240}
\end{figure}

\begin{figure}[tb]
\centerline{
\epsfxsize=7.5cm \epsfbox{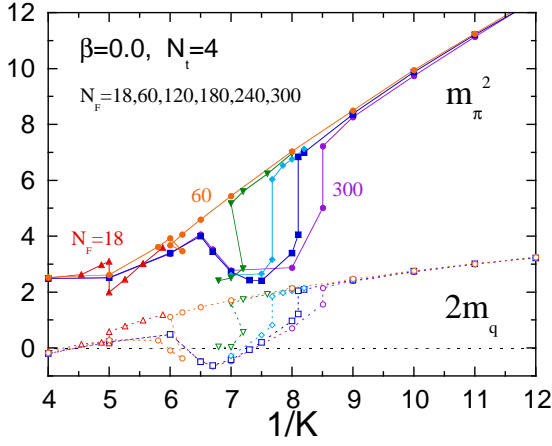}
}
\vspace{-0.1cm}
\caption{
Results of $m_\pi^2$ and $2m_q$ at $\beta=0$ for $N_F=18$--300.
}
\label{B0Nt4}
\end{figure}

\begin{figure}[tb]
\centerline{
\epsfxsize=8cm \epsfbox{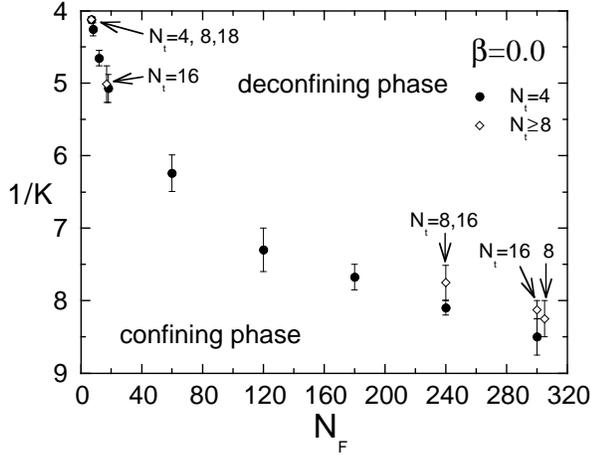}
}
\vspace{-0.1cm}
\caption{
The transition point $1/K_d$ at $\beta=0$ versus $N_F$
for $N_t=4$ and $N_t\ge 8$.
For clarity, data at $N_t=8$ for $N_F=300$ is slightly shifted
to a larger $N_F$ in the figure.
}
\label{Kd.Nf}
\end{figure}

\section{Results of numerical simulations at $g=\infty$}
\label{sec:strong}

We first study the strong coupling limit $\beta=0$ ($g=\infty$).
In a previous work \cite{previo} we have shown the following.
When $N_F \leq 6$, we have only one confined phase from the heavy
quark limit $K=0$ up to the chiral limit $K_c=0.25$.
On the other hand,
for $N_F \geq 7$, we found a strong first order transition at
$K=K_d < K_c$. 
We confirmed that this transition $K_d$ is a bulk phase transition
(transition at zero-temperature) by 
increasing the value of $N_t$ up to $N_t=18$ for $N_F=7$.
When quarks are heavy ($K < K_d$), both the plaquette and the Polyakov
loop are small, and $m_\pi$ satisfies the PCAC relation
$m_\pi^2 \propto m_q$.
Therefore quarks are confined and the chiral
symmetry is spontaneously broken in this phase.
We found that $m_q$ in the confined phase is non-zero at the
transition point $K_d$,
{\it i.e.}, the chiral limit does not belong to this phase.
On the other hand, when quarks are light ($K > K_d$), the plaquette and
the Polyakov loop are large.
In this phase, $m_\pi$ remains large in the chiral limit
and is almost equal to twice the lowest Matsubara frequency $\pi/N_t$.
This implies that the pion state is an almost free two-quark state
and, therefore, quarks are not confined in this phase.
The pion mass is nearly equal to the scalar meson mass,
and the rho meson mass to the axial vector meson mass.
The chiral symmetry is also manifest within
corrections due to finite lattice spacing.

In this paper, we extend the study to larger $N_F$.
In Fig.~\ref{B0.W}, we show the results of the plaquette and Polyakov loop 
at $N_t=4$ for $N_F =7$ -- 300.
Clear first order transitions can be seen at $1/K$ larger than $1/0.25=4$.
We then study the $N_t$ dependence of the results,
as shown in Fig.~\ref{B0.W240} for the cases $N_F=240$ and 300.
We find that,
although the transition point shows a slight shift to smaller $1/K$
when we increase $N_t$ from 4 to 8, the transition stays at
the same point for $N_t \geq 8$.
For $N_F=240$, $1/K_d \simeq 8.1(1)$ at $N_t=4$ and 
$1/K_d = 7.8(2)$ at $N_t=8$ and 16. 
A similar result was reported for $N_F=7$ in \cite{previo}.
We conclude that the transition is a bulk transition
for $N_F \geq 7$.

Figure~\ref{B0Nt4} shows the results of $m_\pi^2$ and $2m_q$,
in the lattice units, at $\beta=0$ for
various numbers of flavors. We clearly see that at the exactly same 
hopping parameter $K=K_d$ where the plaquette makes a gap, the pion mass
and the quark mass also make gaps. When the quark is heavy ($K \le K_d$),
the PCAC relation ($m_\pi^2 \propto m_q$) is well satisfied. 
On the other hand, when the quark mass is smaller than the critical value,
the $1/K$ dependence of the pion mass squared $m_\pi^2$ and the
quark mass $m_q$ looks strange at first sight.
However, when one compares this dependence of $m_q$ with that of the
free quark case shown in Fig.~\ref{Binf}(a), one easily notices
that the $1/K$ dependence is essentially the same as that of the free
quark.
The quark mass vanishes at $1/K \simeq 7$--8 for $N_F=60$, 240, and 300.
This corresponds to the fact that the free quark mass vanishes at $1/K=8$.
The $1/K$ dependence of $m_\pi^2$ is also essentially the same as
that of the free quark case shown in Fig.~\ref{Binf}(b).
We stress that 
the chiral limit where the quark mass vanishes does not exist
in the confined phase.

We show a map of the phase transition
point $K=K_d$ at $\beta=0$ in Fig.~\ref{Kd.Nf}.

\begin{figure}[tb]
\centerline{
a)\epsfxsize=7.8cm\epsfbox{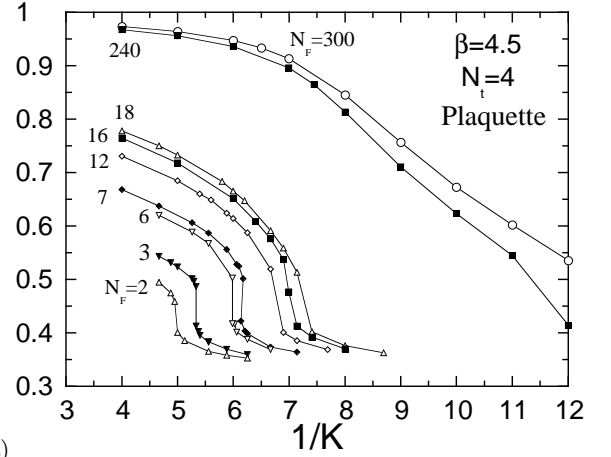}
}
\vspace{2mm}
\centerline{ 
b) \epsfxsize=7.8cm\epsfbox{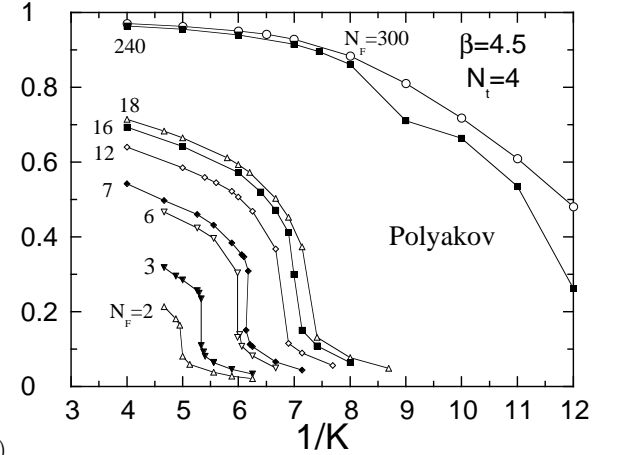}
}
\vspace{-0.1cm}
\caption{
Results of (a) plaquette and (b) Polyakov loop for various $N_F$ obtained at 
$\beta=4.5$ on $N_t=4$ lattices.
}
\label{B4.5wp}
\end{figure}

\begin{figure}[tb]
\centerline{
a)\epsfxsize=7.8cm\epsfbox{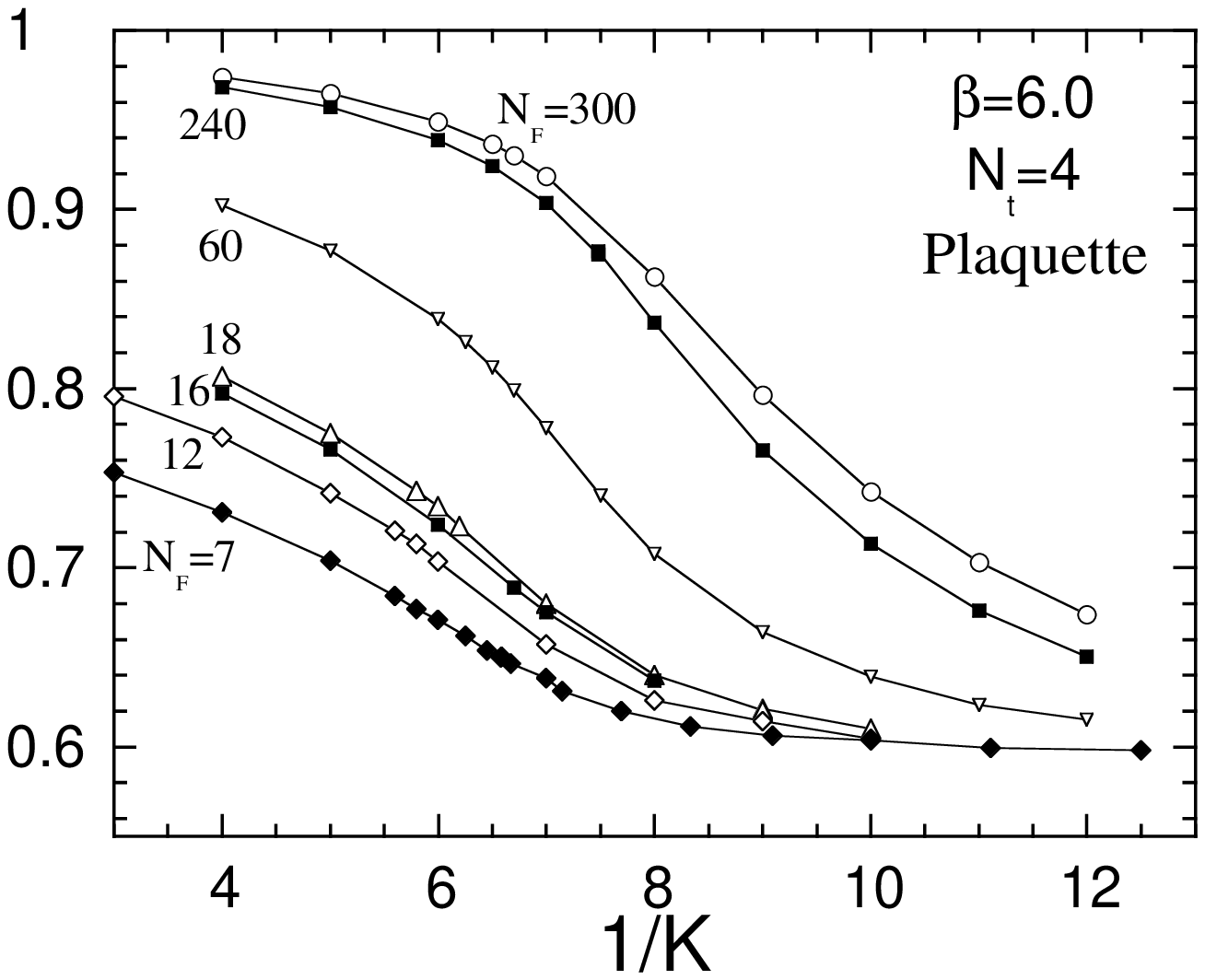}
}
\vspace{2mm}
\centerline{ 
b) \epsfxsize=7.8cm\epsfbox{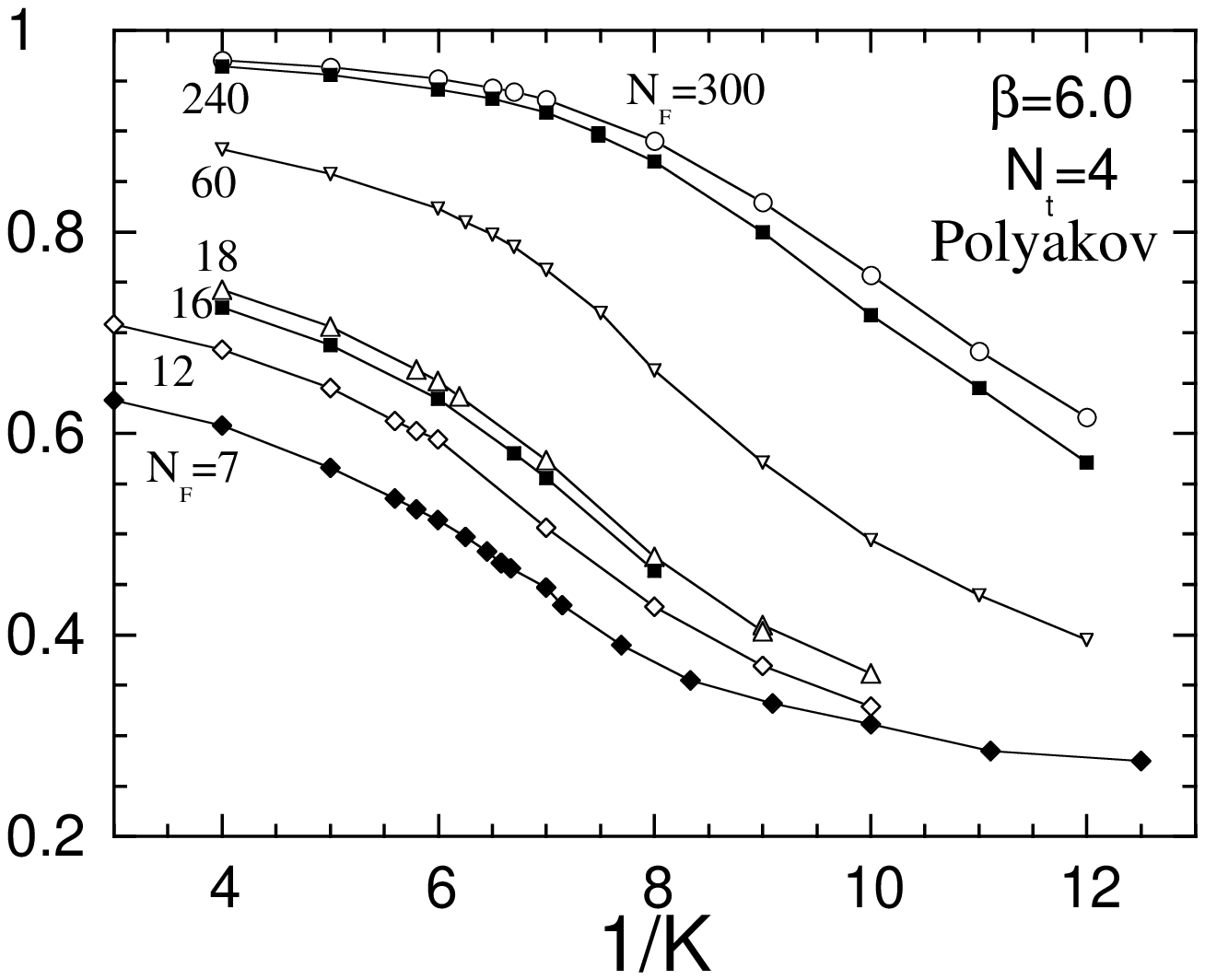}
}
\vspace{-0.1cm}
\caption{
The same as Fig.~\protect\ref{B4.5wp}, but for $\beta=6.0$.
}
\label{B6.0wp}
\end{figure}

\begin{figure}[tb]
\centerline{
a) \epsfxsize=7.5cm\epsfbox{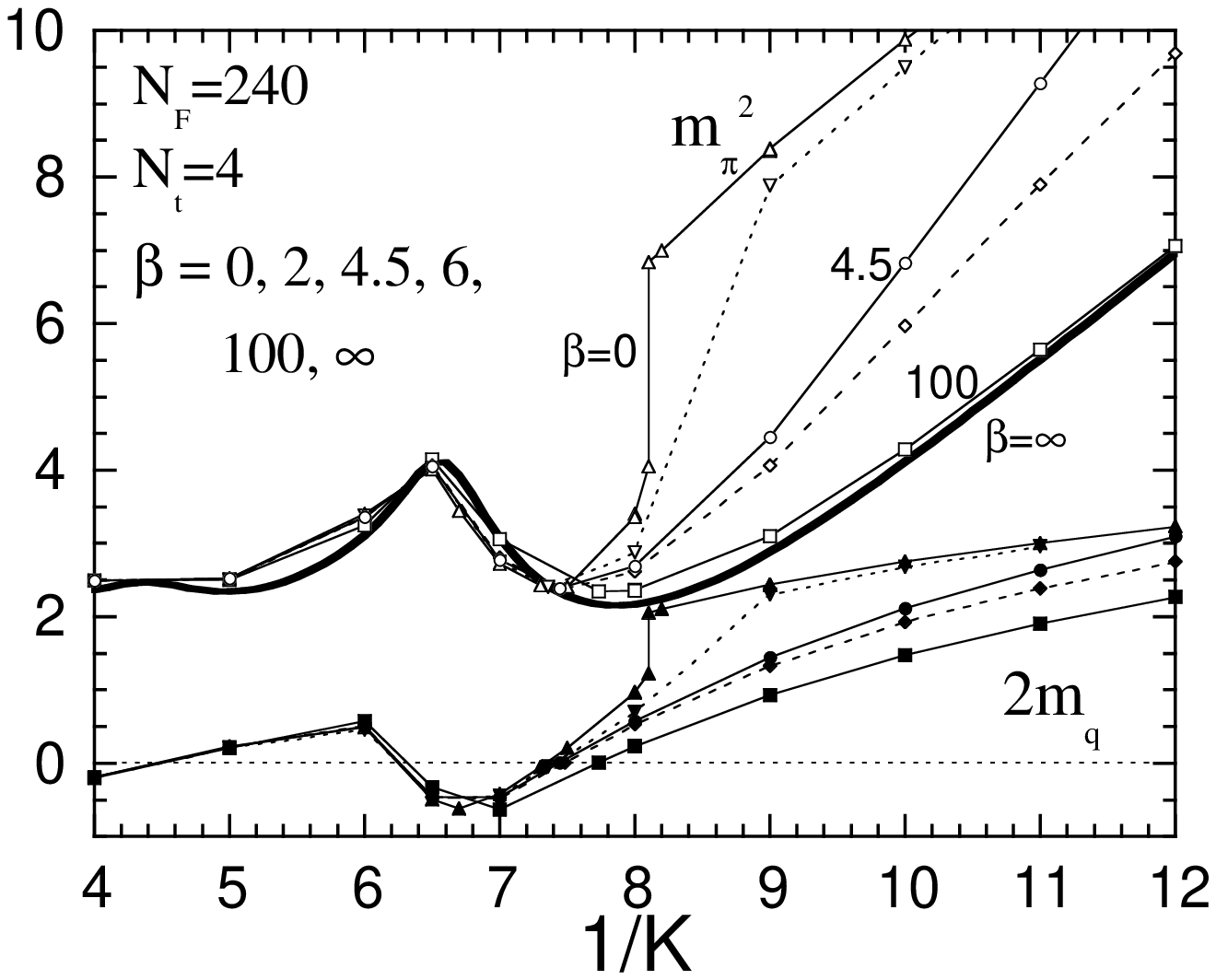}
}
\vspace{2mm}
\centerline{ 
b) \epsfxsize=7.5cm\epsfbox{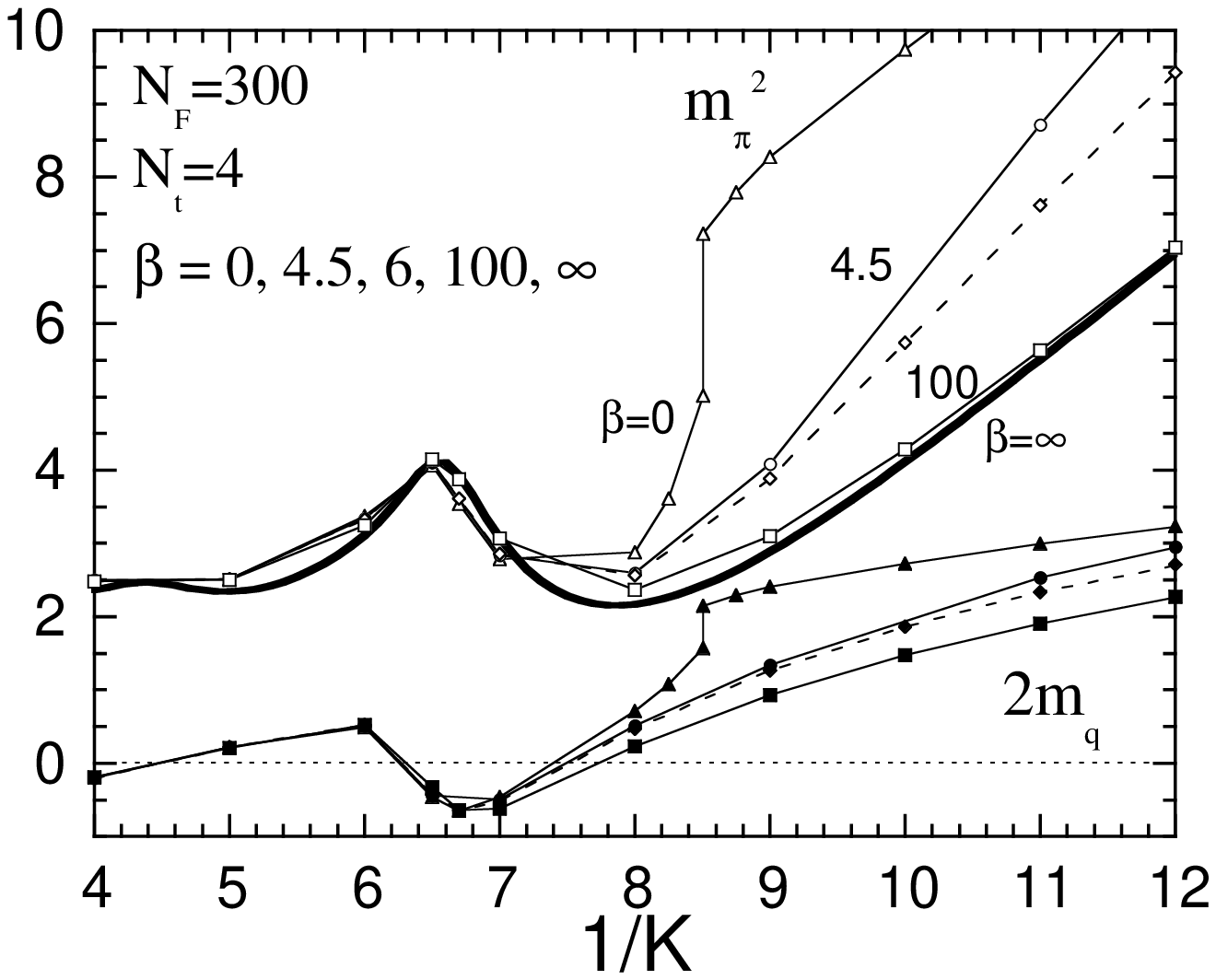}
}
\vspace{-0.1cm}
\caption{
$m_\pi^2$ and 2$m_q$ for (a) $N_F=240$ and (b) 300, 
obtained on $N_t=4$ lattices at various $\beta$.
}
\label{massNf240}
\end{figure}

\begin{figure}[tb]
\centerline{
\epsfxsize=7.5cm\epsfbox{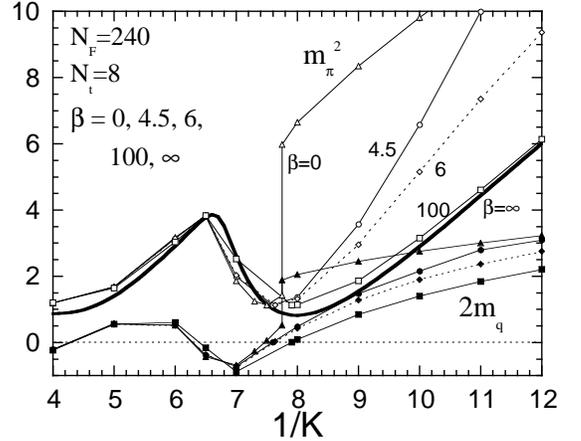}
}
\vspace{-0.1cm}
\caption{
The same as Fig.~\protect\ref{massNf240}, but for $N_F=240$ 
and $N_t=8$.
}
\label{massNf240T8}
\end{figure}

\begin{figure}[tb]
\centerline{
a) \epsfxsize=7.5cm\epsfbox{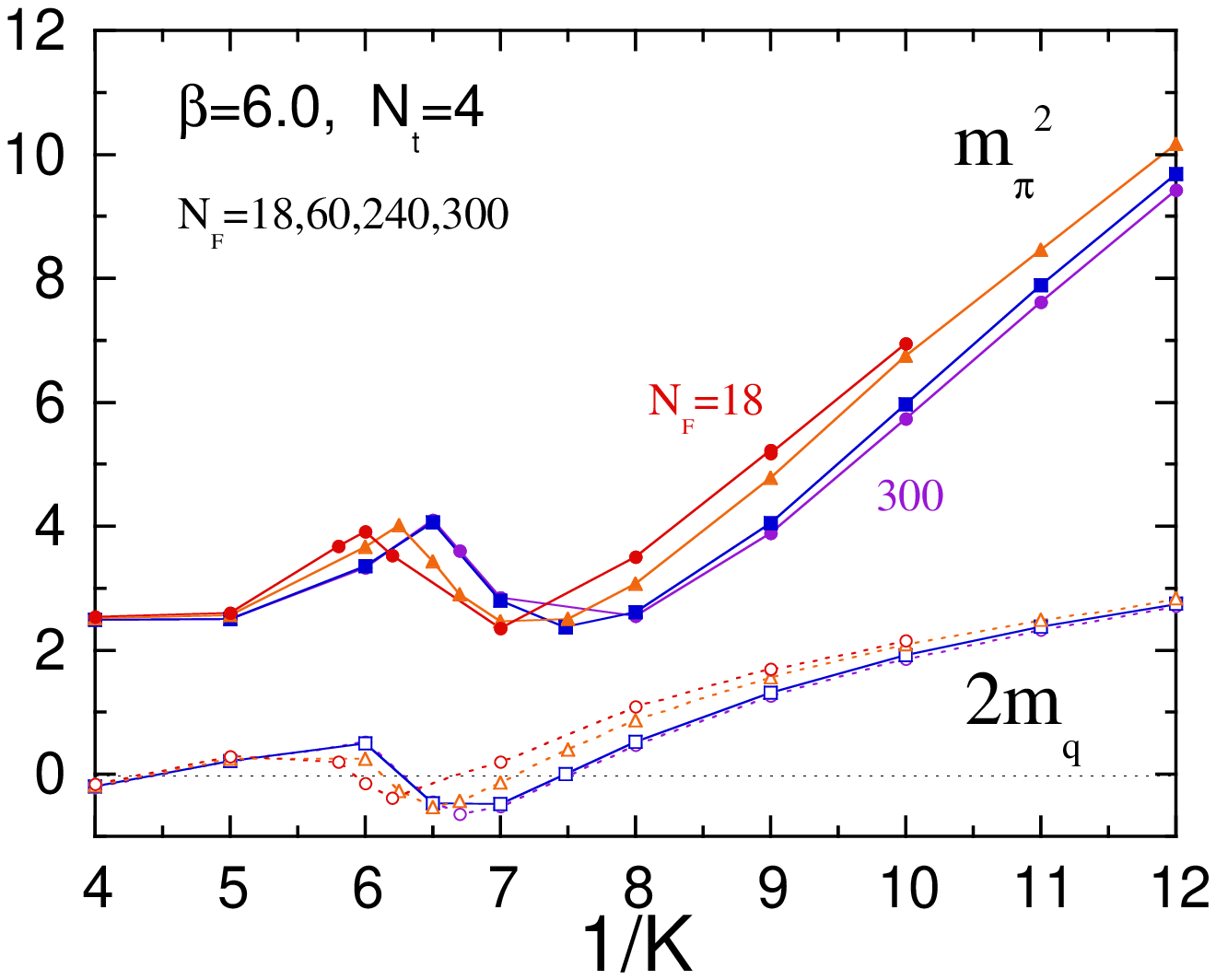}
}
\vspace{2mm}
\centerline{ 
b) \epsfxsize=7.5cm\epsfbox{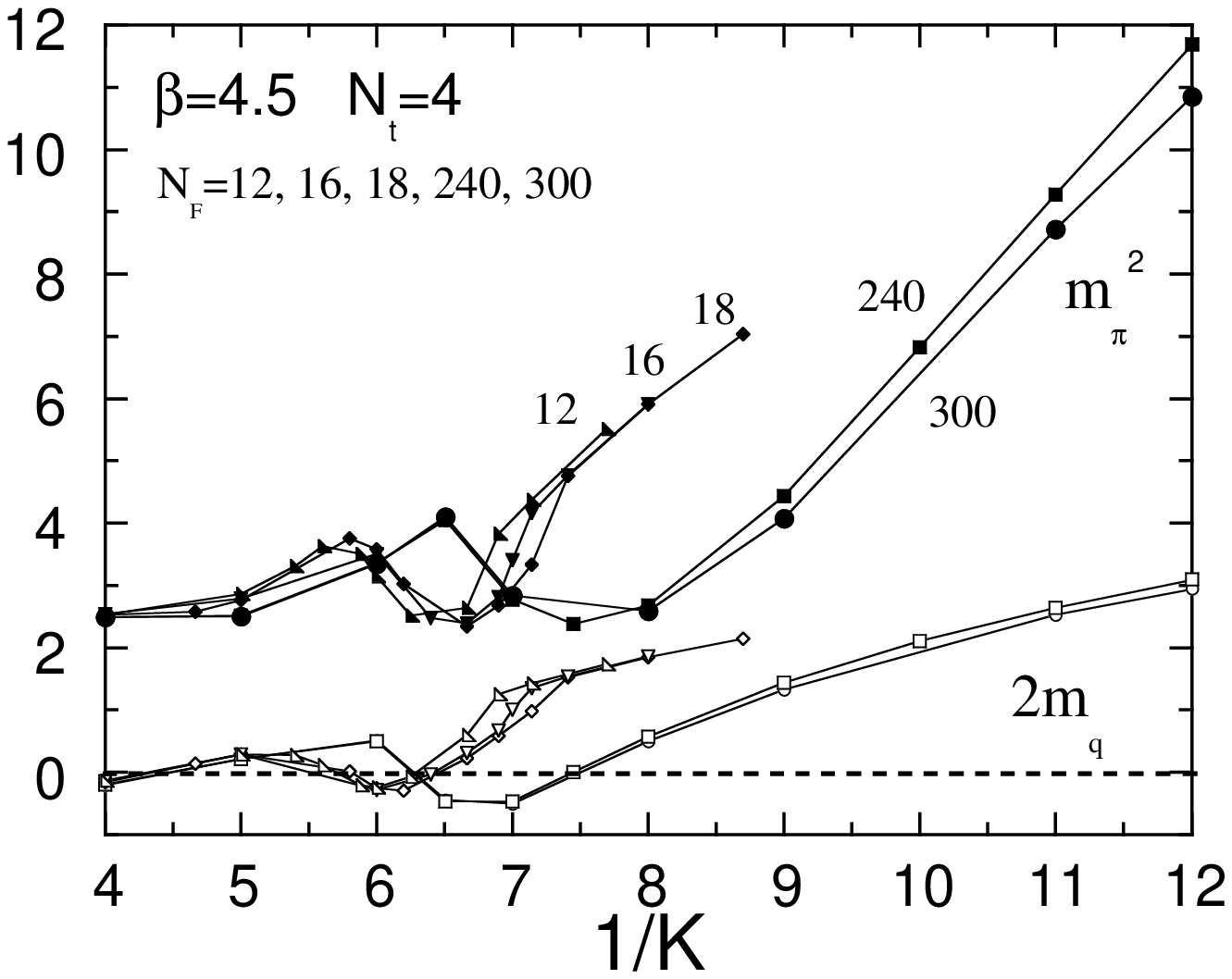}
}
\vspace{-0.1cm}
\caption{
Results of $m_\pi^2$ and 2$m_q$ for large $N_F$ obtained on $N_t=4$
lattices at 
(a) $\beta=6.0$
and
(b) $\beta=4.5$.
}
\label{massNfLarge}
\end{figure}

\begin{figure}[tb]
\centerline{
a) \epsfxsize=7.5cm\epsfbox{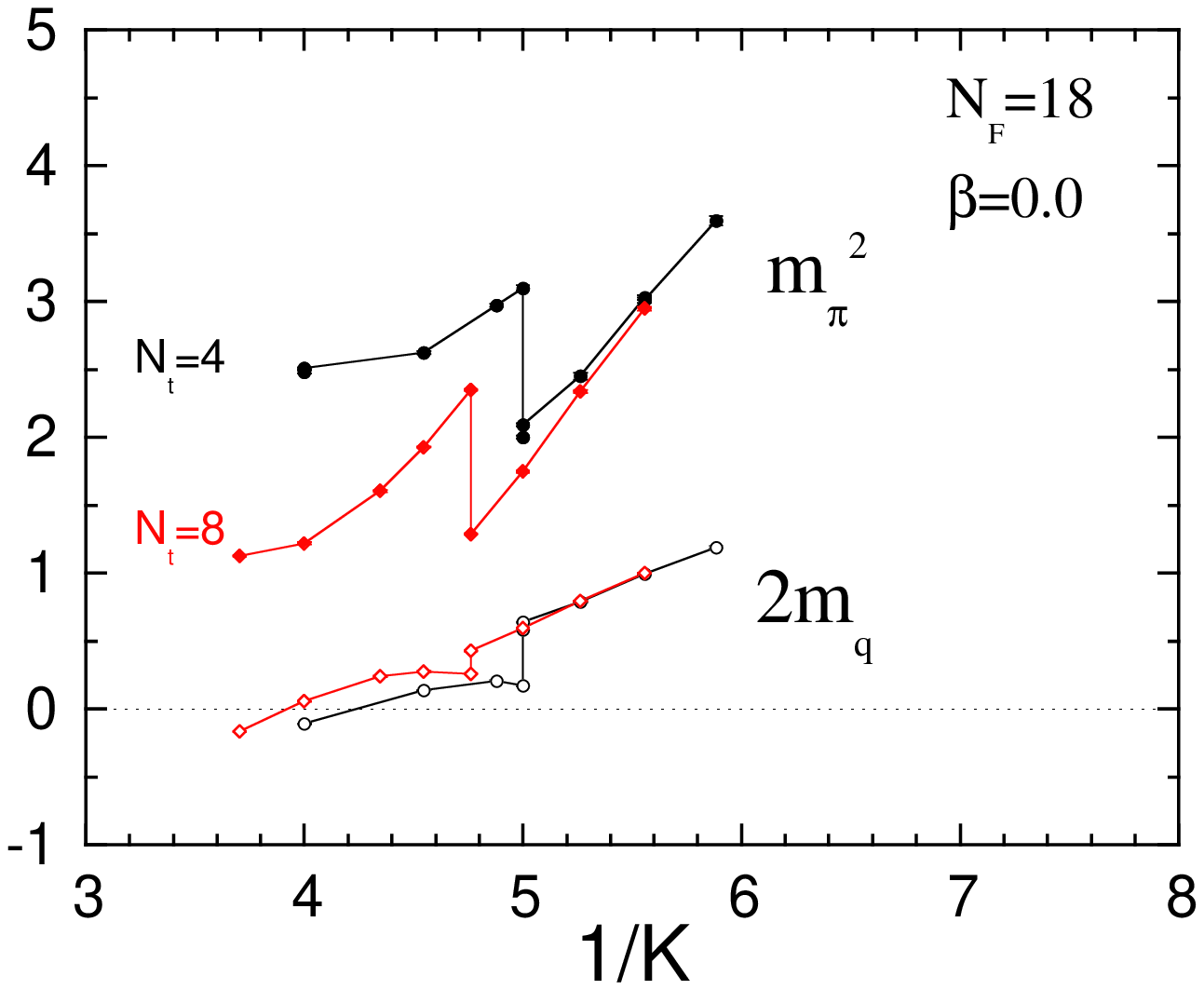}
} 
\vspace{2mm} 
\centerline{
b) \epsfxsize=7.5cm\epsfbox{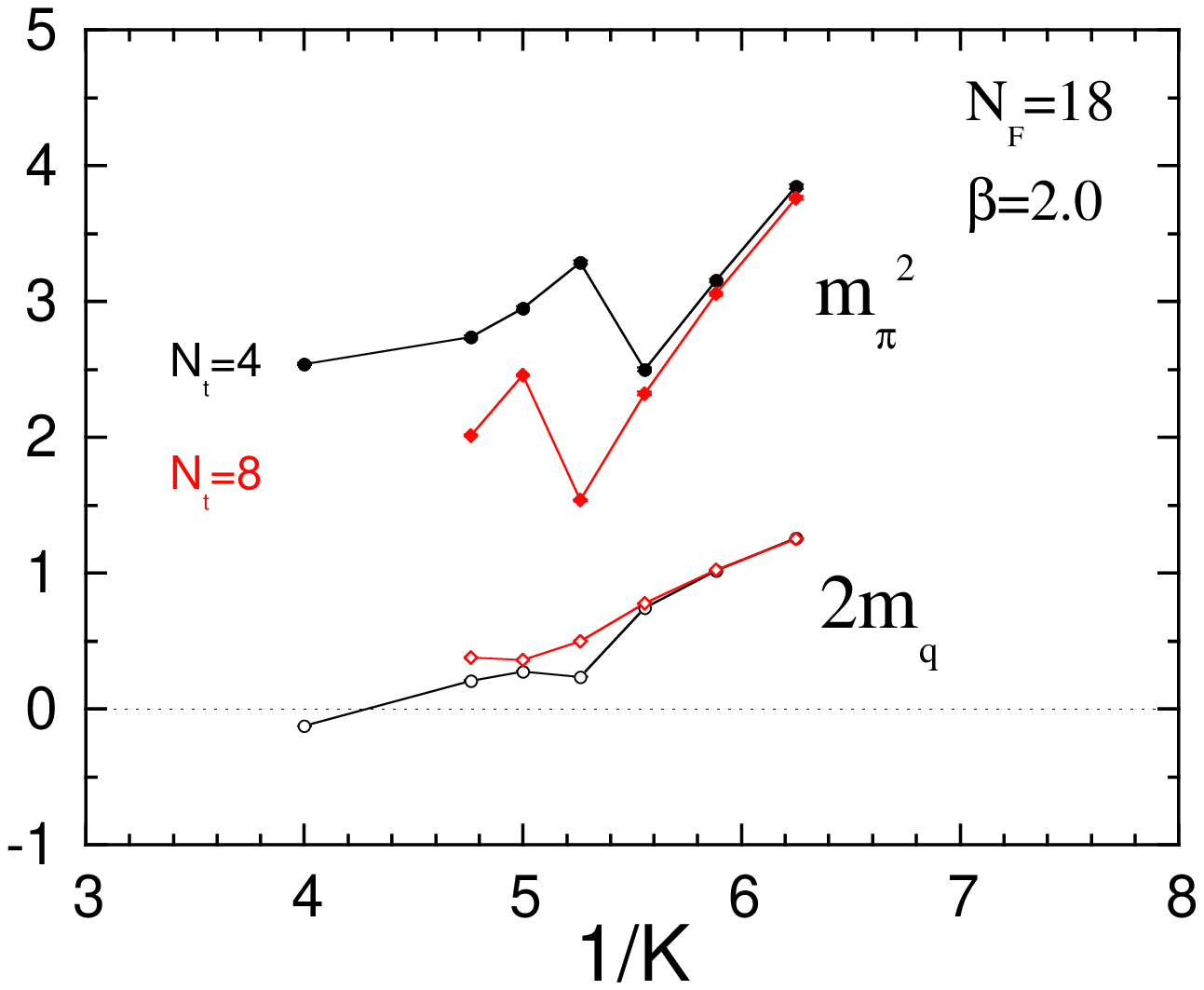}
}
\vspace{-0.1cm}
\caption{
$N_F=18$: 
Results of $m_\pi^2$ and 2$m_q$ versus $1/K$.
(a) $\beta=0.0$
and
(b) $\beta=2.0$,
}
\label{Nf18_1}
\end{figure}

\begin{figure}[tb]
\centerline{
a) \epsfxsize=7.5cm\epsfbox{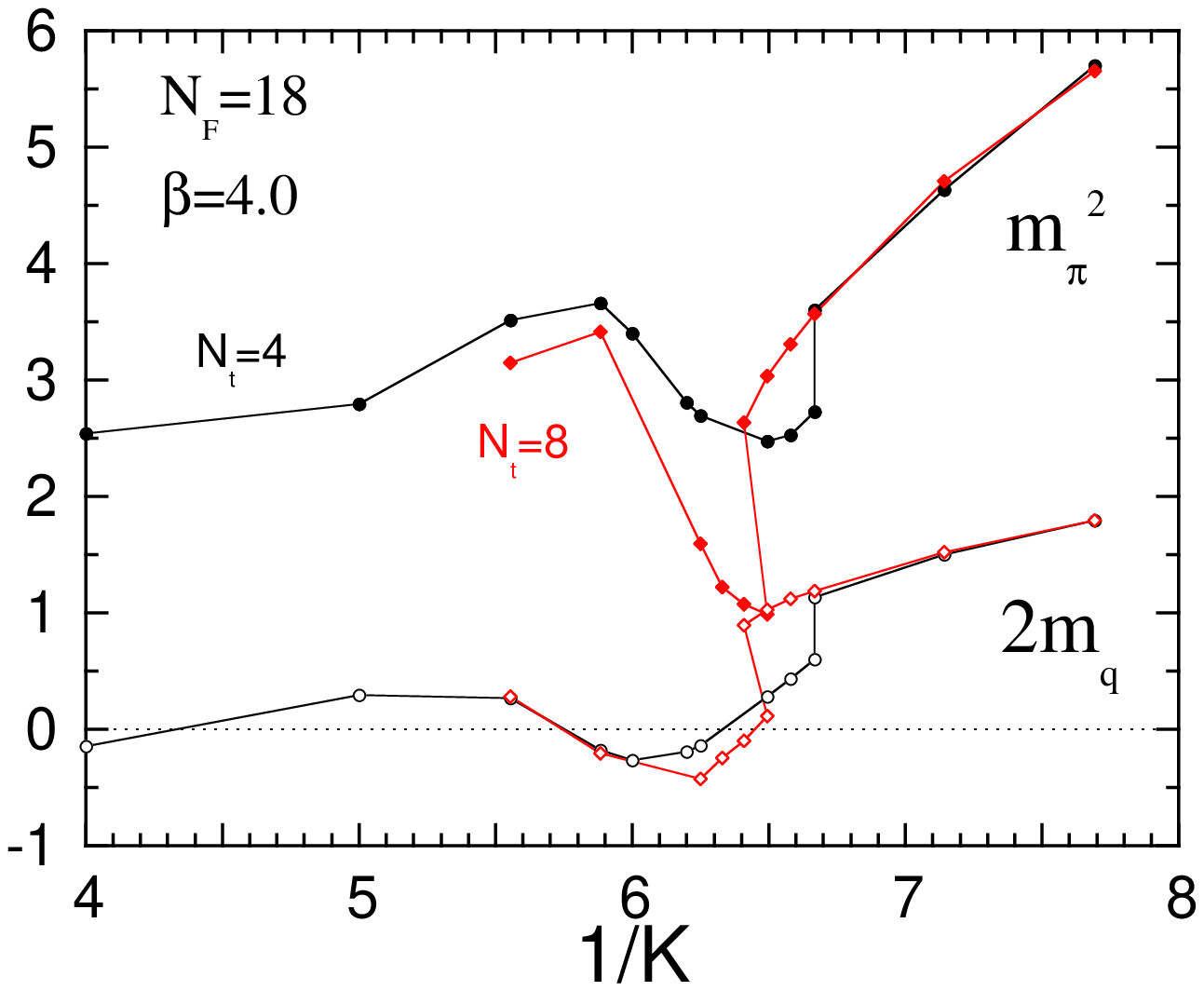}
}
\vspace{2mm} 
\centerline{
b) \epsfxsize=7.5cm\epsfbox{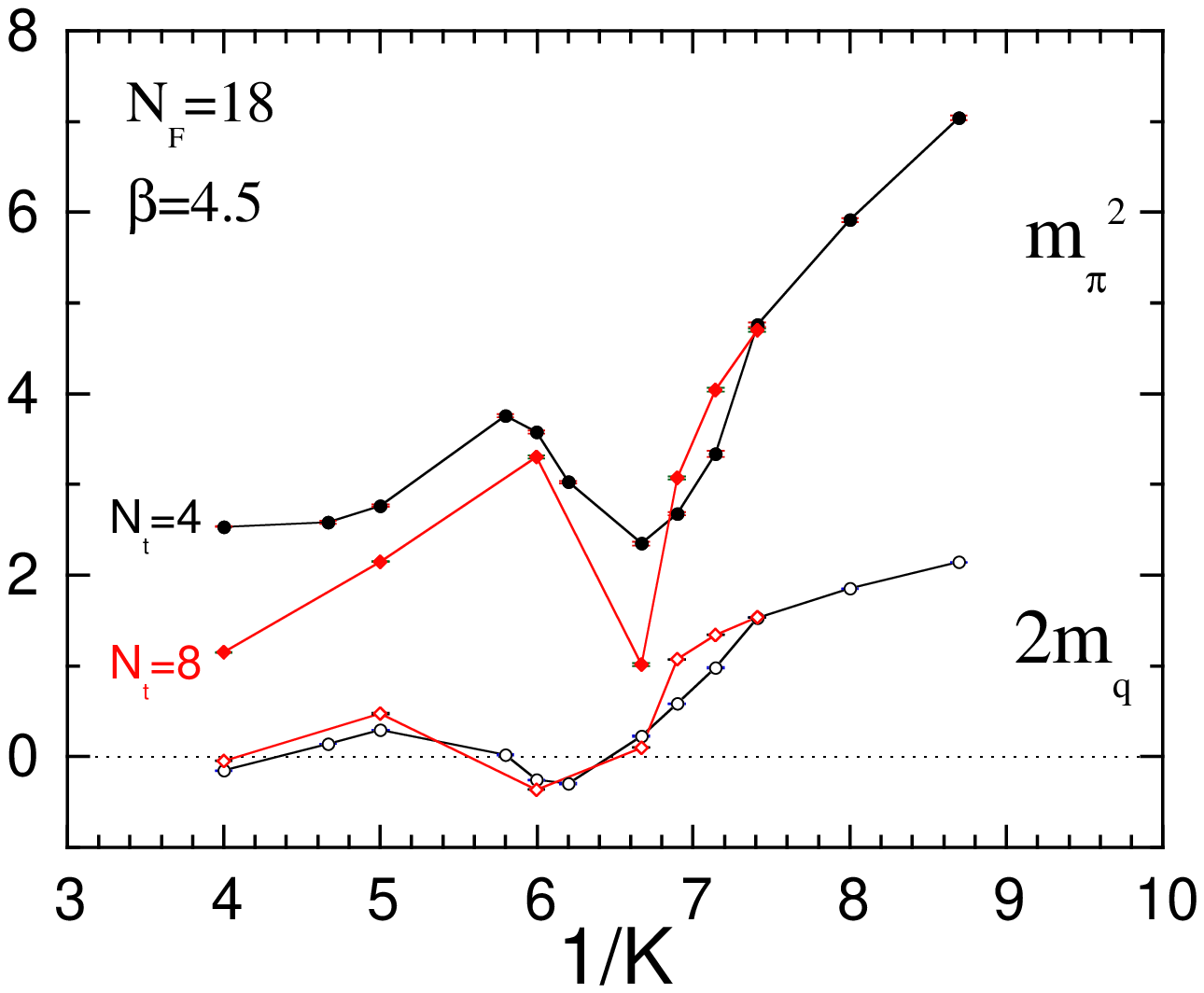}
} 
\vspace{2mm} 
\centerline{  
c) \epsfxsize=7.5cm\epsfbox{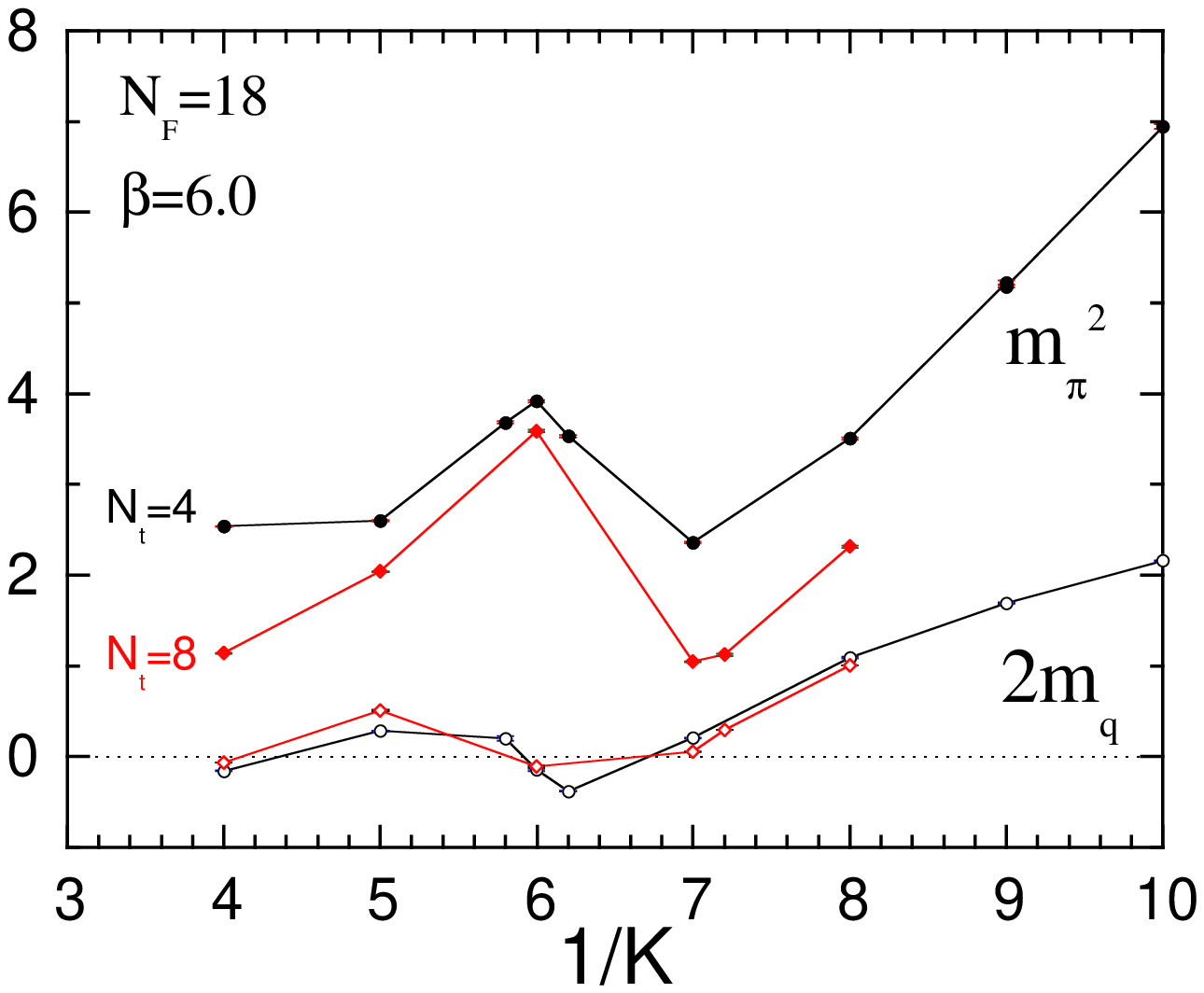}
}
\vspace{-0.1cm}
\caption{
$N_F=18$: 
Results of $m_\pi^2$ and 2$m_q$ versus $1/K$.
(a) $\beta=4.0$,
(b) $\beta=4.5$,
and
(c) $\beta=6.0$.
}
\label{Nf18_2}
\end{figure}

\begin{figure}[tb]
\centerline{
a) \epsfxsize=7.5cm\epsfbox{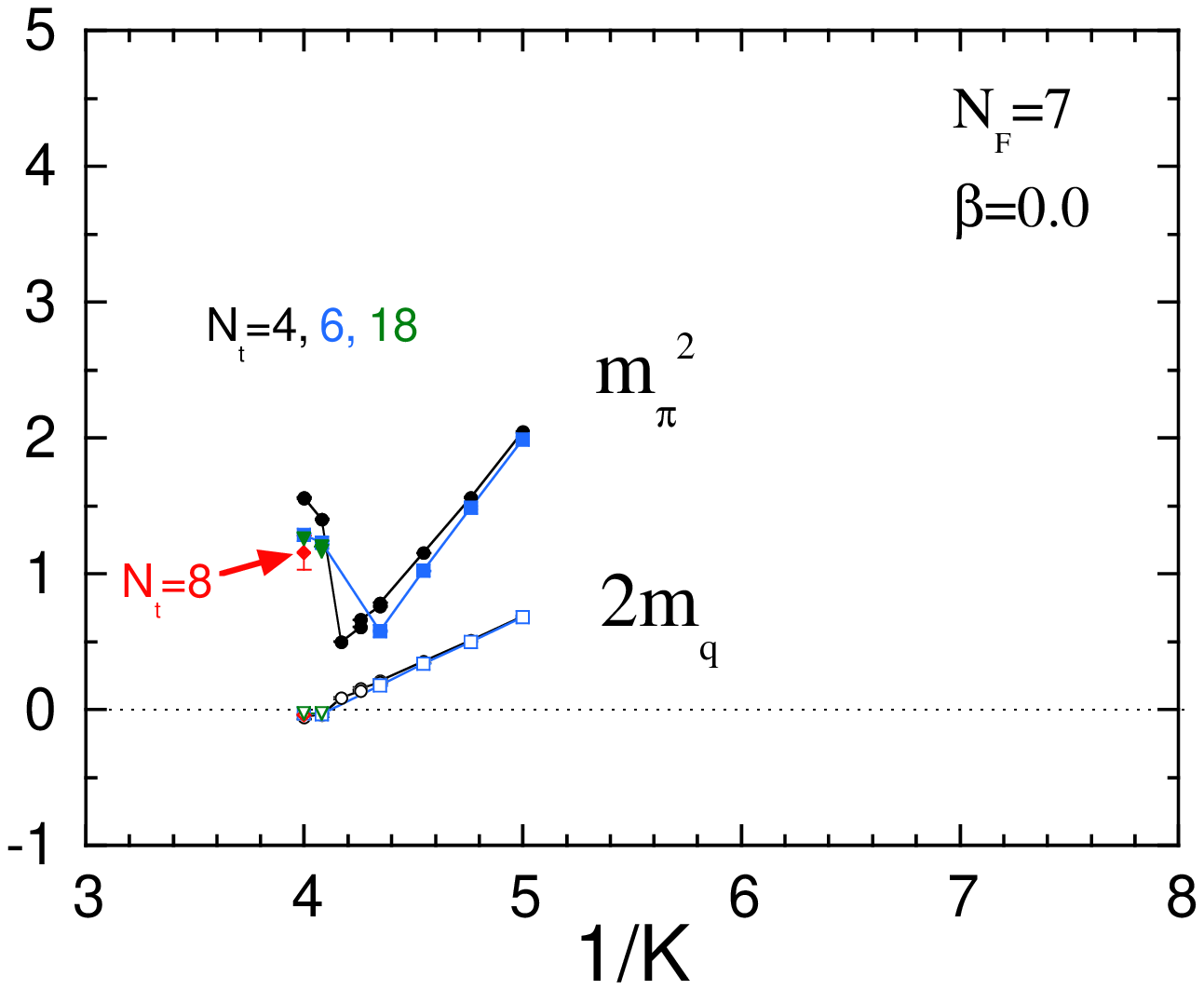}
} 
\vspace{2mm} 
\centerline{
b) \epsfxsize=7.5cm\epsfbox{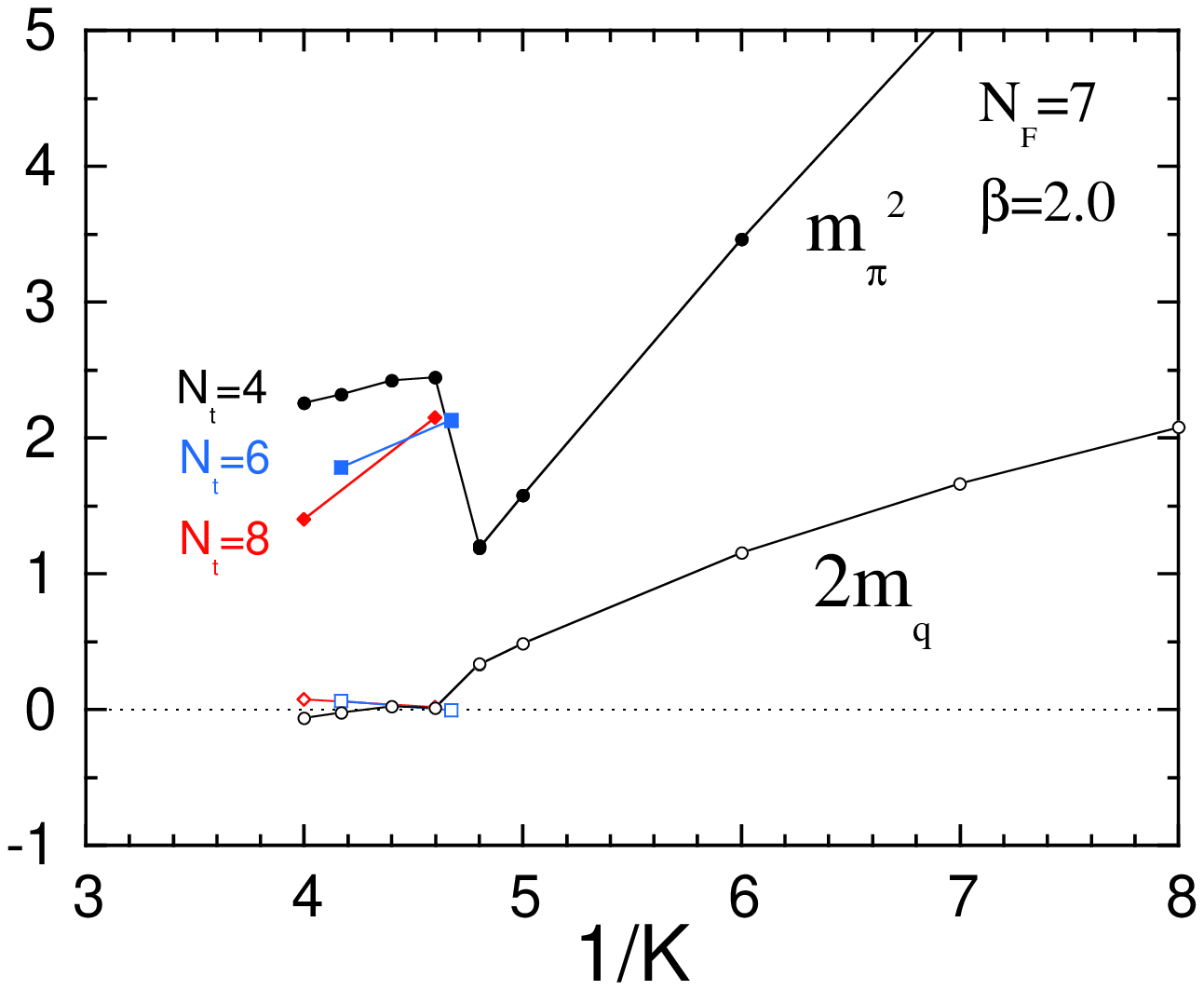}
}
\vspace{-0.1cm}
\caption{
$N_F=7$: 
Results of $m_\pi^2$ and 2$m_q$ versus $1/K$.
(a) $\beta=0.0$,
(b) $\beta=2.0$.
}
\label{Nf7_1}
\end{figure}

\begin{figure}[tb]
\centerline{
a) \epsfxsize=7.5cm\epsfbox{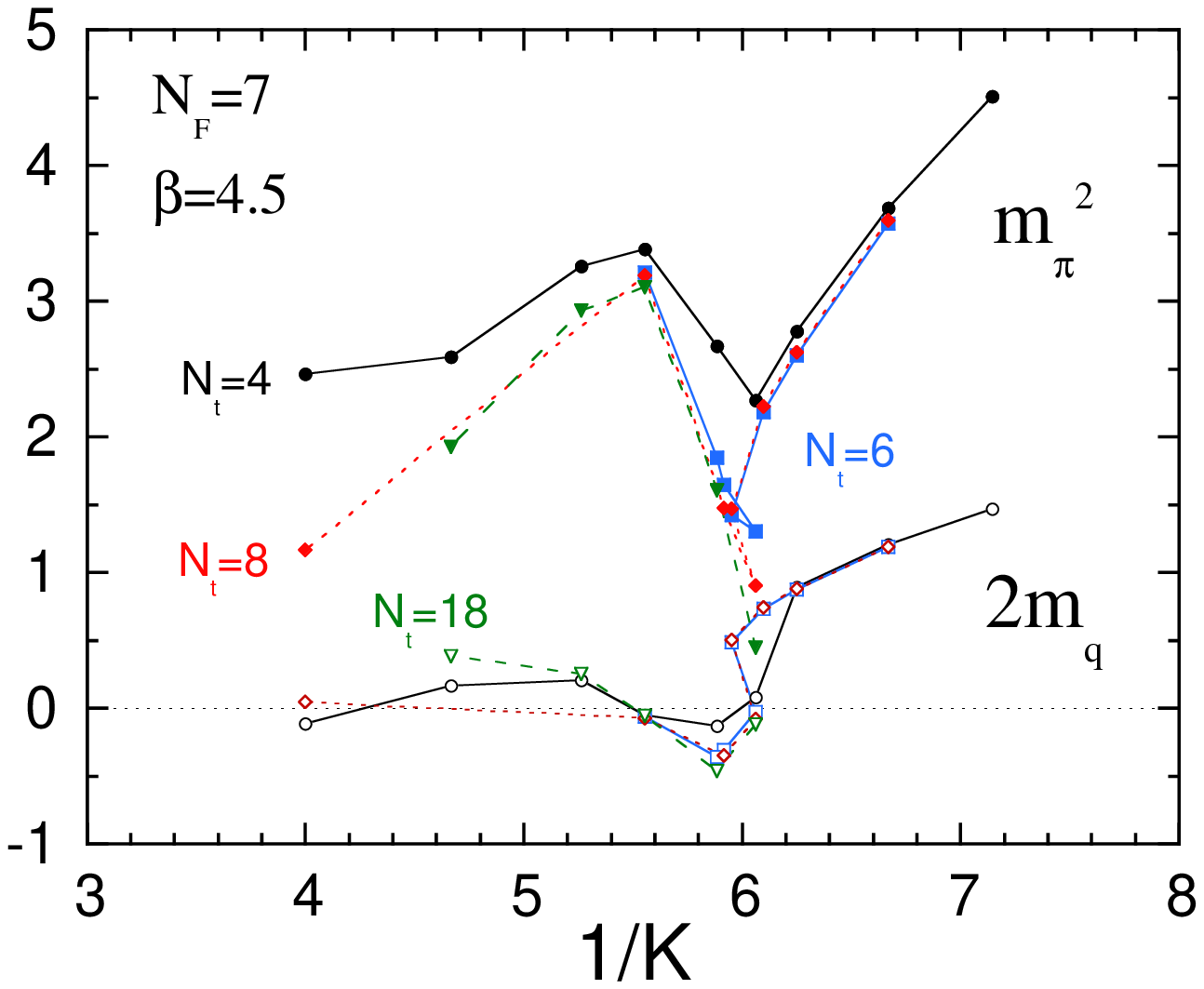}
} 
\vspace{2mm} 
\centerline{  
b) \epsfxsize=7.5cm\epsfbox{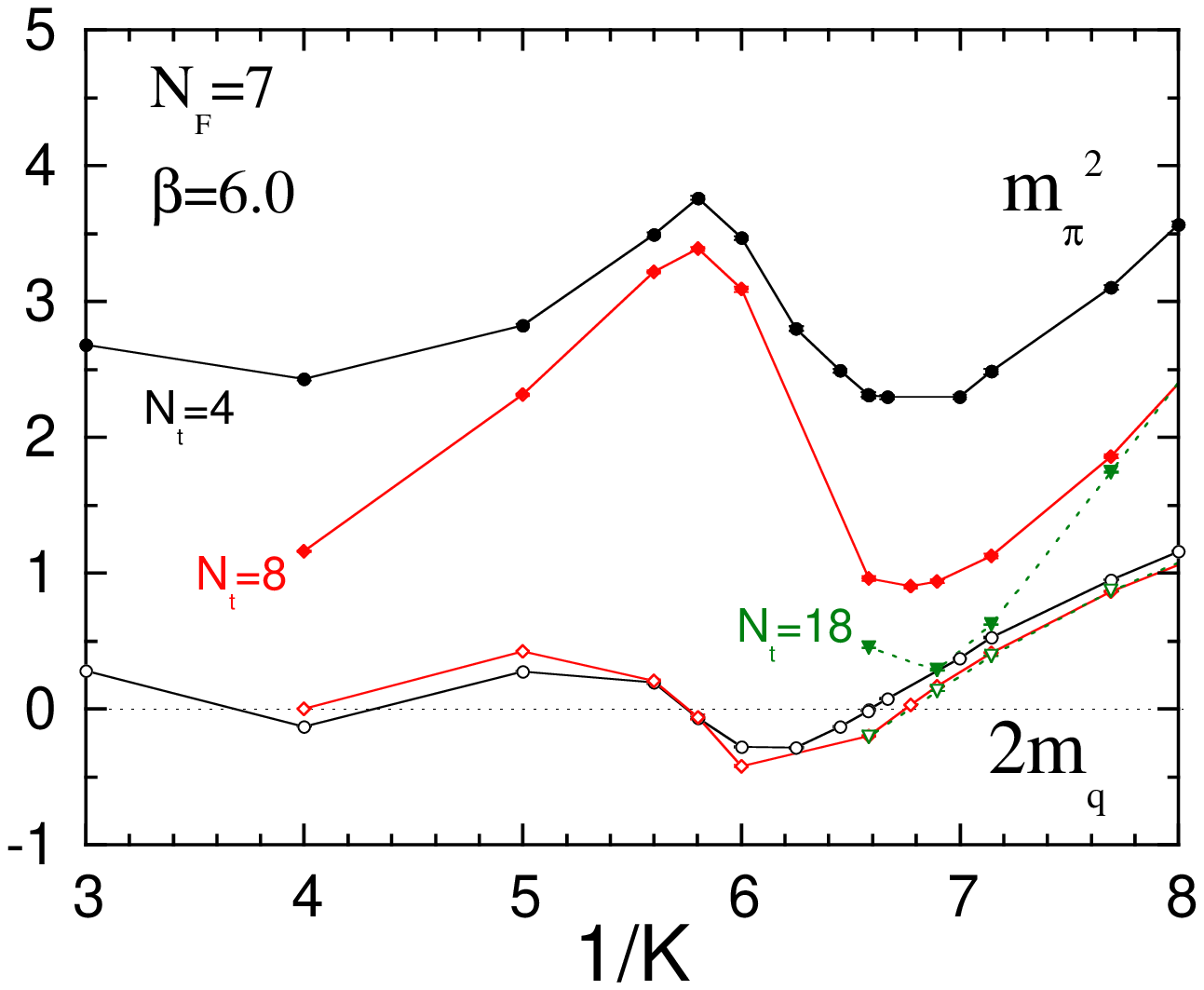}
}
\vspace{-0.1cm}
\caption{
$N_F=7$: 
Results of $m_\pi^2$ and 2$m_q$ versus $1/K$.
(a) $\beta=4.5$,
(b) $\beta=6.0$.
}
\label{Nf7_2}
\end{figure}

\begin{figure}[tb]
\centerline{
a) \epsfxsize=7.5cm\epsfbox{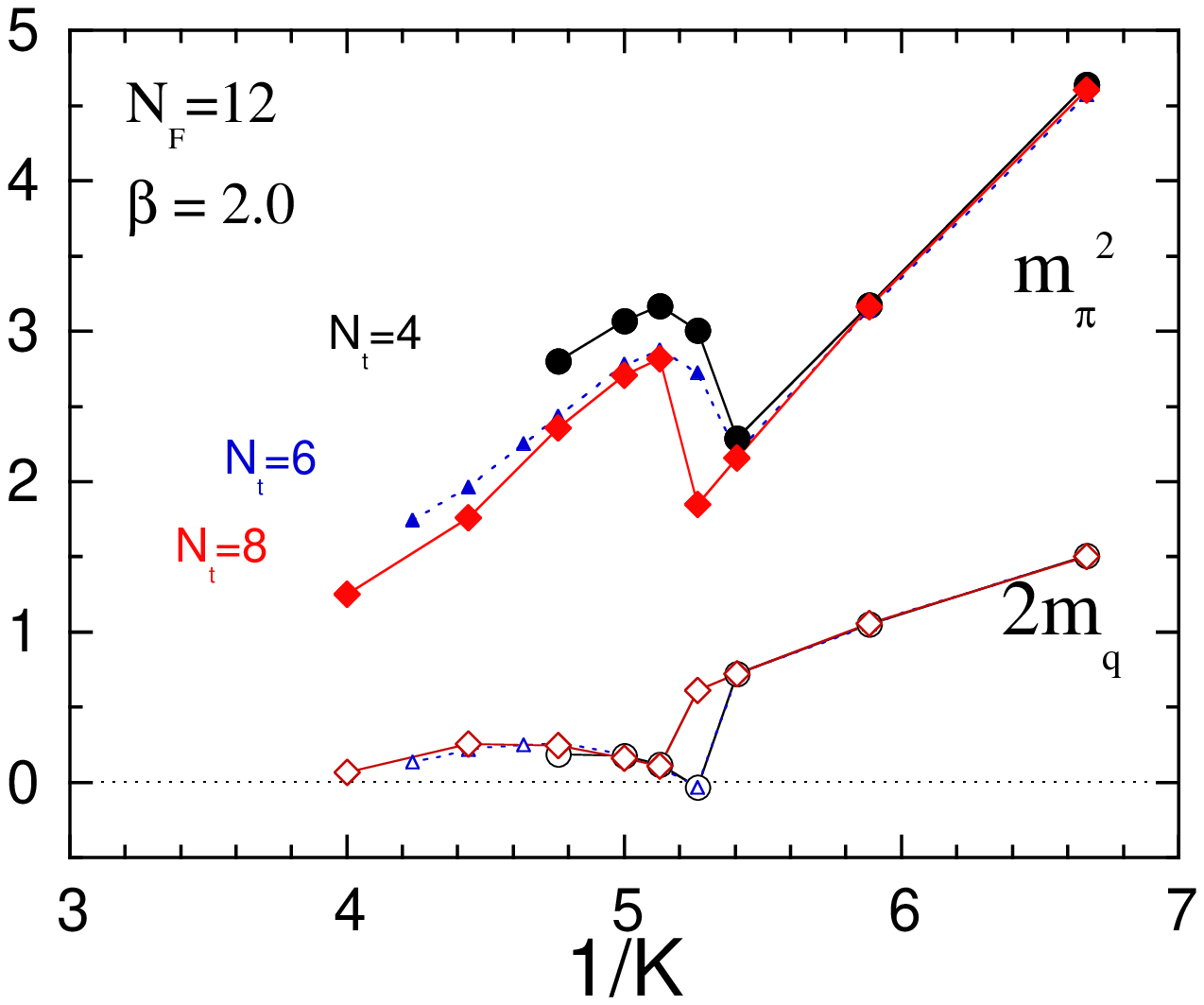}
} 
\vspace{2mm} 
\centerline{  
b) \epsfxsize=7.5cm\epsfbox{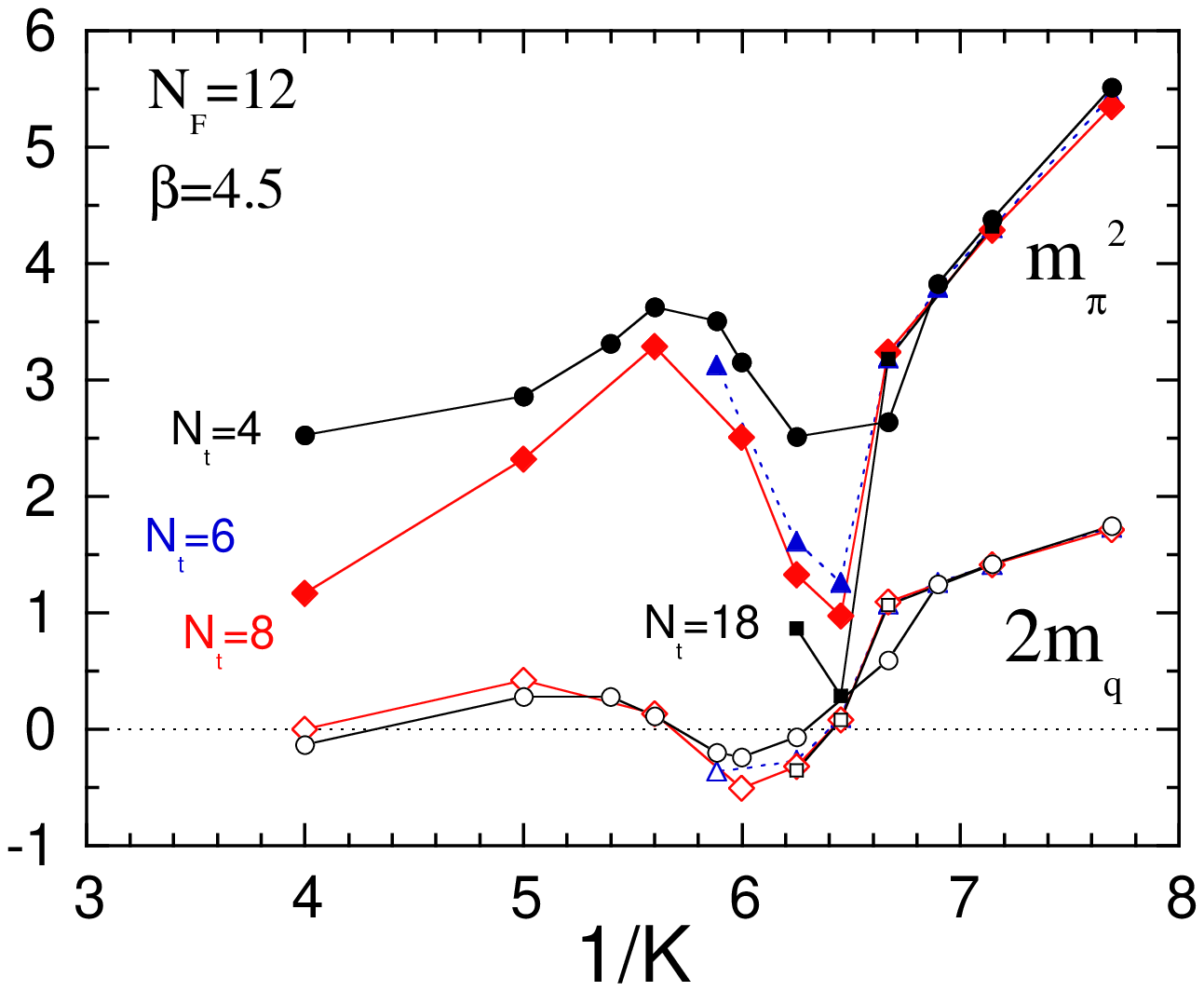}
}
\vspace{-0.1cm}
\caption{
$N_F=12$: 
Results of $m_\pi^2$ and 2$m_q$ versus $1/K$.
(a) $\beta=2.0$,
(b) $\beta=4.5$.
}
\label{Nf12}
\end{figure}

\begin{figure}[tb]
\centerline{
a) \epsfxsize=7.5cm\epsfbox{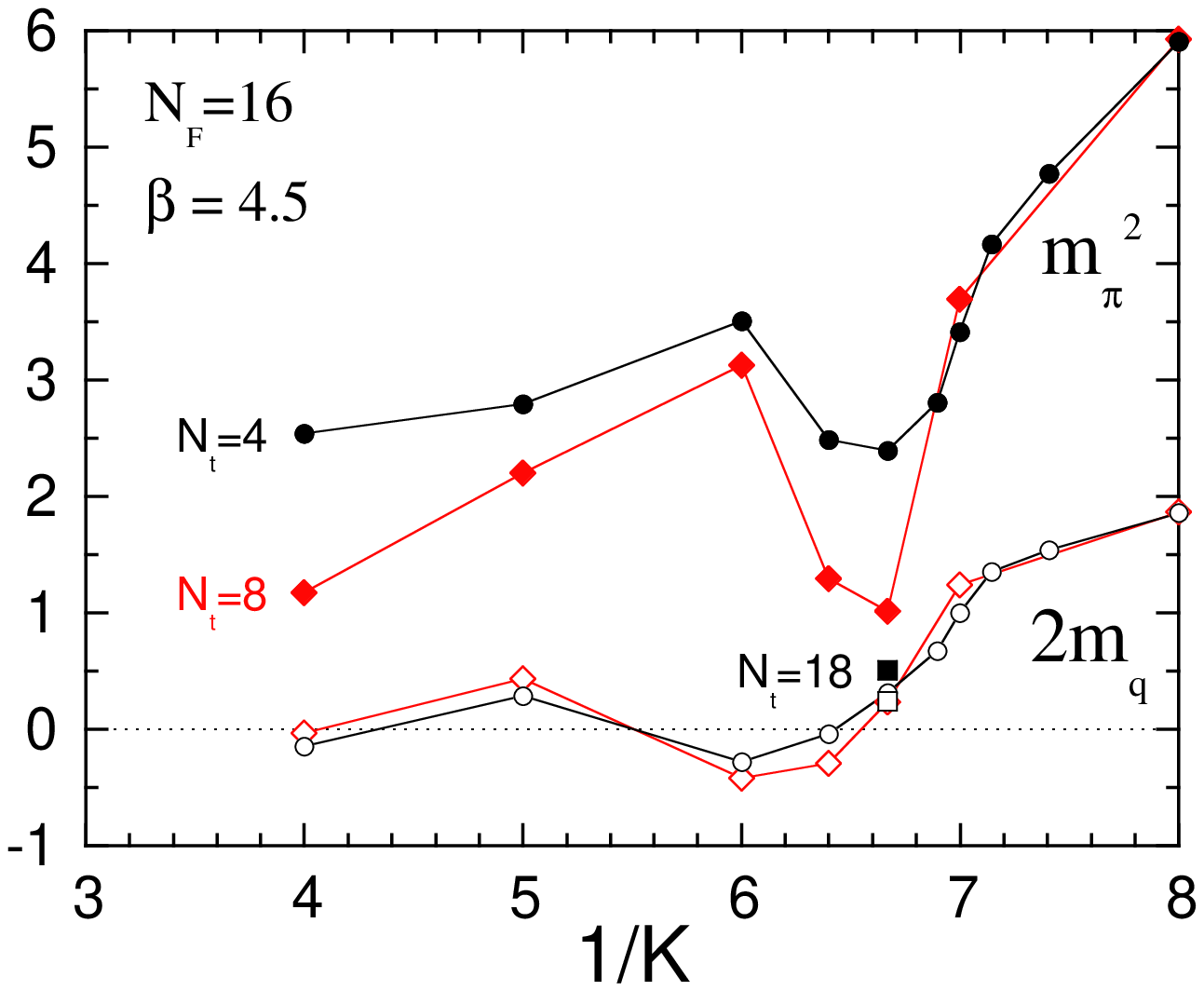}
} 
\vspace{2mm} 
\centerline{  
b) \epsfxsize=7.5cm\epsfbox{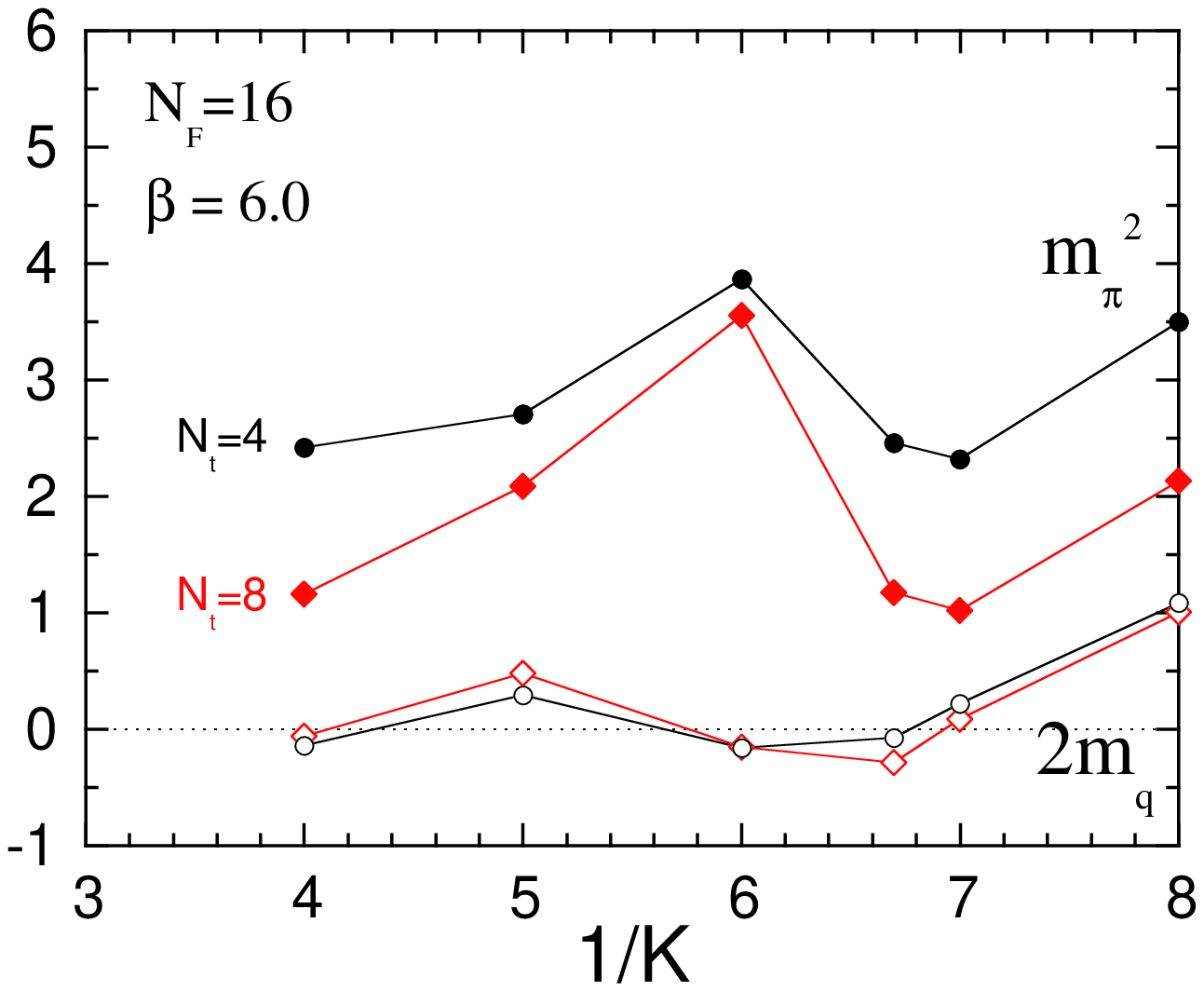}
}
\vspace{-0.1cm}
\caption{
$N_F=16$: 
Results of $m_\pi^2$ and 2$m_q$ versus $1/K$.
(a) $\beta=4.5$,
(b) $\beta=6.0$.
}
\label{Nf16}
\end{figure}

\begin{figure}[tb]
\centerline{
\epsfxsize=7.8cm\epsfbox{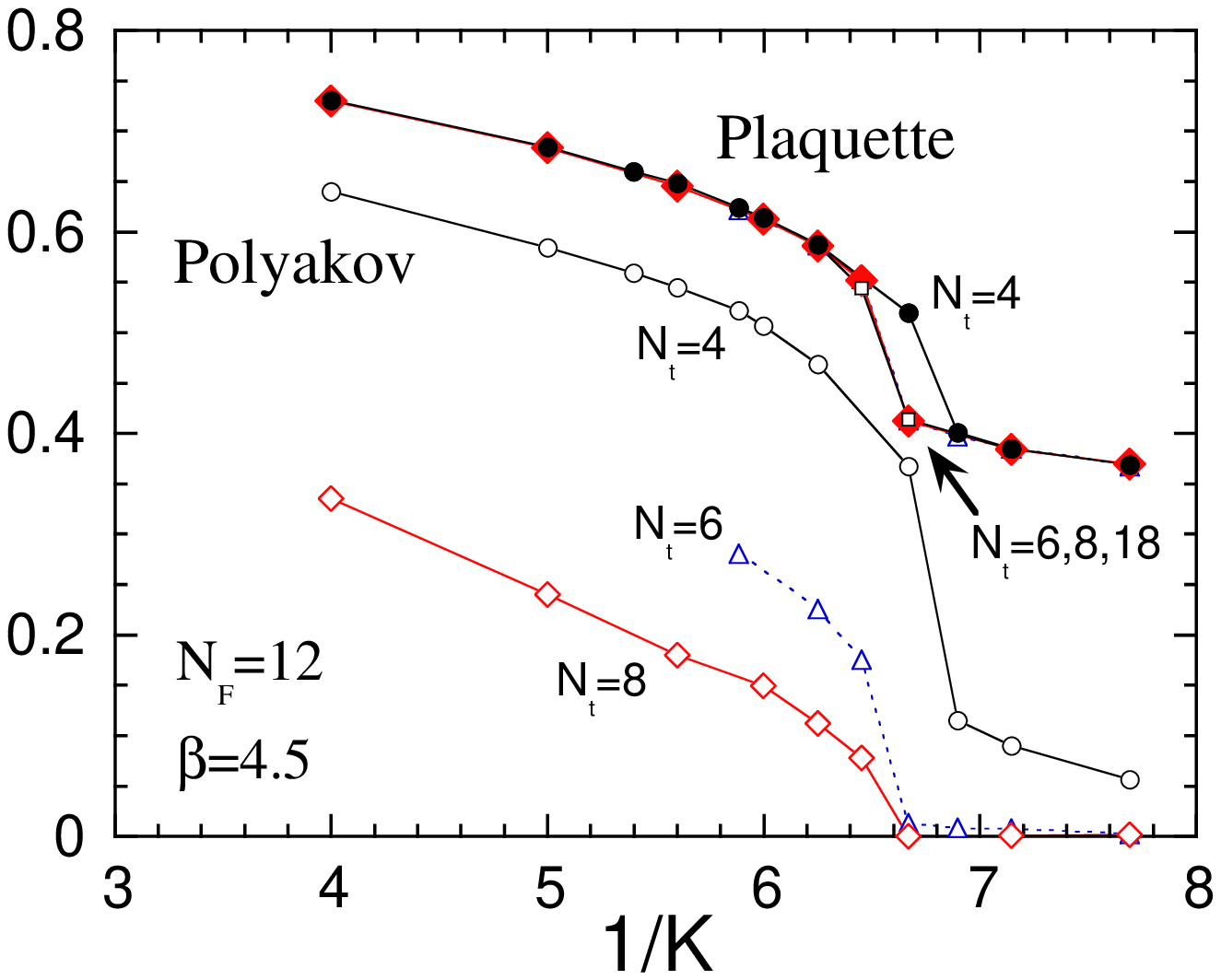}
}
\vspace{-0.1cm}
\caption{
$N_F=12$: 
Plaquette and Polyakov loop as functions of $1/K$ and $N_t$ 
at $\beta=4.5$.
}
\label{Nf12_wp}
\end{figure}

\section{Results of numerical simulations at finite $g$}
\label{sec:finite}

We now extend the study to finite $\beta$ ($g < \infty$).
Results of the plaquette and Polyakov loops at $\beta=4.5$ and 6.0 are plotted 
in Figs.~\ref{B4.5wp} and \ref{B6.0wp}, respectively, for various $N_F$.
We note that at $\beta=4.5$, both the plaquette and Polyakov loop 
shows singular behavior at some quark mass for $N_F \le 18$, in contrast 
to the cases of $N_F \ge 240$.

In order to understand the structure of the phase diagram for each 
number of flavors, from now on we discuss each $N_F$ separately:
We first intensively investigate the cases $N_F=240$
and 300, and then decrease $N_F$,
because we found that the phase structure is quite 
simple for $N_F$ $\gsim$ 240.

\subsection{$N_F=240$ and $300$}

Figure~\ref{massNf240}
shows the results of $m_\pi^2$ and $2m_q$ for $N_F=240$ and 300, obtained 
at various values of $\beta$ on the $N_t=4$ lattice.
A very striking fact is that the shape of $m_\pi^2$ and $2m_q$
as a function of $1/K$
only slightly changes in the deconfined phase,
when the value of $\beta$ decreases from $\infty$
down 0.
Only the position of the local minimum of $m_\pi^2$ at $1/K\simeq8$,
which corresponds to the vanishing point of $m_q$,
slightly shifts toward smaller $1/K$.
The results for $N_F=240$ and 300 are essentially the same, 
except for very small shifts of the transition point
and the minimum point of $m_\pi^2$.
We obtain similar results also for $N_t=8$ (see Fig.~\ref{massNf240T8}).
That is, the massless line in the deconfined phase runs through from
$\beta=\infty$ to $\beta=0$. Thus we obtain the phase diagram shown
in Fig.~\ref{Nf240}.

From the perturbation theory,
the $m_q=0$ point at $\beta=\infty$ is a trivial IR fixed point
for $N_F \ge 17$.
The phase diagram shown in Fig.~\ref{Nf240} suggests that
there are no other fixed points on the $m_q=0$ line at finite $\beta$.
In order to confirm this, we investigate
the direction of the renormalization group (RG) flow along the $m_q=0$
line for $N_F=240$,
using a Monte Carlo renormalization group (MCRG) method.

We make a block transformation for a change of scale factor 2,
and estimate the quantity $\Delta \beta = \beta(2a) - \beta(a)$:
We generate configurations
on an $8^4$ lattice on the $m_q=0$ points at $\beta=0$ and 6.0
and make twice blockings.
We also generate configurations on a $4^4$ lattice
and make once a blocking.
Then we calculate $\Delta\beta$
by matching the value of the plaquette at each step.
From a matching of our data, we obtain $\Delta \beta \simeq 6.5$ at
$\beta=0$ and $10.5$ at $\beta=6.0$.
The value obtained from the two-loop perturbation theory is
$\Delta \beta \simeq 8.8$ at $\beta = 6.0$.
The signs are the same and the magnitudes are comparable.

It is known for the pure SU(3) gauge theory
that one has to make a more careful analysis
using several types of Wilson loop with many blocking steps
to extract a precise value of $\Delta \beta$.
We reserve elaboration of this point and a fine tuning of $1/K$ at each
$\beta$ for future works.
For $N_F=240$, because the velocity of the RG flow is large,
we will be able to obtain the sign and an approximate value of
$\Delta\beta$ by a simple matching.

This result implies that the direction of the RG flow on the $m_q=0$ line
at $\beta=0$ and 6.0 is the same as that at $\beta=\infty$.
This further suggests that there are no fixed points at finite $\beta$.
All of the above imply that the theory is trivial for $N_F=240$.

\subsection{$240 > N_F \ge 17$}

Now we decrease $N_F$ from 240.
As discussed in Sec.~\ref{sec:strong},
the deconfined phase transition point $K_d$ decreases with
decreasing $N_F$ in the strong coupling limit $\beta=0$.
However, except for this shift of the bulk transition point,
the $1/K$ dependence of $m_\pi^2$ and $m_q$ are quite similar
in the deconfined phase when we vary $N_F$ from 300 down to 17,
as shown in Fig.~\ref{B0Nt4}.
The results at $\beta=6.0$ and 4.5 shown in Fig.~\ref{massNfLarge}
indicate that the $1/K$ dependence of $m_\pi^2$ and $m_q$
are almost identical to each
other, except for a small shift toward smaller $1/K$ as $N_F$ is
decreased. 
These facts imply that the structures of the deconfined phase are
essentially identical from $N_F=17$ to 300.

As we show below, a closer examination of the data shows that the massless 
quark line in the deconfined phase hits the phase transition line 
at finite $\beta$ when $N_F$ is not so large as $N_F \le 60$, 
while it runs through from $\beta=\infty$ to $\beta=0$ when $N_F$ is 
very large such as 240 and 300. 

For $N_F=18$ we make simulations at $\beta=0.0$, 2.0, 4.0, 4.5, and 6.0
on $N_t=4$ and 8 lattices, as listed in Table~\ref{tab:param18}.
Results for $m_\pi^2$ and $m_q$ are shown in Figs.~\ref{Nf18_1} 
and \ref{Nf18_2}.
We note that, at $\beta=4.5$, the $m_\pi^2$ and $m_q$ 
show characteristic behavior of massless quarks
around $1/K=6.5$ in the deconfined phase, both on $N_t=4$ and 8 lattices.
Such massless point exists also at $\beta=6.0$, but is absent at 
$\beta=0.0$ and 2.0.

At $\beta=4.0$ we carry out detailed simulations around the transition point.
On the $N_t=4$ lattice, 
the first order deconfining phase transition locates around 
$K=0.150$ ($1/K = 6.67$), 
and the massless quark point exists at $K\simeq 0.158$ ($1/K \simeq 6.33$)
in the deconfined phase. 
When we increase $N_t$ to 8, the phase transition point shifts to 
$K=0.154$--0.156 ($1/K =6.49$--6.41) where 
we observe two-state signals lasting more than 450 trajectories. 
We confirm that the points $K=0.152$ and 0.158 belong to the confined 
and deconfined phase, respectively.
In the deconfined phase the massless quark point exists at 
$K \simeq 0.155$ ($1/K \simeq 6.45$).
See Fig.~\ref{Nf18_2}(a).
From this we conclude that the massless quark line in the deconfined phase
hits the first order bulk phase transition line around $\beta=4.0$.
The phase structure for $N_F=18$ is summarized in Fig.~\ref{Nf12_18}(a).

\subsection{$16 \geq N_F \geq 7$}
\label{sec:nf167}

As stressed several times,
the quark confinement is lost for $N_F \ge 7$ at $\beta=0$ ($g=\infty$).
We now intensively simulate the cases $N_F=7$, 12 and 16 at several 
finite values of $\beta$.
The simulation parameters for $N_F=7$, 12 and 16
are listed in Tables~\ref{tab:param7}, \ref{tab:param12} 
and \ref{tab:param16}.

In Figs.~\ref{Nf7_1} and \ref{Nf7_2},
$m_\pi^2$ and $2m_q$ at selected values of $\beta$ are plotted for $N_F=7$. 
Results for $N_F=12$ and 16 are similar, as shown in Figs.~\ref{Nf12} 
and \ref{Nf16}.

For $N_F=7$, 12 and 16, we find that 
the $1/K$ dependence of $m_\pi^2$ and $2m_q$ in the deconfined phase 
are similar to those shown in Fig.~\ref{massNfLarge} for 
$N_F \geq 17$ at all the values of $\beta$. 
That is, they
are essentially the same as those of a free quark state shown in 
Fig.~\ref{Binf}.
Careful looking at the values of $m_\pi^2$ and $2m_q$
at $\beta=4.5$ for both cases reveals that the massless line hits
the phase transition line around $\beta=4.5$, at $1/K \sim 6.1$ for
$N_F=7$ and at $1/K \sim 6.5$ for $N_F=12$.

Together with the data of plaquette and Polyakov loop 
(see Fig.~\ref{Nf12_wp} as a typical example), 
we obtain the phase diagram shown in Fig.~\ref{Nf12_18}(b) for $N_F=12$.
The phase diagrams for $N_F=7$ and 16 are similar.
We find that the gross feature of these phase diagrams is quite similar 
to the case $N_F=18$ shown in Fig.~\ref{Nf12_18}(a). 

We note again that when $N_F \le 16$, the point $g=0$ is an UV fixed point.
This is in clear difference with the case of $N_F \ge 17$.

\subsection{$N_F \le 6$}

As already reported in ref.~\cite{previo}, 
numerical results show that, at $\beta=0$, the confined phase
extends to the chiral limit. 
Simulations at finite values of $\beta$ show that the phase structure
is similar for $N_F=2$--6 \cite{Stand26}.
(See Table~\ref{tab:param6} for run parameters for $N_F=6$.)
Thus we have the phase diagram shown
in Fig.~\ref{NfSmall}.

\section{Implication for physics}
\label{sec:non-degenerate}

We now extend the discussion to the cases of non-degenerate quarks.
In the case of degenerate masses, the global structure of the phase 
diagram is the same at large $\beta$ for $N_F \geq 7$. 
As far as the quark mass is below a critical value, the system is in 
the deconfined phase. 
Therefore, we expect that
our proposal for the relation of the phase structure and the
number of flavors remain unchanged for non-degenerate cases,
when we redefine $N_F$ as the number of flavors
which satisfies the condition $m_0 < \Lambda_d$:
If there exist more than seven quarks whose masses are lighter than
$\Lambda_d$, quarks should not be confined.
In nature quarks are confined, therefore the number of flavors whose masses
are less than $\Lambda_d$ should be equal or less than six.
The scale $\Lambda_d$ is numerically calculable and will
be obtained in the future.

This conclusion is based on our assumption that the deconfining
phase transition occurs at $m_0^{critical} \sim \xi^{-1}$.
Now we make a comment on the alternative possibility that the deconfining
phase transition occurs at $m_0^{critical} \sim a^{-1}$. 
If this would be correct,
the number of flavors should be equal or less than six
irrespective of their masses.
Since six species of quarks have been
discovered, this implies that there are no more species of quarks.

\section{Conclusions}
\label{sec:conclusions}

We have investigated numerically the phase structure of QCD 
for the general number of flavors. 
Performing a series of simulations for degenerate quark mass cases 
employing the one-plaquette gauge action and the standard Wilson quark
action, 
we have found that when $N_F \ge 7$
there is a line of a first order phase transition between the 
confined phase
and a deconfined phase at a finite current quark mass
in the strong coupling region and the intermediate coupling region.
The massless quark line exists only in the deconfined phase.

Based on these numerical results in the strong coupling limit and 
in the intermediate coupling region,
together with an assumption that the phase transition occurs at 
$m_0^{critical} \sim \xi^{-1}$ in the weak coupling region,
where $\xi$ is a typical correlation length of gluons,
we propose the following phase structure. 
The phase structure crucially depends on the number of flavors, $N_F$, 
whose masses are less than $\Lambda_d \sim \xi^{-1}$ which is
the physical scale which characterizes the phase transition:
When $N_F \ge 17$, there is only
a trivial IR fixed point and therefore the theory in the continuum limit
is free. On the other hand, when $16 \ge N_F \ge 7$,
there is a non-trivial IR fixed point and therefore the theory is
non-trivial with anomalous dimensions, however, without quark confinement.
Theories which satisfy both quark confinement and spontaneous
chiral symmetry breaking in the continuum limit
exist only for $N_F \le 6$.
If there exist more than seven quarks whose masses are lighter than
$\Lambda_d$, quarks should not be confined.
In nature quarks are confined, therefore the number of flavors whose masses
are less than $\Lambda_d$ should be equal or less than six.
The scale $\Lambda_d$ is numerically calculable and will
be obtained in the future.

We have discussed that, only from a theoretical argument, 
we cannot exclude an alternative possibility that the deconfining
phase transition occurs at $m_0^{critical} \sim a^{-1}$. 
This corresponds to the case $\Lambda_d =\infty$.
If this would be correct, the total number of flavors
should be equal or less than six irrespective of their masses: 
Since six species of quarks have been 
discovered, this implies that there are no more species of quarks. 
This conclusions is based on an additional assumption that
QCD only is responsible to quark confinement. 
If some other interactions would also affect the quark confinement 
at some scale $\Lambda_{new}$,
the number of quarks, whose masses are smaller than $\Lambda_{new}$,
should be six, in this case.

Which of $m_0^{critical} \sim a^{-1}$ or $m_0^{critical} \sim \xi^{-1}$ 
is realized can be investigated by numerical simulations in future.

\section*{Acknowledgement}

We thank Sinya Aoki and Akira Ukawa for valuable discussions.
This work is in part supported by Grant-in-Aids of Ministry of Education,
Science and Culture (Nos.\ 626001 and 0242003).

\section*{Appendix: SU(2) QCD}

\begin{figure}[tb]
\centerline{
a) \epsfxsize=7.5cm\epsfbox{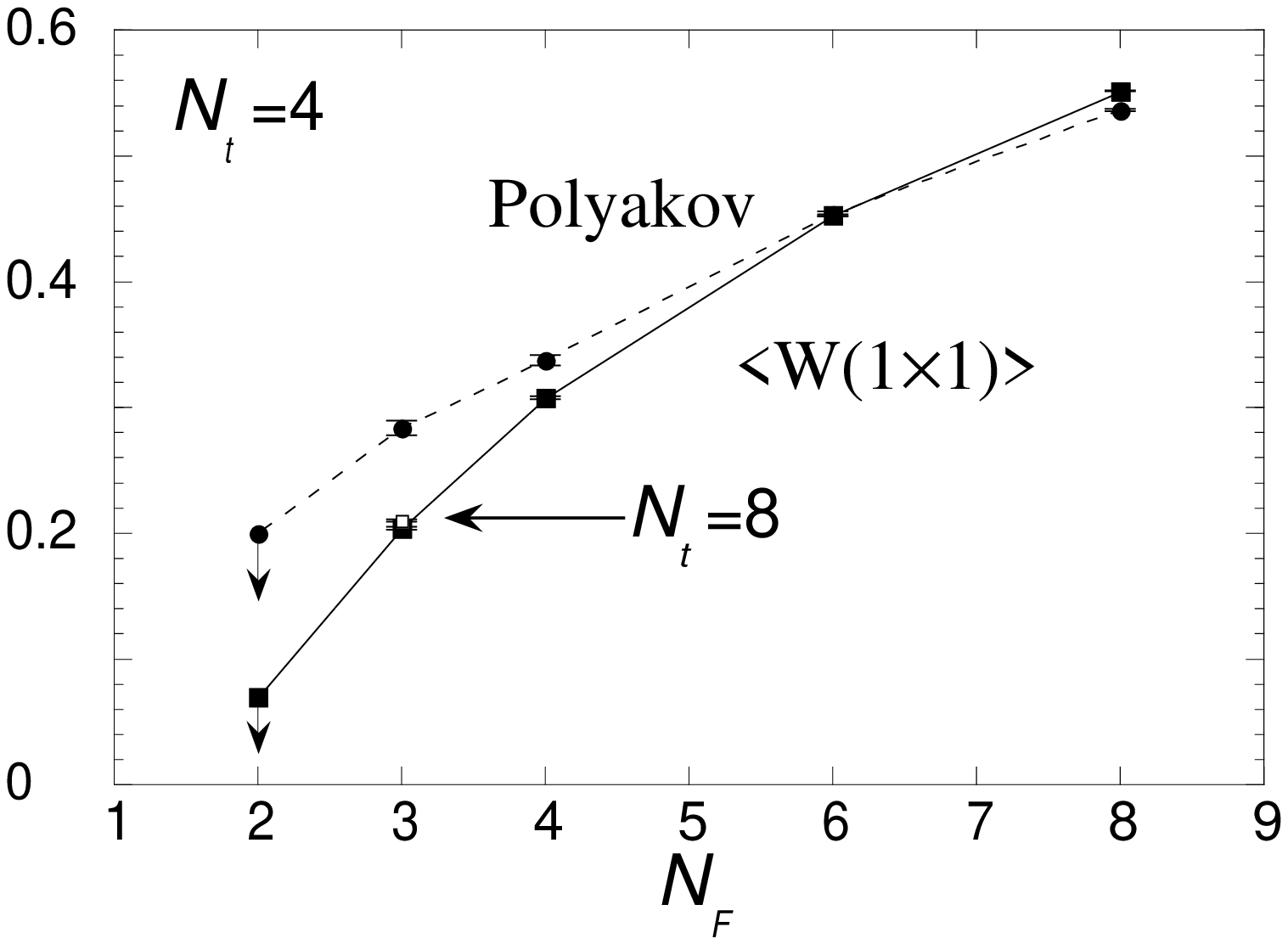}
} 
\vspace{2mm} 
\centerline{  
b) \epsfxsize=7.5cm\epsfbox{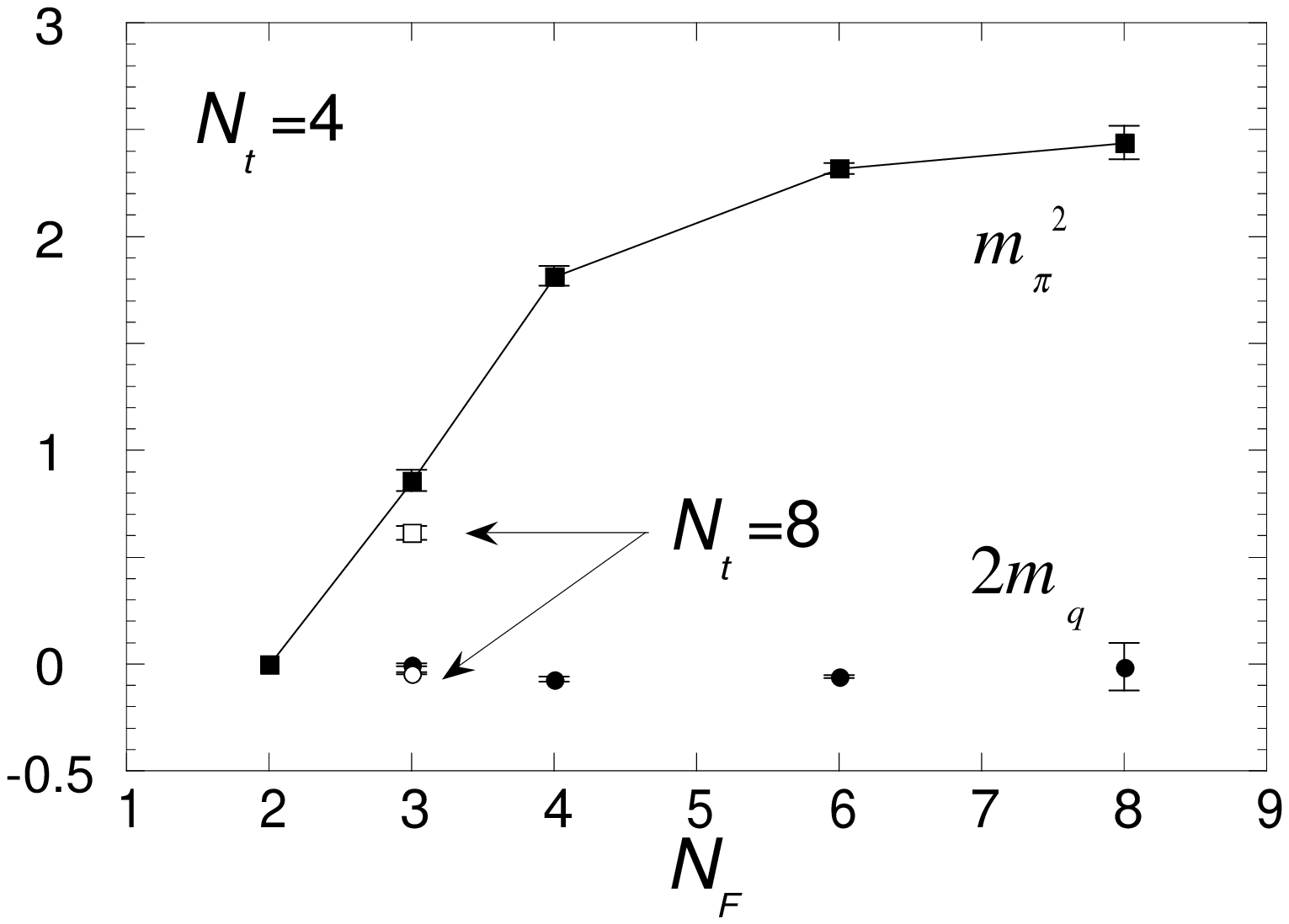}
}
\vspace{-0.1cm}
\caption{
Results of physical quantities at $g=\infty$ and $K=K_c$ for SU(2) QCD.
(a) the plaquette and Polyakov loop, and
(b) $m_\pi^2$ and 2$m_q$.
}
\label{SU2}
\end{figure}

\begin{table}
\caption{Simulation parameters for SU(2) QCD at $\beta=0.0$
performed on $8^2\times10\times N_t$ lattices with $N_t=4$ and 8.
The number of configurations, sampled every 5 trajectories, are 
10--22, except for the $N_F=2$ runs at $K=0.25$ which are not 
thermalized.}
\begin{ruledtabular}
\begin{tabular}{ccll}
$N_F$ & $N_t$ & $\kappa$ & $\Delta\tau$ \\
\hline
2 & 4  & 0.25 & 0.01, 0.005 \\
\hline
3 & 4  & 0.21, 0.22, 0.23, 0.24, 0.25 & 0.01 \\
  & 8  & 0.25 & 0.01 \\
\hline
4 & 4  & 0.25 & 0.01 \\
\hline
6 & 4  & 0.25 & 0.01 \\
\hline
8 & 4  & 0.16, 0.17, 0.18, 0.19, 0.20, \\
  &    & 0.21, 0.23, 0.25 & 0.01 \\
\end{tabular}
\label{tab:paramSU2}
\end{ruledtabular}
\end{table}

As an extension of the color SU(3) QCD case, we study color SU(2) QCD
in the strong coupling limit. We note that the SU(2) beta function
has a characteristics similar to the SU(3) one: In the case of SU(2),
the asymptotic freedom is lost when $N_F \geq 11$ and the 2-loop result
for the critical $N_F$ is 5, instead of 17 and 8, respectively, 
for SU(3) \cite{previo}. The smaller numbers for critical $N_F$ are natural
because the confining force is weaker in SU(2).

Adopting the standard plaquette gauge action and the Wilson quark action, 
we have performed a series of simulations in the strong coupling limit
following the strategy of the SU(3) case.
Our simulation parameters are compiled in Table~\ref{tab:paramSU2}.

First, performing a simulation for $N_F=8$ and $N_t=4$, we confirm that
the deconfining transition occurs at $K=0.2$--0.21 and that the 
chiral limit $K=K_c=0.25$ is in the deconfined phase. 
Decreasing $N_F$ gradually at $K=K_c$, we find that the chiral limit
remains in the deconfined phase down to $N_F=3$.
When we further decrease $N_F$ to 2, the number of the inversion $N_{inv}$
in CG iterations shows a rapid increase with molecular-dynamics time
and finally exceeds 10,000 in clear contrast with small numbers of 
$O(10^2)$ for $N_F \geq 3$.
We conclude that the chiral limit is in the confined phase for $N_F=2$.
Fig.~\ref{SU2}(a) shows our results of physical quantities versus $N_F$ at 
$K=K_c$.
To study the $N_t$ dependence of the transition, we simulate on an
$N_t=8$ lattice for the critical case $N_F=3$. The stability of the 
result, shown in Fig.~\ref{SU2}(b), confirms that the deconfining 
transition we observe is a bulk phase transition.

At finite coupling constants we may expect results similar to those
of SU(3). Thus, we conjecture in parallel to the SU(3) case
that, when $N_F \geq 11$, the theory is free. On the other hand,
when $10 \geq N_F \geq 3$, the theory is non-trivial with anomalous
dimensions without quark confinement.
Theories which satisfies both quark confinement and spontaneous 
chiral symmetry breaking exists only for $N_F \leq 2$.



\end{document}